\documentclass[useAMS,usenatbib]{mn2e}
\usepackage{amssymb}
\usepackage{graphicx}
\usepackage{lscape}
\usepackage{rotating}
\def \curlB {\vec{\nabla}\times (e^\nu \vec{B})}

\begin{document}

\title[Isolated neutron stars and magneto-thermal evolution models.]
{Unifying the observational diversity of isolated neutron stars via magneto-thermal evolution models.}
\author[D.~Vigan\`o et al.]{D.~Vigan\`o$^1$\thanks{E-mail: daniele.vigano@ua.es}, N.~Rea$^2$, J.A.~Pons$^1$, R.~Perna$^3$, 
D.N. Aguilera$^4$\thanks{Now at German Aerospace Center (DLR) - Bremen, Germany}, J.A.~Miralles$^1$\\
$^1$ Departament de F\'isica Aplicada, Universitat d'Alacant, Ap. Correus 99, 03080 Alacant, Spain\\
$^2$ Institute of Space Sciences (CSIC--IEEC), Campus UAB, Faculty of Science, Torre C5-parell, E-08193 Barcelona, Spain\\
$^3$ Department of Astrophysical and Planetary Sciences and JILA, University of Colorado, 440 UCB, Boulder, 80309, USA\\
$^4$ Laboratorio Tandar, Comisi\'on Nacional de Energ\'ia At\'omica, Av. Gral Paz 1499, 1630 San Martin, Buenos Aires, Argentina \\
}

\date{}
\maketitle

\label{firstpage}

\begin{abstract}
  
Observations of magnetars and some of the high magnetic field pulsars
have shown that their thermal luminosity is systematically higher than
that of classical radio-pulsars, thus confirming the idea that
magnetic fields are involved in their X-ray emission. Here we present the
results of 2D simulations of the fully-coupled evolution of temperature and magnetic field in neutron
stars, including the state-of-the-art kinetic coefficients and, for the first time, the important effect of the Hall term. After gathering
and thoroughly re-analysing in a consistent way all the best available
data on isolated, thermally emitting neutron stars, we compare our
theoretical models to a data sample of 40 sources.  We find that our
evolutionary models can explain the phenomenological diversity of
magnetars, high-B radio-pulsars, and isolated nearby neutron stars by
only varying their initial magnetic field, mass and envelope
composition. Nearly all sources appear to follow the expectations of the
standard theoretical models. Finally, we discuss the expected outburst rates and the evolutionary links between different
classes. Our results constitute a major step towards the grand
unification of the isolated neutron star zoo.

\end{abstract}

\begin{keywords}
stars: neutron -- stars: magnetic field -- X-rays: stars
\end{keywords}

%%%%%%%%%%
\section{Introduction}
The great advances in X-ray observations during the last several
decades have improved our understanding of the physics of neutron stars (NSs). 
Isolated NSs, as seen in X-rays, are phenomenologically quite heterogeneous. 
The high X-ray luminosities, long periods, and the occurrence of bursts and outbursts characteristic of 
Anomalous X-ray Pulsars (AXPs) and Soft Gamma Repeaters (SGRs) \citep{mereghetti08},
are interpreted as different facets of the restless dynamics of a strong magnetic field in
these (typically) young sources, in agreement with the
magnetar model \citep{thompson95}. 
Many radio-loud Rotation-Powered Pulsars (RPPs) are also detected in
X-rays \citep{becker09}. Their emission is usually powered by the conversion of a fraction of
their rotational energy into non-thermal radiation; however, in a few cases,
they also show thermal emission. In particular, high-B pulsars
\citep{ng11} appear to connect the X-ray-quiet standard radio-pulsars
with the very active magnetars, showing intermediate
luminosities, and sporadic magnetar-like activity. 
The radio-quiet X-ray Isolated NSs (XINSs) are a class of relatively old,
nearby cooling NSs, with the cleanest detected thermal emission and a
relatively large magnetic field. Last, Central Compact Objects (CCOs, \citealt{gotthelf13}) represent a  handful of puzzling, radio-quiet sources which in some cases combine a very weak external magnetic field with a relatively large
luminosity and evidence for anisotropic surface temperature distribution.

Although isolated NSs have been divided into all these
observational classes for historical reasons, a unifying vision considers them as
different manifestations of the same underlying physics
\citep{kaspi10}. In this context, one of the main theoretical tasks is
to explain the varied phenomenology of their X-ray emission.
X-ray spectra carry precious information about the surface temperature
and the physics driving the cooling of the NSs. The detected
X-ray flux, if accompanied by a reliable distance measurement, leads
to an estimate of the bolometric luminosity, which can be confronted
with cooling models to infer properties of dense matter in the
NS interior. In addition, timing properties (period and period derivative) provide
us with information about the rotational energy loss, which is
believed to be regulated mainly by the dipolar component of the
external magnetic field. 

The theory of NS cooling has been
developed since short after the first hints of X-ray emission from the
surface of NSs in the Sixties (e.g., \citealt{morton64,chiu64,tsuruta65}). During
the following decades, microphysical inputs have been refined to build
state-of-the-art 1D cooling models (see \citealt{yakovlev81,nomoto86,page90,vanriper91,page92,pethick92}, and
references within for pioneering results), but the effects of magnetic
fields were barely taken into account \citep{miralles98,page00}. \cite{page04,page09} proposed
the so-called minimal cooling scenario, in which all the necessary
known microphysical ingredients are consistently included
(specific heat, conductivities, effects of superfluidity, neutrino emission processes), but fast neutrino-cooling
processes (direct URCA by nuclear or exotic matter) are excluded. The
minimal cooling paradigm has been shown to be consistent with the luminosity of the
weakly magnetised NSs ($\lesssim 10^{13}$ G), but
magnetars and some of the high-B pulsars show a relatively large X-ray luminosity, thus requiring the presence
of additional heat sources, likely linked to magnetic field evolution.

The evolution of the magnetic field in NSs has been studied extensively by a number of authors 
\citep{goldreich92,geppert02,hollerbach02,hollerbach04,cumming04,arras04,pons07a,gonzalez10,glampedakis11}. 
In the solid crust of NSs, the magnetic field, besides being dissipated by the Joule effect, evolves due to the
Lorentz force acting on the electron fluid, the so-called Hall term. 
When this term dominates, electric currents are squeezed 
in a smaller volume and move through the crust, creating small scale structures and allowing an 
interplay between poloidal and toroidal components of the magnetic field.
The Hall term plays an important role in the evolution of large magnetic
fields, $B\gtrsim 10^{14}$ G, and the complexity of Hall dynamics has limited previous studies. 
In \cite{aguilera08b}, a 2D cooling code followed the evolution of the temperature, for a given analytical prescription of a simplified, homologous decay of the magnetic field.
In \cite{pons07b}, a spectral code follows the evolution of the magnetic field, including both the
resistive and the Hall terms, but the temperature evolution was described
by an isotropic, analytical cooling law. 
However, to describe the evolution of strongly magnetised NSs, the coupling between temperature and magnetic 
field must be taken into account. The first 2D simulations
of the fully coupled magneto-thermal evolution \citep{pons09} included only Ohmic dissipation and, only recently,
we have overcome numerical difficulties in the treatment of the important Hall term for large magnetic fields. In \cite{vigano12a} we presented the
first finite-difference, 2D magneto-thermal code able to manage
arbitrarily large B intensities while consistently treating the Hall
term. This is crucial to test whether dissipation of currents in the NS crust in realistic models  
can explain the observational data. 

In this paper, we present results for the magnetic field and temperature evolution of NSs by means of a state-of-the-art code for the computation of the 2D magneto-thermal evolution. We couple the code described in \cite{aguilera08b} for the temperature evolution with the relativistic version of the recently developed code of magnetic field evolution presented by \cite{vigano12a}. This allows us to study in detail the expected properties of objects
of any age. We apply the code to a systematic study of the magneto-thermal evolution of NSs,
revising the results from previous studies \citep{aguilera08b,pons09}
and comparing them with the observations of different classes of
isolated NSs. For a meaningful comparison of our models with the best
available data, we have re-analized the most accurate observations in the {\em
  XMM-Newton} and {\em Chandra} archives for all isolated neutron
stars (among radio pulsars, high-B pulsars, magnetars, XINSs, and CCOs)
with a detected (non-controversial) surface thermal emission.

In \S\ref{sec:observations}, we summarize the 
spectral data of the isolated NSs that we have analysed, presenting an updated table for
their timing and spectral properties. In
\S\ref{sec:theory}, we recall the basic equations and the state-of-the-art
microphysics setup. In \S\ref{sec:results}, we present the results of
the simulations, focusing on the evolution of magnetic field, surface
temperature, and luminosity. In \S\ref{sec:unification}, we compare
timing and spectral properties of a sample of magnetised sources, in
the spirit of providing a unified vision of isolated NSs. Conclusions
are presented in \S\ref{sec:conclusion}.

%%%%%%%%%%%%%%%%%  TABLE 1  %%%%%%%%%%%%%%%%%%%%%%%%%%%%%
% \begin{sidewaystable*}
\begin{table*}
\tiny
\begin{center}
\centering
\begin{tabular}{l l c c c c c c c l}
\hline
\hline
Source & assoc./nick & Class & $P$ & $\log(\dot{P})$ & $\log(\dot{E}_{rot})$ & $\log(B_p)$ & $\log(\tau_c)$ & $\log(\tau_k)$ & method\\
& & & [s] & & [erg/s] & [G] &  [yr] &  [yr] &\\
\hline
 CXOU J185238.6+004020 &      SNR         Kes79 &  CCO &  0.105 &  -17.1 &   32.5 &   10.8 &    8.3 &    3.7-3.9 &      SNR\\
        1E 1207.4-5209 &      SNR   G296.5+10.0 &  CCO &  0.424 &  -16.7 &   31.1 &   11.3 &    8.5 &    3.4-4.3 &      SNR\\
         RX J0822-4300 &      SNR       PuppisA &  CCO &  0.112 &  -17.0 &   32.4 &   10.8 &    8.3 &    3.5-3.6 &      SNR\\
  CXO J232327.9+584842 &      SNR          CasA &  CCO &     -  &     -  &     -  &     -  &     -  &        2.5 &hist./SNR\\
        PSR J0538+2817 &      SNR          S147 &  RPP &  0.143 &  -14.4 &   34.7 &   12.2 &    5.8 &  $\sim$4.6 &      SNR\\
          PSR B1055-52 &                        &  RPP &  0.197 &  -14.2 &   34.5 &   12.3 &    5.7 &         -  &          -\\
        PSR J0633+1746 &      aka       Geminga &  RPP &  0.237 &  -14.0 &   34.5 &   12.5 &    5.5 &         -  &          -\\
          PSR B1706-44 &                        &  RPP &  0.102 &  -13.0 &   36.5 &   12.8 &    4.2 &         -  &          -\\
          PSR B0833-45 &      SNR          Vela &  RPP &  0.089 &  -12.9 &   36.8 &   12.8 &    4.1 &    3.7-4.2 &      SNR\\
          PSR B0656+14 &      SNR       Monogem &  RPP &  0.385 &  -13.3 &   34.6 &   13.0 &    5.0 &  $\sim$4.9 &      SNR\\
          PSR B2334+61 &      SNR    G114.3+0.3 &  RPP &  0.495 &  -12.7 &   34.8 &   13.3 &    4.6 &  $\sim$4.0 &      SNR\\
        PSR J1740+1000 &                        &  RPP &  0.154 &  -11.7 &   37.4 &   13.6 &    3.1 &         -  &          -\\
        PSR J0726-2612 &                        &   HB &  3.440 &  -12.5 &   32.4 &   13.8 &    5.3 &         -  &          -\\
        PSR J1119-6127 &      SNR    G292.2-0.5 &   HB &  0.408 &  -11.4 &   36.4 &   13.9 &    3.2 &    3.6-3.9 &      SNR\\
        PSR J1819-1458 &                   RRAT &   HB &  4.263 &  -12.2 &   32.5 &   14.0 &    5.1 &         -  &          -\\
        PSR J1718-3718 &                        &   HB &  3.378 &  -11.8 &   33.2 &   14.2 &    4.5 &         -  &          -\\
       RX J0420.0-5022 &                        & XINS &  3.450 &  -13.6 &   31.4 &   13.3 &    6.3 &         -  &          -\\
       RX J1856.5-3754 &                        & XINS &  7.055 &  -13.5 &   30.5 &   13.5 &    6.6 &    5.5-5.7 &  pr.motion\\
       RX J2143.0+0654 &      aka       RBS1774 & XINS &  9.428 &  -13.4 &   30.3 &   13.6 &    6.6 &         -  &          -\\
       RX J0720.4-3125 &                        & XINS &  8.391 &  -13.2 &   30.7 &   13.7 &    6.3 &    5.8-6.0 &  pr.motion\\
       RX J0806.4-4123 &                        & XINS & 11.370 &  -13.3 &   30.2 &   13.7 &    6.5 &         -  &          -\\
       RX J1308.6+2127 &      aka       RBS1223 & XINS & 10.310 &  -13.0 &   30.6 &   13.8 &    6.2 &    5.9-6.1 &  pr.motion\\
       RX J1605.3+3249 &                        & XINS &     -  &     -  &     -  &     -  &     -  &  $\sim$5.7 &  pr.motion\\
           1E 2259+586 &      SNR        CTB109 &  MAG &  6.979 &  -12.3 &   31.7 &   14.1 &    5.4 &    4.0-4.3 &      SNR\\
           4U 0142+614 &                        &  MAG &  8.689 &  -11.7 &   32.1 &   14.4 &    4.8 &         -  &          -\\
  CXO J164710.2-455216$^*$ &  cluster           Wd1 &  MAG & 10.611 &  -12.0 &   31.5 &   14.3 &    5.2 &         -  &          -\\
         XTE J1810-197$^*$ &                        &  MAG &  5.540 &  -11.1 &   33.3 &   14.6 &    4.1 &         -  &          -\\
        1E 1547.0-5408$^*$ &      SNR  G327.24-0.13 &  MAG &  2.072 &  -10.3 &   35.3 &   14.8 &    2.8 &         -  &          -\\
        1E 1048.1-5937$^*$ &                        &  MAG &  6.458 &  -10.6 &   33.5 &   14.9 &    3.7 &         -  &          -\\
    CXOU J010043.1-721 &                    SMC &  MAG &  8.020 &  -10.7 &   33.2 &   14.9 &    3.8 &         -  &          -\\
 1RXS J170849.0-400910 &                        &  MAG & 11.003 &  -10.7 &   32.7 &   15.0 &    4.0 &         -  &          -\\
 CXOU J171405.7-381031$^*$ &      SNR        CTB37B &  MAG &  3.825 &  -10.2 &   34.6 &   15.0 &    3.0 &  $\sim$3.7 &      SNR\\
           1E 1841-045 &      SNR         Kes73 &  MAG & 11.782 &  -10.4 &   33.0 &   15.1 &    3.7 &    2.7-3.0 &      SNR\\
         SGR 0501+4516 &      SNR           HB9 &  MAG &  5.762 &  -11.2 &   33.1 &   14.6 &    4.2 &  $\sim$ 4  &          -\\
           SGR 1627-41 &      SNR    G337.0-0.1 &  MAG &  2.595 &  -10.7 &   34.6 &   14.7 &    3.3 &  $\sim$3.7 &      SNR\\
           SGR 0526-66 &      SNR      N49(LMC) &  MAG &  8.054 &  -10.4 &   33.5 &   15.0 &    3.5 &  $\sim$3.7 &      SNR\\
           SGR 1900+14$^*$ &  cluster               &  MAG &  5.200 &  -10.0 &   34.4 &   15.1 &    3.0 &    3.6-3.9 &  pr.motion\\
           SGR 1806-20$^*$ &  cluster       W31 &  MAG &  7.602 &   -9.6 &   34.4 &   15.5 &    2.6 &    2.8-3.0 &  pr.motion\\
         SGR 0418+5729 &                        &  MAG &  9.078 &  -14.4 &   29.3 &   13.1 &    7.6 &         -  &          -\\
    Swift J1822.3-1606$^*$ &                        &  MAG &  8.438 &  -13.1 &   30.7 &   13.7 &    6.2 &         -  &          -\\
\hline
\hline
\end{tabular}
\end{center}
\caption[Timing properties and age estimates]{Timing properties and age estimates of our sample of isolated
  NSs. Here $\dot{E}_{rot}=3.9\times 10^{46}\dot{P}/P^3$ erg/s is the rotational energy loss and
  $B_p=6.4\times 10^{19}(P\dot{P})^{1/2}$~G is the magnetic field strength at the pole, assuming that rotational
  energy losses are dominated by dipolar magnetic torques. Sources with
  multiple/variable $\dot{P}$ values in the literature are labelled
  with a $^*$. For references, see the ATNF pulsar catalog$^2$, the McGill magnetar catalog$^3$,
  and our online catalog$^1$. In the text, we denote individual sources by short names or nicknames.}
\label{tab:timing}
% \end{sidewaystable*}
\end{table*} 
%%%%%%%%%%%%%%%%%%%%%%%%%%%%%%%%%%%%%%%%%%

%%%%%%%%%%
\section{Data on cooling neutron stars}\label{sec:observations}

To confront theoretical cooling models with observational data, we need
to know simultaneously the age and some quantity related to the
thermal emission from the NS surface (luminosity or
temperature). The number of sources for which both measures are
available is limited, and in most cases subject to large
uncertainties. In this section, we
will discuss in detail the sample of selected sources (see
Table\,\ref{tab:spectral} for the complete list), which includes:

\noindent
$\bullet$ Four CCOs \citep{gotthelf13}, including the very young NS in Cassiopeia A,
and the only three CCOs with a measured value of $P$ and $\dot{P}$. The discrepancy between
the CCOs characteristic age and the estimated age of their SNRs, together with
their very slow spin-down, indicates that magnetic braking is much
less effective than in other classes, and their spin period is
very close to the natal one. We have ignored the other CCOs
candidates since they have spectral information with poor
statistics and/or a very uncertain age of the associated SNR. \\

\noindent
$\bullet$ Eight rotation powered pulsars, including the Vela pulsar
and the so-called three  Musketeers (PSR~B0656, PSR~B1055 and
the $\gamma$-ray-loud, radio-quiet Geminga; \citealt{deluca05}). We
have excluded most of the young pulsars, many of which associated with
pulsar wind nebulae (see
\citealt{becker09} for an observational review of X-ray pulsars),
since in those cases data are compatible with non-thermal emission
powered by the rotational energy loss, which is orders of magnitude larger
than their X-ray luminosity (i.e. Crab pulsar and RX
J0007.0+7303 in SNR CTA1; \citealt{caraveo10}). We also exclude several old pulsars \citep{zavlin04} with thermal emission
from a tiny hot spot (few tens of m$^2$), since the temperature of the
small hot spots is probably unrelated to the
cooling history of the NS.\\

\noindent
$\bullet$ The seven X-ray nearby Isolated NSs (XINSs) known
as {\it the magnificent seven}
\citep{haberl07,turolla09,kaplan09b}. All of them have good spectra,
and in most cases well determined timing properties and good
distance determinations (often with direct parallax
measurements). \\

\noindent
$\bullet$ Four high-B radio-pulsars (HB, see \citealt{ng11} for a
review), with inferred magnetic fields $B_{\rm p} \sim 10^{13}-10^{14}$ G and
good quality spectra. We have included the only RRAT detected so far
in X-ray (PSR~J1819, \citealt{mclaughlin07}). We have excluded
the magnetar-like pulsar PSR~J1846--0258 since during quiescence its
X-ray emission does not show a significant thermal component
\citep{ng08,livingstone11b}, and it is orders of magnitude smaller
than its rotational energy loss.\\

\noindent
$\bullet$ Seventeen magnetars (MAG; comprising both AXPs and SGRs)
have good quiescent spectra. Among the four most recently discovered
magnetars with measured timing properties, we have included the last
available observations after the outburst decay of Swift J1822
\citep{rea12,scholz12} and SGR~0418 \citep{rea10,rea13}, which
are supposedly close to quiescence. Instead, we have excluded SGR
1833--0832 and Swift J1834.9--0846, discovered during an outburst and
which have not yet been detected in quiescence. \\

We now discuss separately the timing and spectral properties of our
sample. All the data presented in the following subsections,
with links to abundant references, can also be found in our website\footnote{{\rm http://www.neutronstarcooling.info/}}. We plan to
periodically update and extend it.

%%%%%%%%%%% TABLE 2 %%%%%%%%%%%%%%%%%%%%%%%%%%
\begin{table*}
\tiny
\begin{center}
\begin{tabular}{l c c c c}
\hline
\hline
Source & Date Obs. & Obs.ID (sat.) & Exposure & Cts. \\
 & & & [ks] & [$10^3$] \\
\hline
  CXOU J185238.6+004020 &    2008-10-11 &    0550670601     (XMM) &  25.7 &     2.6\\
         1E 1207.4-5209 &    2002-08-04 &    0155960301     (XMM) &  74.6 &    90.2\\
          RX J0822-4300 &    2009-12-18 &    0606280101     (XMM) &  24.1 &    30.1\\
   CXO J232327.9+584842 &    2006-10-19 &          6690 (Chandra) &  61.7 &     9.1\\
         PSR J0538+2817 &    2002-03-08 &    0112200401     (XMM) &  10.0 &     4.3\\
           PSR B1055-52 &    2000-12-15 &    0113050201     (XMM) &  51.8 &    28.4\\
         PSR J0633+1746 &    2002-04-04 &    0111170101     (XMM) &  56.7 &    44.8\\
           PSR B1706-44 &    2002-03-13 &    0112200701     (XMM) &  28.4 &     5.1\\
           PSR B0833-45 &    2006-04-27 &    0153951401     (XMM) &  71.6 &  1150.0\\
           PSR B0656+14 &    2001-10-23 &    0112200101     (XMM) &   6.0 &    39.8\\
           PSR B2334+61 &    2004-03-12 &    0204070201     (XMM) &  26.8 &     0.3\\
         PSR J1740+1000 &    2006-09-28 &    0403570101     (XMM) &  28.6 &     2.4\\
         PSR J0726-2612 &    2011-06-15 &         12558 (Chandra) &  17.9 &     1.0\\
         PSR J1119-6127 &    2003-06-26 &    0150790101     (XMM) &  41.9 &     0.4\\
         PSR J1718-3718 &    2010-08-10 &         10766 (Chandra) &    10 &     0.0\\
         PSR J1819-1458 &    2008-03-31 &    0505240101     (XMM) &  59.2 &     6.8\\
        RX J0420.0-5022 &    2003-01-01 &    0141751001     (XMM) &  18.0 &     0.7\\
        RX J1856.5-3754 &    2011-10-05 &    0412601501     (XMM) &  18.2 &    46.7\\
        RX J2143.0+0654 &    2004-05-31 &    0201150101     (XMM) &  15.2 &    21.9\\
        RX J0720.4-3125 &    2003-05-02 &    0158360201     (XMM) &  39.7 &   105.7\\
        RX J0806.4-4123 &    2003-04-24 &    0141750501     (XMM) &  14.3 &    13.7\\
        RX J1308.6+2127 &    2003-12-30 &    0163560101     (XMM) &  17.3 &    33.0\\
        RX J1605.3+3249 &    2003-01-17 &    0671620101     (XMM) &  22.3 &    54.9\\
            1E 2259+586 &    2002-06-11 &    0038140101     (XMM) &  34.6 &   321.2\\
            4U 0142+614 &    2004-03-01 &    0206670101     (XMM) &  36.8 &  1780.0\\
   CXO J164710.2-455216 &    2006-09-16 &    0404340101     (XMM) &  40.4 &     1.7\\
          XTE J1810-197 & 2009-09-05/23 &06059902-3-401     (XMM) &  45.2 &    21.1\\
         1E 1547.0-5408 &    2004-02-08 &    0203910101     (XMM) &   6.4 &     0.6\\
         1E 1048.1-5937 &    2005-05-25 &    0147860101     (XMM) &  42.7 &   142.6\\
     CXOU J010043.1-721 &    2001-11-21 &    0018540101     (XMM) &  58.1 &     8.7\\
  1RXS J170849.0-400910 &    2003-08-28 &    0148690101     (XMM) &  31.1 &   212.5\\
  CXOU J171405.7-381031 &    2010-03-17 &    0606020101     (XMM) &  50.9 &    12.1\\
            1E 1841-045 &    2002-10-07 &    0013340201     (XMM) &   4.4 &    14.1\\
          SGR 0501+4516 &    2009-08-30 &    0604220101     (XMM) &  37.8 &    32.3\\
            SGR 1627-41 &    2008-09-25 &    0560180401     (XMM) & 105.0 &     3.1\\
            SGR 0526-66 &    2009-07-31 &         10808 (Chandra) &  28.7 &     5.1\\
            SGR 1900+14 &    2004-04-08 &    0506430101     (XMM) &  45.3 &    25.0\\
            SGR 1806-20 &    2005-10-04 &    0164561401     (XMM) &  22.8 &    27.5\\
          SGR 0418+5729 &    2012-08-25 &    0693100101     (XMM) &  63.1 &     0.5\\
     Swift J1822.3-1606 &    2012-09-08 &    0672283001     (XMM) &  20.2 &    17.4\\
\hline
\hline
\end{tabular}
\caption{Log of the observations by {\em XMM-Newton}/EPIC-pn and {\em Chandra}/ACIS used in this paper.}
\label{tab:log}
\end{center}
\end{table*}
%%%%%%%%%%%%%%%%%%%%%%%%%%%%%%%%%%%%%%%%%%%%%%%%%%%%%

%%%%%%%%%%%%%%%%%%%%%%%%%%%%%%%%%%
\subsection{On timing properties and age estimates.}

If both the spin period ($P$) and the period derivative ($\dot{P})$ of
the source are known, the characteristic age $\tau_c=P/2 \dot{P}$
can be used as an approximation to the real age, with which it
coincides only if the initial period was much shorter than the current
value and the magnetic field has been constant during the entire NS
life. Unfortunately this is not the most common situation, and
usually, for middle-aged and old objects, $\tau_c$ is found to be
larger than the real age, when the latter has been obtained by kinematic measurements (hereafter, kinematic age $\tau_k$). When the object is located in a supernova remnant, the kinematic age can be inferred by studying the expansion of the
nebula (see \citealt{allen04} for a review with particular attention to
the magnetar associations). For a few other nearby sources (e.g., some
XINSs and few magnetars; \citealt{tetzlaff11,tendulkar12}), the proper
motion, with an association to a birth place, can give the kinematic age.
We have collected the most updated and/or reliable available
information on timing properties and kinematic age from the
literature, the ATNF
catalog\footnote{http://www.atnf.csiro.au/people/pulsar/psrcat/}
\citep{manchester05}, and
the McGill online magnetar
catalog\footnote{http://www.physics.mcgill.ca/$\sim$pulsar/magnetar/main.html}.
We present in Table~\ref{tab:timing} all the sources in our sample,
with their well established associations \citep{gaensler01}, the known timing properties
with the characteristic age and the alternative estimate for the age,
when available. The inferred value of the surface, dipolar magnetic
field at the pole $B_p$, assuming the standard dipole-braking formula,
is shown as well.

We note that, for RPPs timing properties are stable over a time span
of tens of years, and $P$ and $\dot{P}$ are precisely measured, but
for magnetars the timing noise is much larger. In some cases,
different values of $\dot{P}$ have been reported, differing
even by one order of magnitude (see for instance Table 2 of the online
McGill catalog). Consequently, we should take these values with
caution, especially for the most extreme objects (largest values of
$\dot{P}$, see \S~\ref{sec:unification} for further discussion).

%%%%%%%%%%%%%%%%% TABLE 3 %%%%%%%%%%%%%%%%%%%%%%%%%
\begin{table*}
\tiny
\begin{center}
\begin{tabular}{l c c c c c c c c c c}
\hline
\hline
Source & $\log(f_X)$ & $d$  & $kT_{bb}$ & $R_{bb}$ & $kT_{nsa/rcs}$ & $R_{nsa}$ & $\log(L)$ & best fit & $kT_{cool}$ & $\log(L_{cool})$ \\
 & [erg/cm$^2$s] & [kpc] & [eV] & [km] & [eV] & [km]  & [erg/s] & model & [eV] & [erg/s] \\
\hline
 CXOU J185238.6+004020 &  -12.3 &                   7.1 &  440 &  0.9 &  290 &  3.0 &   33.5-33.7 &      BB/nsa &  < 100 &  <33.1\\
        1E 1207.4-5209 &  -11.8 & 2.1$^{+ 1.8}_{- 0.8}$ &  190 &  9.6 &  145 &  7.4 &   33.0-34.0 &    BB*/nsa* &  <  60 &  <32.2\\
         RX J0822-4300 &  -11.3 &          2.2$\pm 0.3$ &  400 &  1.7 &  204 &  6.4 &   33.5-33.7 &      BB/nsa &  <  90 &  <32.9\\
  CXO J232327.9+584842 &  -11.8 & 3.4$^{+ 0.3}_{- 0.1}$ &  450 &  1.7 &  288 &  2.7 &   33.4-33.6 &      BB/nsa &  < 110 &  <33.3\\
        PSR J0538+2817 &  -12.1 &          1.3$\pm 0.2$ &  160 &  2.6 &   -  &   -  &   32.7-32.9 &       BB+PL &  <  50 &  <31.9\\
          PSR B1055-52 &  -13.4 &         0.73$\pm0.15$ &  190 &  0.3 &   -  &   -  &   32.2-32.6 &      2BB+PL &     70 &      -\\
        PSR J0633+1746 &  -12.5 &0.25$^{+0.22}_{-0.08}$ &  140 &  0.1 &   -  &   -  &   31.6-32.5 &      2BB+PL &     42 &      -\\
          PSR B1706-44 &  -12.1 & 2.6$^{+ 0.5}_{- 0.6}$ &  160 &  3.3 &   -  &   -  &   31.7-32.1 &       BB+PL &  <  60 &  <32.2\\
          PSR B0833-45 &  -10.5 &         0.28$\pm0.02$ &  120 &  5.0 &   80 &  9.4 &   32.1-32.3 & (BB/nsa)+PL &  <  40 &  <31.5\\
          PSR B0656+14 &  -12.6 &         0.28$\pm0.03$ &  100 &  2.4 &   -  &   -  &   32.7-32.8 &      2BB+PL &     50 &      -\\
          PSR B2334+61 &  -14.0 & 3.1$^{+ 0.2}_{- 2.4}$ &  160 &  1.1 &   86 &  7.9 &   30.7-32.1 &      BB/nsa &  <  50 &  <31.9\\
        PSR J1740+1000 &  -13.8 &                   1.4 &  170 &  0.4 &   68 &  7.8 &   32.1-32.2 &     2BB/nsa &     78 &      -\\
        PSR J0726-2612 &  -14.0 &                   1.0 &   90 &  4.6 &   -  &   -  &   32.1-32.5 &          BB &  <  40 &  <31.5\\
        PSR J1119-6127 &  -13.0 &          8.4$\pm 0.4$ &  270 &  1.5 &   -  &   -  &   33.1-33.4 &          BB &  < 120 &  <32.9\\
        PSR J1819-1458 &  -12.6 &                   3.6 &  130 & 12.3 &   -  &   -  &   33.6-33.9 &          BB &     -  &      -\\
        PSR J1718-3718 &  -13.2 & 4.5$^{+ 5.5}_{- 0.0}$ &  190 &  2.0 &   -  &   -  &   32.8-33.5 &          BB &  <  90 &  <32.9\\
       RX J0420.0-5022 &  -17.8 &                  0.34 &   50 &  3.4 &   -  &   -  &   30.9-31.0 &          BB &     -  &      -\\
       RX J1856.5-3754 &  -14.4 &         0.12$\pm0.01$ &   63 &  4.1 &   -  &   -  &   31.5-31.7 &          BB &     -  &      -\\
       RX J2143.0+0654 &  -13.1 &                  0.43 &  107 &  2.3 &   -  &   -  &   31.8-31.9 &          BB &     -  &      -\\
       RX J0720.4-3125 &  -13.3 &0.29$^{+0.03}_{-0.02}$ &   84 &  5.7 &   -  &   -  &   32.2-32.4 &          BB &     -  &      -\\
       RX J0806.4-4123 &  -13.4 &                  0.25 &  101 &  1.2 &   54 &  8.2 &   31.2-31.4 &    BB*/nsa* &     -  &      -\\
       RX J1308.6+2127 &  -12.1 &                  0.50 &   94 &  5.0 &   -  &   -  &   32.1-32.2 &         BB* &     -  &      -\\
       RX J1605.3+3249 &  -13.0 &                  0.10 &   99 &  0.9 &   56 &  5.3 &   30.9-31.0 &      BB/nsa &     -  &      -\\
           1E 2259+586$^\dagger$ &  -10.3 &          3.2$\pm 0.2$ &  400 &  2.9 &  120 &   -  &   35.0-35.4 &      RCS+PL &  < 120 &  <33.4\\
           4U 0142+614$^\dagger$ &   -9.8 &          3.6$\pm 0.5$ &  400 &  6.5 &  290 &   -  &   35.4-35.8 &      RCS+PL &  < 150 &  <33.8\\
  CXO J164710.2-455216 &  -12.2 & 4.0$^{+ 1.5}_{- 1.0}$ &  330 &  0.6 &  150 &   -  &   33.1-33.6 &         RCS &  < 120 &  <33.4\\
         XTE J1810-197 &  -11.7 &          3.6$\pm 0.5$ &  260 &  1.9 &   -  &   -  &   34.0-34.4 &         2BB &    116 &      -\\
        1E 1547.0-5408 &  -11.5 &          4.5$\pm 0.5$ &  520 &  0.3 &  100 &   -  &   34.3-34.7 &         RCS &  < 150 &  <33.8\\
        1E 1048.1-5937 &  -10.8 &          2.7$\pm 1.0$ &  640 &  0.6 &  370 &   -  &   33.8-34.5 &         RCS &  < 100 &  <33.1\\
    CXOU J010043.1-721 &  -12.5 &         60.6$\pm 3.8$ &  350 &  9.2 &  300 &   -  &   35.2-35.5 &         RCS &     -  &      -\\
 1RXS J170849.0-400910$^\dagger$ &  -10.4 &          3.8$\pm 0.5$ &  450 &  2.1 &  320 &   -  &   34.8-35.1 &      RCS+PL &  < 130 &  <33.6\\
 CXOU J171405.7-381031 &  -11.4 &         13.2$\pm 0.2$ &  540 &  1.6 &  340 &   -  &   34.9-35.2 &         RCS &  < 180 &  <34.1\\
           1E 1841-045$^\dagger$ &  -10.4 & 9.6$^{+ 0.6}_{- 1.4}$ &  480 &  5.0 &  270 &   -  &   35.2-35.5 &      RCS+PL &  < 200 &  <34.3\\
         SGR 0501+4516 &  -11.3 & 1.5$^{+ 1.0}_{- 0.5}$ &  570 &  0.2 &  110 &   -  &   33.2-34.0 &         RCS &  < 100 &  <32.9\\
           SGR 1627-41 &  -11.6 &         11.0$\pm 0.2$ &  450 &  2.0 &  280 &   -  &   34.4-34.8 &      RCS+PL &  < 300 &  <34.9\\
           SGR 0526-66 &  -12.0 &         49.7$\pm 1.5$ &  480 &  3.6 &  320 &   -  &   35.4-35.8 &         RCS &  < 200 &  <34.3\\
           SGR 1900+14$^\dagger$ &  -11.1 &         12.5$\pm 1.7$ &  390 &  4.4 &  330 &   -  &   35.0-35.4 &      RCS+PL &  < 150 &  <33.8\\
           SGR 1806-20$^\dagger$ &  -10.6 &13.0$^{+ 4.0}_{- 3.0}$ &  690 &  2.0 &  390 &   -  &   35.1-35.5 &      RCS+PL &  < 250 &  <34.7\\
         SGR 0418+5729$^\circ$ &  -14.0 &                   2.0 &  320 &  0.1 &   -  &   -  &   30.7-31.1 &          BB &  <  40 &  <31.5\\
    Swift J1822.3-1606$^\circ$ &  -11.5 &          1.6$\pm 0.3$ &  540 &  0.2 &  300 &   -  &   32.9-33.2 &         RCS &  <  70 &  <32.5\\
\hline
\hline
\end{tabular}
\end{center}
$^\circ$ The source has been recently discovered in outburst and it
could have not reached the quiescence yet at the time of the latest
available observation.\\
$^\star$ Absorption line(s) {\tt gabs} included in the fit.\\
$^\dagger$ Hard tail detected in quiescence hard X-ray spectra
($E\gtrsim 20$ keV).
\caption{Emission properties of the thermally emitting neutron
  stars. $f_X$ is the 1--10\,keV band unabsorbed flux
  according to the indicated best fit
  model (the {\tt bbodyrad} model flux is chosen whenever multiple
  models are equally compatible with the data). $kT_{bb}$ and $R_{bb}$
are the temperature and radius inferred by the {\tt bbodyrad} model.
$kT_{nsa}$ is the
  temperature inferred by the {\tt nsa} model for the weakly
  magnetised cases ($B\lesssim 5\times 10^{13}$ G) with acceptable
  associated radius $R_{nsa}$, also indicated. $kT_{rcs}$ is the
  temperature inferred by the {\tt RCS} model for strongly magnetised
  sources. $L$ is the bolometric luminosity from the
  thermal component(s) of the fit, and assuming the indicated distance
  (whose references are listed in the online table). The range of $L$
  includes both statistical and distance errors; for strongly absorbed
  sources (i.e., most of magnetars) a minimum arbitrary factor of
  $50\%$ uncertainty is assumed to account for systematical
  model-dependent uncertainty. $kT_{cool}$ is either the lower temperature
  for models including 2 BB, compatible with emission from the entire
  surface, or the upper limit for cases showing emission from a small spot
  $R_{bb}\sim$ few km. In the latter case, $L_{cool}$ is the associated upper
  limit to the ``hidden`` thermal luminosity. See text for
  details on the spectral models. All radii, temperatures and luminosities are the values as measured by a distant observer.
}
\label{tab:spectral}
\end{table*}

%%%%%%%%%%%%%%%%%%%%%%%%%%%%%%%%%%%%%%%%%%%%%%%%%%%%%%%%%%%%%%

\subsection{On luminosities and temperatures from spectral data analysis.}

Luminosities and temperatures can be obtained by spectral analysis,
but it is usually difficult to determine them accurately. The
luminosity is always subject to the uncertainty in the distance
measurement, while the inferred effective temperature depends on the choice
of the emission model (blackbody vs. atmosphere models, composition,
condensed surface, etc.), and it carries large theoretical uncertainties in
the case of strong magnetic fields.  We often find that more than one
model can fit equally well the data, without any clear, physically
motivated preference for one of them, with inferred effective
temperatures differing up to a factor of two.  Photoelectric
absorption from interstellar medium further constitutes a source of
error in temperature measurements, since the value of the hydrogen
column density $N_H$ is correlated to the temperature value obtained
in spectral fits. Different choices for the absorption model and the
metal abundances can also yield different results for the temperature.
In addition, in the very common case of the presence of inhomogeneous
surface temperature distributions, only an approximation with two or
three regions at different temperatures is usually employed.

Last, in the case of data with few photons and/or strong absorption features,
the temperature is poorly constrained by the fit, adding a large
statistical error to the systematic one. For all of these reasons, the
temperatures inferred by spectral fits can hardly be directly
compared to the {\it physical surface temperatures} extracted from
cooling codes.

Because of the above considerations, the luminosity constitutes a
better choice to compare data and theoretical models.  Since it is an
integrated quantity, it averages effects
of anisotropy and the choice of spectral model. The main uncertainty on
the luminosity is often due to the poor knowledge of the source
distance. In the worst cases, the distance is known within an error of
a few, resulting in up to one order of magnitude of uncertainty in the
luminosity. In addition, the interstellar absorption acts
predominantly in the energy band in which most of the middle age NSs
emit ($E\lesssim 1$ keV). For this reason, hottest
(magnetars) or closest (XINSs) sources are easier to
detect. Similarly to the case of the temperature, the choice of
different models of absorption and chemical abundances can yield
additional systematic errors on the luminosity. However, for the worst
cases, the relative error is about $30\%$, making it usually a
secondary source of error compared with the distance.

\subsection{Data reduction}

In order to control the systematic errors in the luminosity and spectral
parameters, we have homogeneously re-analysed all the data, extracted
directly from the best available observations from {\em Chandra} or
{\em XMM--Newton}.  In Table~\ref{tab:log} we list the log of all the
observations we used.
These observations have been selected with the following criteria.
\begin{enumerate}
 \item For comparable exposure times of a given source, we always preferred
the {\em XMM--Newton} observation, given the larger collecting area
with respect to {\em Chandra}, hence resulting in a more detailed
spectrum. Instead we used {\em Chandra} data for sources which have bright
nebulae.
 \item For variable sources we took the longest available observation
during the quiescent state of the NS.
 \item We excluded those objects which were fitted
equally well without the addition of a thermal component, hence the
latter not being statistically significant in the data.
\end{enumerate}

We processed all {\em XMM--Newton}  observations \citep{jansen01}
listed in Table~\ref{tab:log} using SAS version 11, and
employing the most updated calibration files available at the time the
reduction was performed (November 2012). Standard data screening
criteria are applied in the extraction of scientific products, and we
used only the EPIC-pn camera data. We extracted the source
photons from a circular region of 30\arcsec radius, and a region of
the same size was chosen for the background in the same CCD but as far
as possible from the source position.\footnote{The only exception is
PSR~B1055, for which the observation was performed in timing mode, and
the extraction region is a box.} We restricted our analysis to photons
having PATTERN$\leq$4 and FLAG=0.

The {\em Chandra} data we used in this work were all taken with the
Advanced CCD Imaging Spectrometer (ACIS-S; \cite{garmire03}. Data were
analyzed using standard cleaning
procedures\footnote{http://asc.harvard.edu/ciao/threads/index.html}
and {\tt CIAO} version 4.4. Photons were extracted from a circular
region with a radius of 3\arcsec around the source position, including
more than 90\% of the source photons, and background was extracted
from a region of the same size, far from the source position.

\subsection{Data analysis}

All spectra were grouped in order to optimize the signal to noise in
each spectral bin, but even for the
fainter objects we grouped the spectra in order to have at least 25
counts per spectral bin, and reliably use the chi-squared statistic to
assess the goodness of the fits. The response matrices were built
using ad-hoc bad-pixel files built for each observation.

For the spectral analysis we used the {\tt XSPEC} package (version
12.4) for all fittings, and the {\tt phabs} photoelectric absorption
model with the \cite{anders89} abundances, and the \cite{balucinska92}
photoelectric cross-sections. We usually restricted our spectral
modeling to the 0.3--10\,keV energy band, unless the source was such
that a smaller energy range was needed for the spectral analysis. We
always excluded bad channels when needed.

Depending on the source, we either used a blackbody model ({\tt bbodyrad}) alone, or added a power law ({\tt power}) and/or a second
blackbody component if statistically significant. When physically
motivated, we also fitted data with an NS atmosphere model
({\tt nsa}: \citealt{zavlin96,pavlov95}), always fixing the NS mass to
$M=1.4\,M_\odot$, the radius to $R=10$ km and $B$ to the closest value
to the inferred surface value $B_p$, see Table~\ref{tab:spectral}.

We find that for CCOs the atmosphere models provide slightly better
fits and radii closer to the typical NS values.
In the case of RPPs, they can all be described by thermal
component(s), with an additional power law. In particular, the three
Musketeers (Geminga, PSR~B0656 and PSR~B1055) are well fit by a double blackbody plus a power law. An
atmosphere model (plus a power law) fits Vela better than a blackbody
plus power law, and it is compatible with emission from the entire
surface. However, in several other cases atmosphere models provide
unphysical emitting radii. On the other hand, XINSs X-ray spectra are well fit 
by a single blackbody model, with absorption features in a few cases; 
in a couple of cases an atmosphere model can fit the data as well.

For magnetars we have also used a resonant Compton scattering model ({\tt
RCS}: \citealt{rea08,lyutikov06}), adding a power law component when needed. The lack of a unifying description of the
magnetosphere, consistently dealing with
both magnetic field configuration and the complicated interaction
between magnetospheric plasma and photons, has led to simplified models
for the resonant Compton scattering processes. The {\tt RCS} model is
a plane-parallel model, where the physical parameters are the surface
temperature, the velocity of the electron gas and the optical
depth. Successively, \cite{nobili08a,nobili08b} have taken into
account RCS processes in a mathematically consistent, globally twisted
magntospheric configuration \citep{thompson02}. In this case, the
physical parameters are the homogeneous surface temperature, a global
twist of the magnetosphere (which provides different intensity of
currents) and the bulk electron velocity, assumed constant through the
magnetosphere. Note that neither the last assumption, nor the globally
twisted geometry are physically motivated. Other configurations,
qualitetively consisting of a confined bundle of currents with
velocity of electrons decreasing outward, are expected
\citep{beloborodov09}. \cite{beloborodov13} progressed further,
presenting the state-of-the-art
models for a self-consistent magnetospheric solution.
The main practical differences between these models are the physical
parameters inferred from the fits (twist, temperature, viewing and
magnetic angles), due to different hypotheses and approximations used.
However, note that we are interested in the unabsorbed flux, which is
an integrated, almost model-independent quantity. For this reason, we
use the public {\tt RCS} model, implemented in the {\tt XSPEC}
package, to account for the soft X-ray tail.
In the {\tt RCS} model, the seed photons originate from the
surface. However, part of the energy of the detected flux ultimately
comes from the magnetospheric plasma, sustained by the large magnetic
energy and dominated by pair-production. How to disentangle the
surface and magnetospheric components is an open question. Lacking
any better choice, we assume a thermal origin for the flux
estimated with the {\tt RCS} model, discarding the flux of a second power
law component, should the latter be needed to reproduce the spectra at
very high energy.

Since our spectral modeling is aimed at a more reliable and
homogeneous comparison with
the theoretical models, we report in Table~\ref{tab:spectral} both
the total 1--10\,keV unabsorbed flux, and the bolometric luminosity
(L) of the thermal component(s) of the spectral fit. Note that, for
rotational powered pulsars for which a strong non-thermal component is
present, the luminosity we quote includes the blackbody component(s) only.
Another quite common feature is the small size of the emitting region,
typically a few km.
Since spectra of most sources are strongly absorbed, this would hide
possible cool components from an extended region.
For this reason, we  also report in Table~\ref{tab:spectral}
the maximum temperature, and associated luminosity, that a $\approx
10$ km radius NS would
have while still being compatible with the lack of detection. These
estimates rely on the particular spectral model we have chosen, and on
the distance. They are indicative of the amount of possible hidden
flux. For some strongly absorbed magnetars, this contribution could be
in principle of the same order of magnitude as the flux detected from
the visible hot spot (see column $L_{cool}$ in the table).

Last, note that luminosities derived for the whole sample of objects
span about five
orders of magnitude, making the relative errors on luminosity
much less substantial than those on the temperature. This further
justifies our choice of taking into account luminosity instead of
temperature to compare
our cooling models to observations.

%%%%%%%%%%
\section{Magneto-thermal evolution.}\label{sec:theory}

Building on previous works \citep{page04,pons07b,aguilera08b,page09,pons09} we have updated and extended
our magneto-thermal evolution code in two major ways: the proper treatment of the important Hall term in the
induction equation describing the magnetic field evolution \citep{vigano12a}, and updated microphysics inputs.
This allows us to follow the long-term evolution of magnetized NSs, a necessary step to understand 
timing and spectral properties of isolated NSs at different ages.  In this section,
we briefly summarize the equations, the method, and the updated ingredients of the simulations.

%%%%%%%%%%
\subsection{Basic equations.}

For our purposes, the small structural deformations induced by rotation and the magnetic field can be safely neglected.
To include general relativistic effects, we consider the standard static metric
\begin{equation}\label{eq:metric}
 ds^2 = - c^2 e^{2\nu(r)}dt^2 + e^{2\lambda(r)}dr^2 + r^2 d\Omega^2 ,
\end{equation}
where $e^\nu(r)$ is the lapse function that accounts for redshift corrections, $e^{\lambda(r)} = (1 - 2Gm(r)/c^2r)^{-1/2}$ is the space curvature factor, and $m(r)$ is the enclosed gravitational mass within radius $r$.

The star thermal evolution is described by the following energy balance equation \begin{equation}
\label{eq:heat_balance}
c_v e^\nu\frac{\partial T}{\partial t} - \vec{\nabla}\cdot[e^\nu \hat{\kappa}\cdot\vec{\nabla}(e^\nu T)] = e^{2\nu}(-{\cal Q}_\nu + {\cal Q}_h)
\end{equation}
where $c_v$ is the volumetric heat capacity, $\hat{\kappa}$ is the thermal conductivity tensor, ${\cal Q}_\nu$ are the
energy loss by neutrino emission per unit volume, and ${\cal Q}_h$ is the Joule heating rate per unit volume.

The conduction of heat becomes anisotropic under the presence of a strong magnetic field, which also controls the heating rate and, secondarily, affects the rate of a few neutrino processes \citep{aguilera08b}. This is the reason why, as shown in previous studies, the thermal evolution is coupled to the magnetic field evolution. In the crust, ions form a Coulomb lattice, while electrons are relativistic, degenerate and can almost freely flow, providing the currents that sustain the magnetic field. The evolution of the system is governed by the Hall induction equation which, using the same notation as in \cite{pons09}, has
the form:
\begin{equation}\label{eq:induction}
\frac{\partial \vec{B}}{\partial t} = - \vec{\nabla}\times \left[ \frac{c^2}{4\pi \sigma} \curlB + \frac{c }{4\pi e n_e} \left[ \curlB \right] 
\times \vec{B}   \right]~
\end{equation}
where the conductivity $\sigma$ takes into account all the (strongly temperature-dependent) electron processes. The first term on the right
hand side accounts for Ohmic dissipation, while the second term is the Hall effect. The magnetization parameter $\omega_B\tau_e\equiv \frac{\sigma B}{c e n_e}$ (where $\omega_B=eB/m^*_ec$ is the gyration frequency of electrons, with $\tau_e$ and $m^*_e$ are the relaxation time and effective mass of electrons), affects both the evolution of magnetic field and temperature. In the regime where $\omega_B\tau_e\gg 1$ (strong magnetic fields, $\gtrsim 10^{14}$ G, and temperatures $\lesssim 5\times 10^8$ K, see \citealt{pons07a,aguilera08b,pons09,vigano12a} for more details), the conductivity tensor $\hat{\kappa}$ becomes very anisotropic, the Hall term dominates, and the induction equation acquires a hyperbolic
character. The Ohmic and Hall timescales vary by orders of magnitude
within the crust and during the evolution, depending strongly on
density, temperature, and magnetic field intensity and curvature.

%%%%%%%%%%%%%%%%%%%%%%%%%%%%%%%%%%%%%%%%%%
\begin{figure*}
 \centering
\includegraphics[width=.5\textwidth]{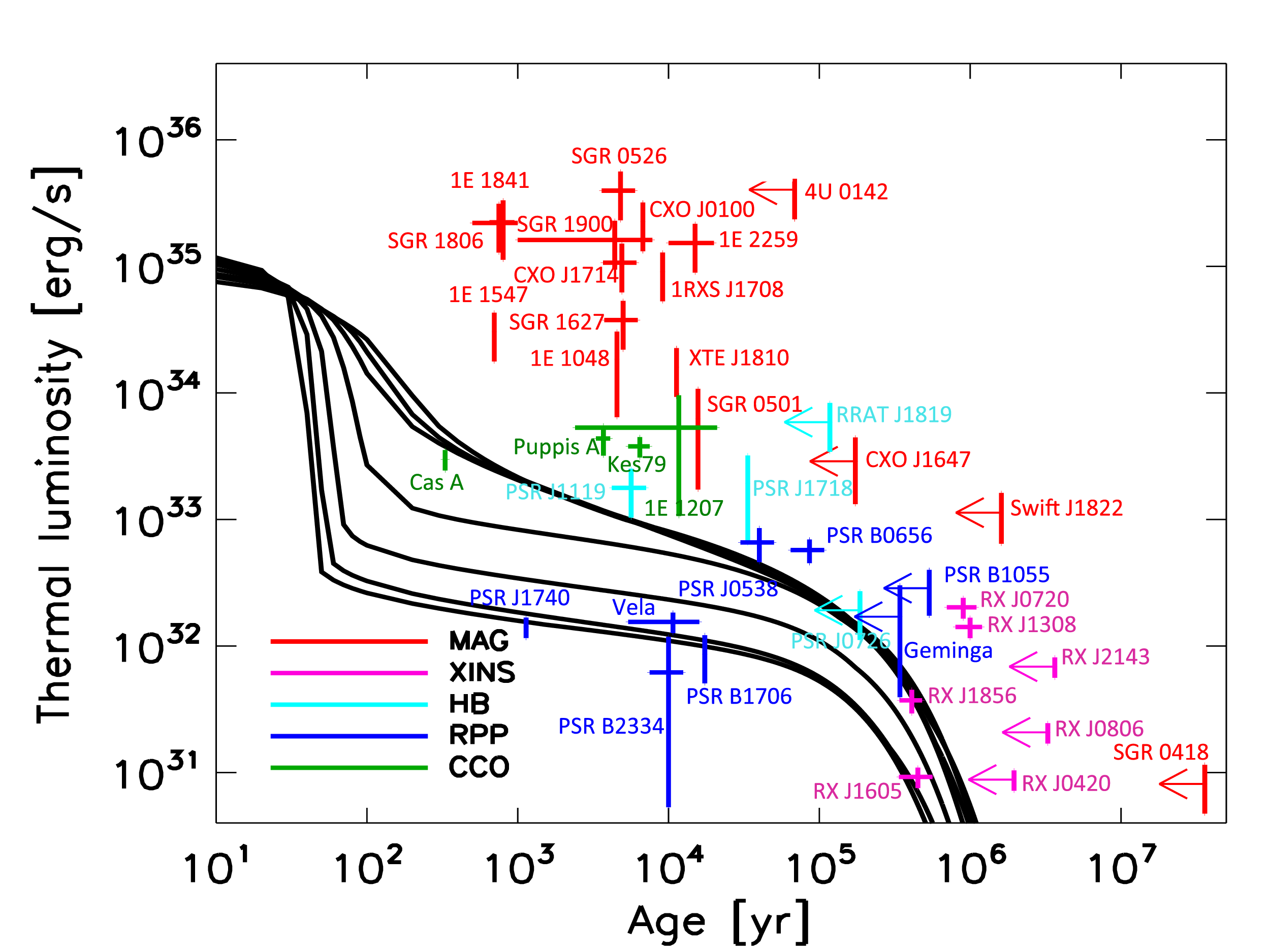}
\hspace{-0.5cm}
\includegraphics[width=.5\textwidth]{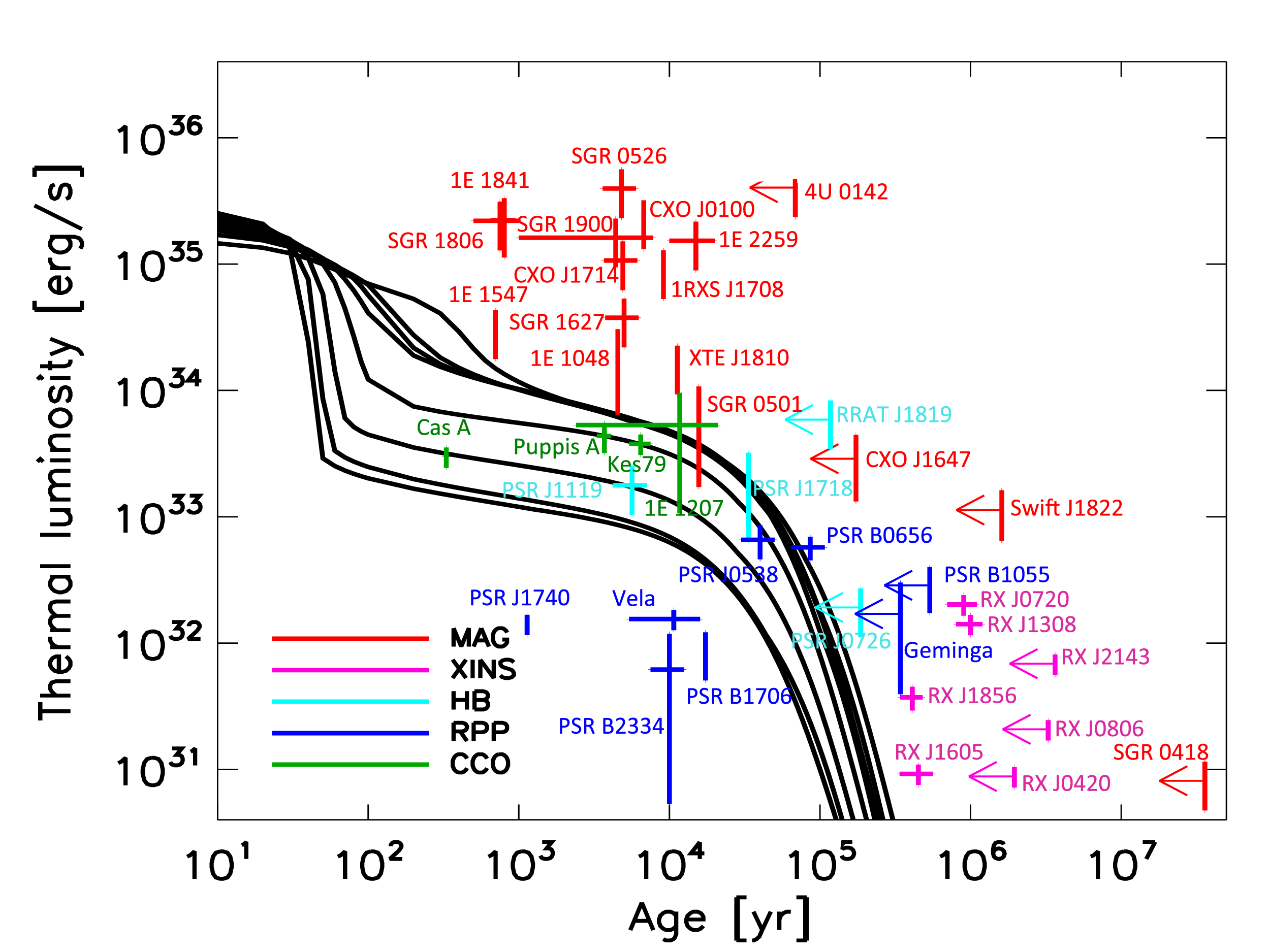}
\caption{Luminosity vs. age for $B=0$ NS models. We show cooling curves for 
8 masses (lines from top to bottom 1.10, 1.25, 1.32, 1.40, 1.48, 1.60, 1.70 and 1.76 $M_\odot$) compared with
data presented in Tables \ref{tab:timing} and \ref{tab:spectral}. 
The left panel corresponds to models with iron envelopes and the right panel to models with light-element envelopes. 
Arrows label sources for which $\tau_c\gtrsim 50$ kyr, and no kinematic age is available, so that the real age is expected to be shorter. 
An uncertainty of $50\%$ has been arbitrarily taken for the kinematic age when error estimates have not been found in the literature.}
 \label{fig:b0}
\end{figure*}
%%%%%%%%%%%%%%%%%%%%%%%%%%%%%%%%%%%%%%%%%%

To overcome the limitations of previous hybrid methods (spectral in
angles, finite-differences in the radial direction) used in
\cite{pons07b} and \cite{pons09}, we use a new finite-difference code
described in \cite{vigano12a}. The main effect of the Hall term is to
transfer part of the magnetic energy from large to small scales, and
between poloidal and toroidal components.  In the case of strong
toroidal components, it also leads to the formation of discontinuities
of the tangential components of the magnetic field, i.e. current
sheets, where the dissipation is strongly enhanced
\citep{vigano12a}. This directly affects the thermal evolution through
the term ${\cal Q}_h$ in Eq.~(\ref{eq:heat_balance}).

The core of NSs (or at least a fraction of its volume) is thought to be a type II superconductor. The dynamics of the magnetic field in the core is not clearly understood. Here, Ohmic dissipation acts over long timescales ($\gtrsim$ Gyr) due to the very high conductivity of nuclear matter, but other mechanisms such as the interplay between flux tubes and vortices, magnetic buoyancy, or ambipolar diffusion may operate to expel magnetic flux from the core on timescales shorter or comparable to the cooling timescale ($\lesssim 10^5$) yr. The detailed study of these mechanisms goes beyond the purpose of this paper, in which we mostly consider models with the magnetic field confined into the crust or, for the models with magnetic fields permeating the core, we only include Ohmic dissipation. This implies that the core magnetic field is basically frozen during cooling timescales.

%%%%%%%%%%
\subsection{Microphysics.}

The microphysical inputs that enter in the magneto-thermal evolution
equations, Eqs.~(\ref{eq:heat_balance}, \ref{eq:induction}), are the
thermal conductivity, the specific heat, the neutrino emissivities,
and the electrical conductivity. For a complete review of the physics of the crust,
with the thermal and transport properties, see \cite{chamel08,page12}.

Unless it is explicitly mentioned, we use the same
inputs as in \cite{aguilera08b} and \cite{pons09}, and references therein. In our
current version of the code, we have updated some of these ingredients
as follows. For the specific heat, we use the latest version of the
equation of state for a strongly magnetized, fully ionized electron-ion
plasma \citep{potekhin10}\footnote{Code publicly available at\\ {\rm
    http://www.ioffe.ru/astro/EIP/index.html.}}. For the thermal and
electrical conductivities in the crust we use the latest {\tt FORTRAN}
routines which include quantizing effects and thermal averaging
corrections accounting for partial degeneration of electrons, and the
effect of the finite size of nuclei in the inner crust\footnote{{\rm
    http://www.ioffe.ru/astro/conduct/}}.
We have adopted the functional form of the superfluid gaps proposed in Table 1 of \cite{ho12}. In addition, the contribution to the thermal conductivity from superfluid phonons \citep{aguilera09} has been suppressed, according
to recent results which indicate that entrainment between superfluid
neutrons and nuclei is larger than expected, resulting in an
increased effective mass of the nuclei in the inner crust
\citep{chamel12a,chamel12b}.

We note that, even for $10^{15}$ G, including quantizing effects in
the crust does not substantially affect the long-term evolution of the
star. However, it may have a strong effect on the envelope models, as
we discuss later. It should also be noted that the electrical
conductivity of the NS crust at low temperatures is
dominated by the interaction with the atomic nuclei of the crystal
lattice. The phonon Umklapp process is modeled by an analytical
formula including the exponential reduction of the scattering rate
\citep{gnedin01}. \cite{chugunov12} shows that, at low
temperatures ($T< 10^7$ K), such formula overestimates the electrical
conductivity by several orders of magnitude. However, here we are
interested in the evolution of a NS during the first $\sim$
Myr of its life, when its temperature is still relatively high and the
used approximation is still valid.

Electron-impurity collisions become the dominant process at low
temperatures $T\lesssim 10^8$ K (\citealt{shternin06} and
refs. within). Therefore, the conductivity at late times is almost
independent of the temperature and is determined by the impurity
content of the crust. This is usually parametrised by the quadratic
deviation of the atomic number $Q_{imp}= \sum_i x_i ( Z_i^2 - \langle
Z^2 \rangle )$, which is highly uncertain. In isolated NS,
especially in magnetars which remain warmer for longer times, $Q_{imp}$
in the outer crust is expected to be low, unlike in accreting NSs in
binary systems, where the outer crust is being continuously
replenished by newly processed nuclei.

For the inner crust, things are different. First, the crust is for the most part an elastic solid, comprising a Coulomb
lattice of {\it normal} spherical nuclei. However, in the innermost
layers near the crust/core boundary, because of the large effect of
the Coulomb lattice energy, cylindrical and planar geometries can
occur, both as nuclei and as bubbles \citep{ravenhall83}. These phases
are collectively named nuclear pasta (by analogy to the shape of
spaghetti, maccaroni and lasagna).  More sophisticated molecular
dynamics simulations \citep{horowitz05,horowitz08} have shown that it
may be unrealistic to predict the exact sizes and shapes of the pasta
clusters, and that the actual shape of the pasta phase can be very
amorphous, with a very irregular distribution of charge. This is
expected to have implications on the transport properties, in
particular on the electrical resistivity. Other simulations indicate that the inner crust could be amorphous and heterogeneous in nuclear charge \cite{jones04}.  The expected values of $Q_{imp}$, coming from the crust formation process alone, could be of the order $O(10)$, providing large values of resistivity.

To include the effect of pasta phases or disorder, we will extend the impurity parameter formalism to the
inner crust in a simplified way. We set $Q_{imp}=0.1$, except in the pasta region (the crust where $\rho>6\times10^{13}$ g/cm$^3$), where we
will let the parameter vary between $Q_{imp}=1-100$. Recently we have discussed in detail the influence of the pasta phase or disorder on the expected evolution of timing properties \citep{pons13}.

%%%%%%%%%%%%%%%%%%%%%%%%%%%%%%%%%%%%%%%%%%
\begin{figure*}
 \centering
\includegraphics[width=.25\textwidth]{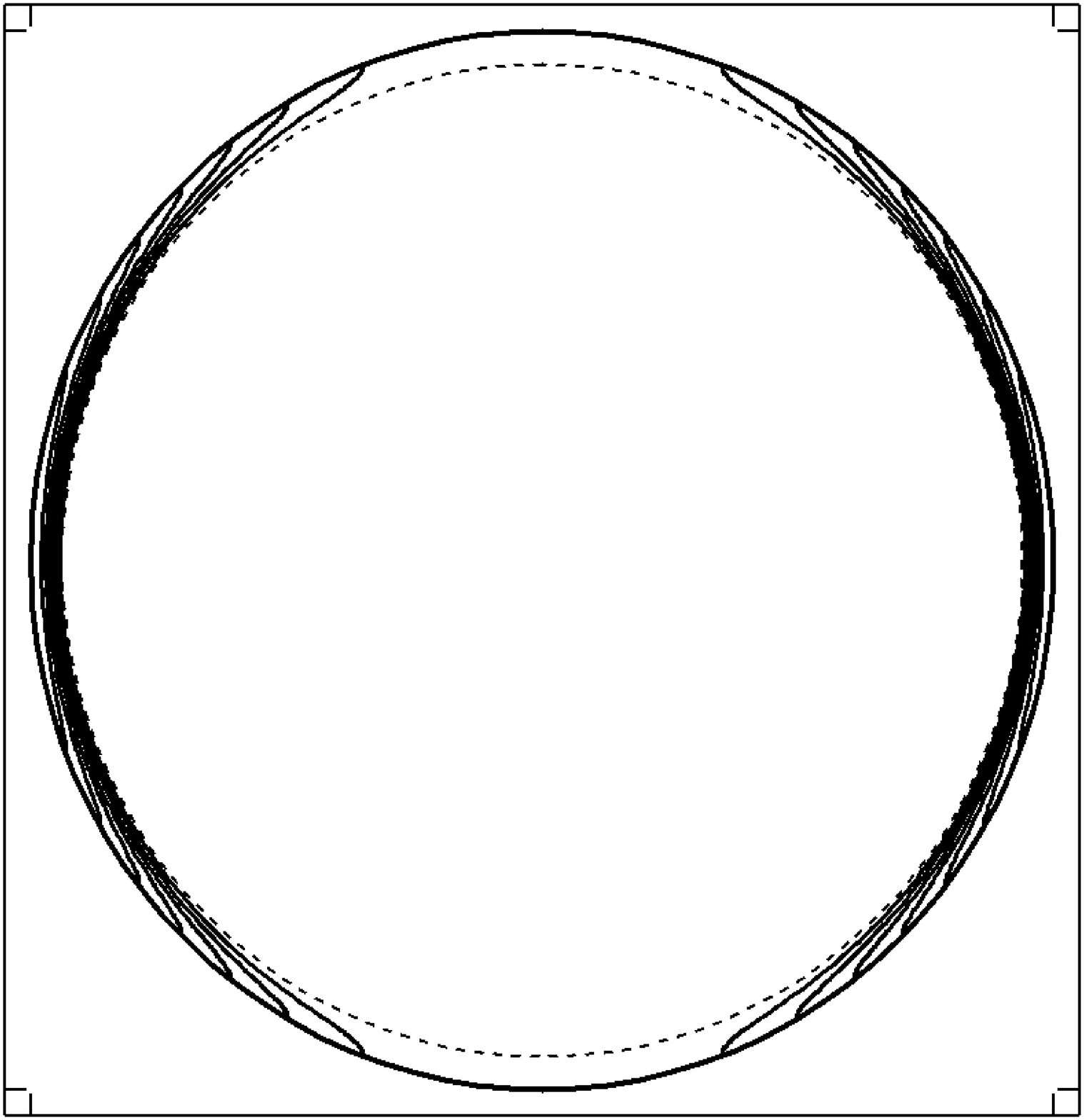}
\includegraphics[width=.25\textwidth]{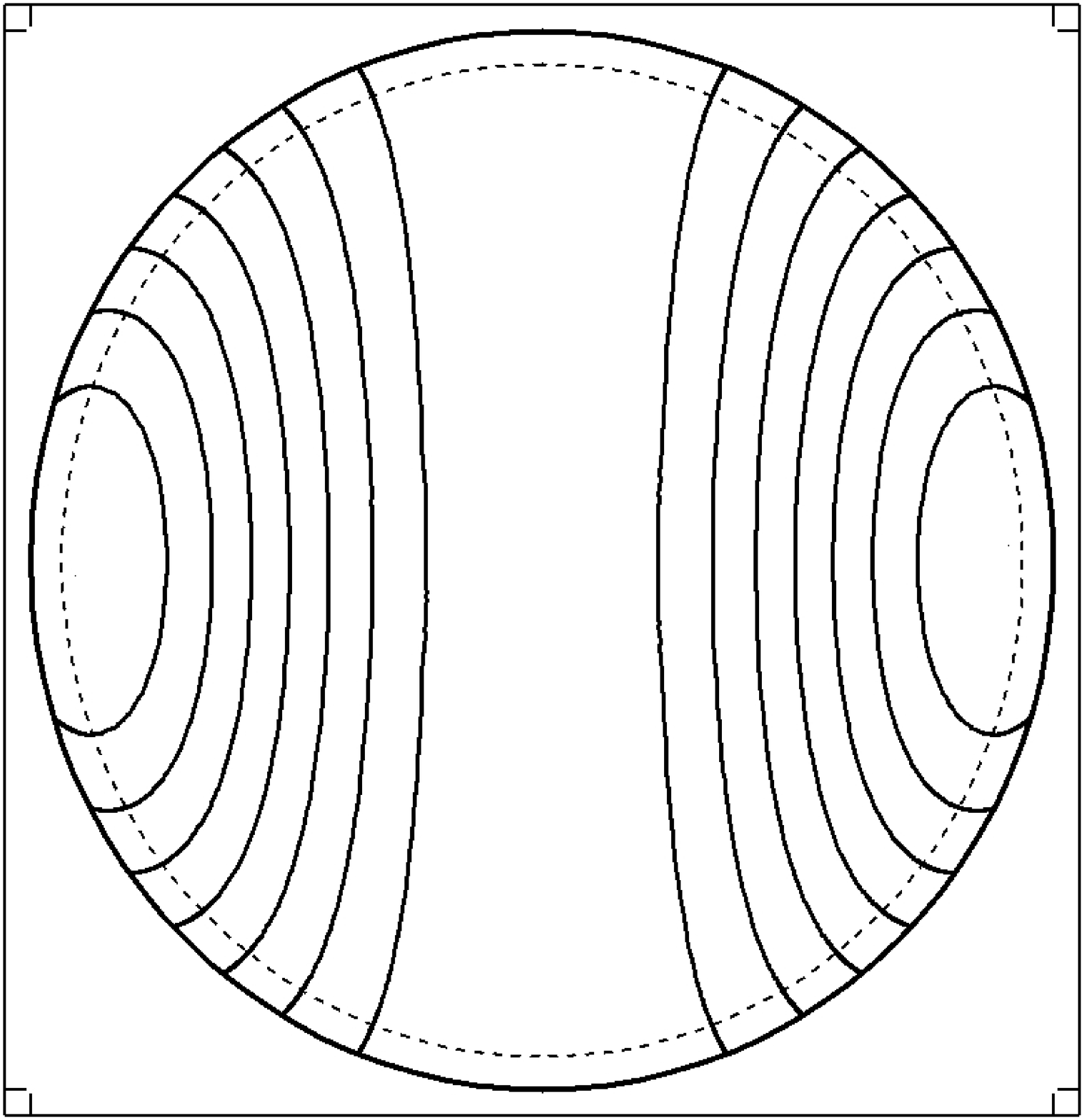}
\includegraphics[width=.25\textwidth]{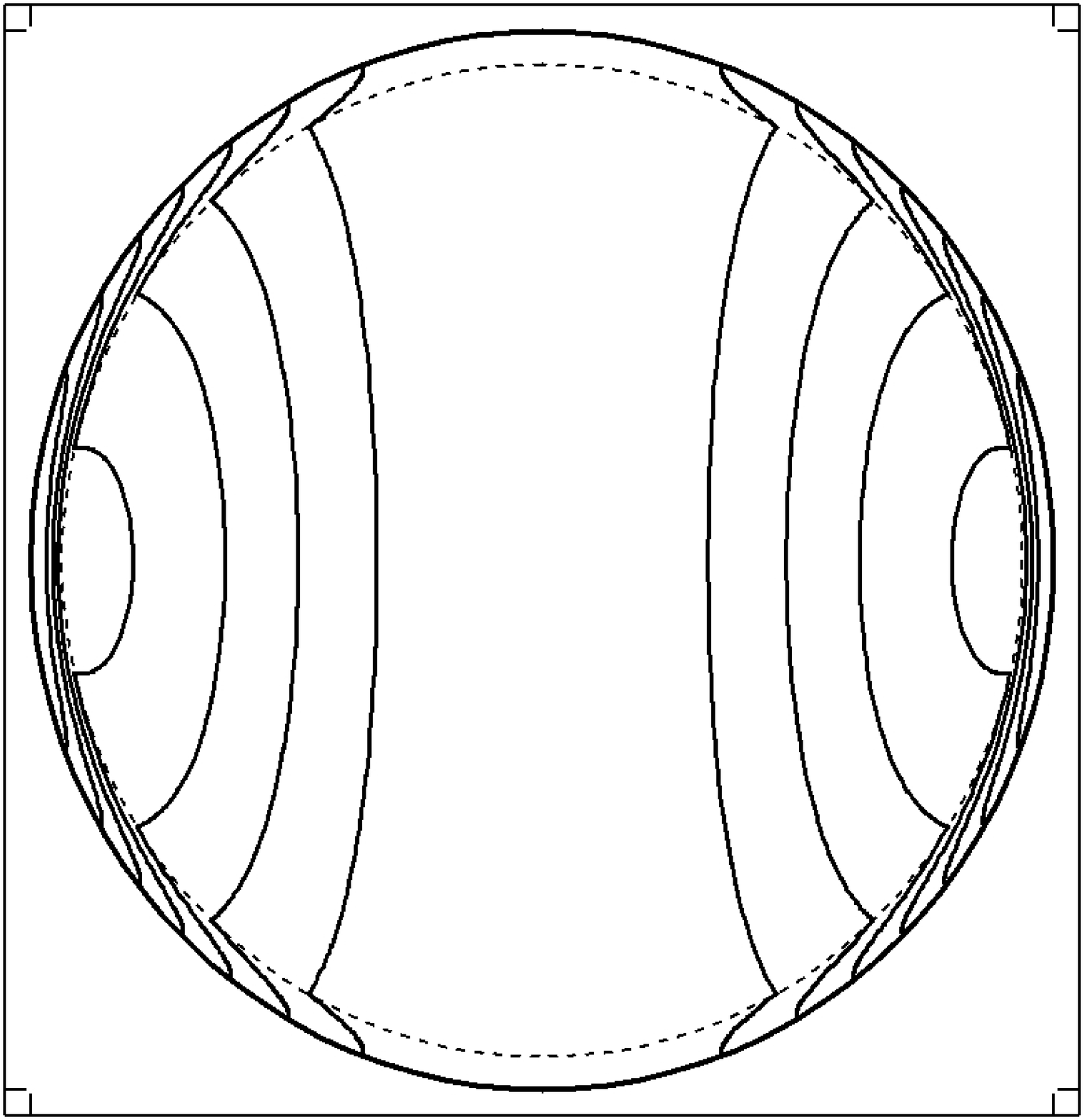}
\caption{Representative configurations of the initial magnetic field geometries. {\it Left:} type A, crustal confined magnetic field; {\it center:} type B, core-extended magnetic field; {\it right:} type C, hybrid. Solid lines show the poloidal magnetic field lines in a meridional plane.}
 \label{fig:initial_b}
\end{figure*}
%%%%%%%%%%%%%%%%%%%%%%%%%%%%%%%%%%%%%%%%%%

%%%%%%%%%%
\subsection{The envelope.}

The envelope (so-called {\it the ocean}) is the liquid layer above the
solid crust and below the atmosphere.  It comprises approximately the
outermost hundred meters of the star, and it is the region with the
largest temperature gradient. Due to its relatively low density, its
thermal relaxation timescale is much shorter than that of the crust,
which makes any attempt to perform cooling simulations in a numerical
grid that includes both regions simultaneously computationally too
expensive. The widely used approach is to use fits to stationary
envelope models to obtain a phenomenological relation between the
temperature at the bottom of the envelope, $T_b$, and the effective
temperature $T_s$.  This $T_b-T_s$ relation depends on the surface
gravity $g_s$, the envelope composition, and the magnetic field strength and
orientation.

After the pioneer work by \cite{gudmundsson83}, further
1D, plane-parallel simulations have followed the same approach,
including neutrino processes and the presence of magnetic field, which
affects the transport of heat, causing the heat flux to become anisotropic
\citep{potekhin01,potekhin07}.  Therefore, the geometry of the magnetic field is
important, since regions permeated by radial magnetic field lines
are thermally connected to the interior, while zones with tangential
magnetic field (equator in the dipolar case) are thermally
insulated. \cite{pons09} revised the envelope model with a 2D code,
allowing for meridional transport of heat, finding that the most
important effect is the attenuation of the anisotropy. For
consistency, in this paper we have built our own envelope models,
revising the fits done in section 3 of \cite{pons09} with the updated
microphysical inputs. We have placed the outer boundary of our crust
at $\rho=3\times 10^{10}$ g cm$^{-3}$, and solved different stationary
envelope models to obtain the $T_b-T_s$ relations that are then used
as an external boundary condition in our simulations.

The luminosity seen by an observer at infinity is obtained by
assuming blackbody emission at the temperature $T_s$ from each patch
of the NS surface. It is not the purpose of this paper to study other
emission models, such as as magnetic atmospheres or condensed surface
models. These do not change appreciably the relation between $T_b$ and the emitted flux
\citep{potekhin07} but can be important to interpret the temperature inferred by a spectral fit (see \S~\ref{sec:observations}).

%%%%%%%%%%%%%%%%%%%%%%%
\begin{figure*}
 \centering
\includegraphics[width=.3\textwidth]{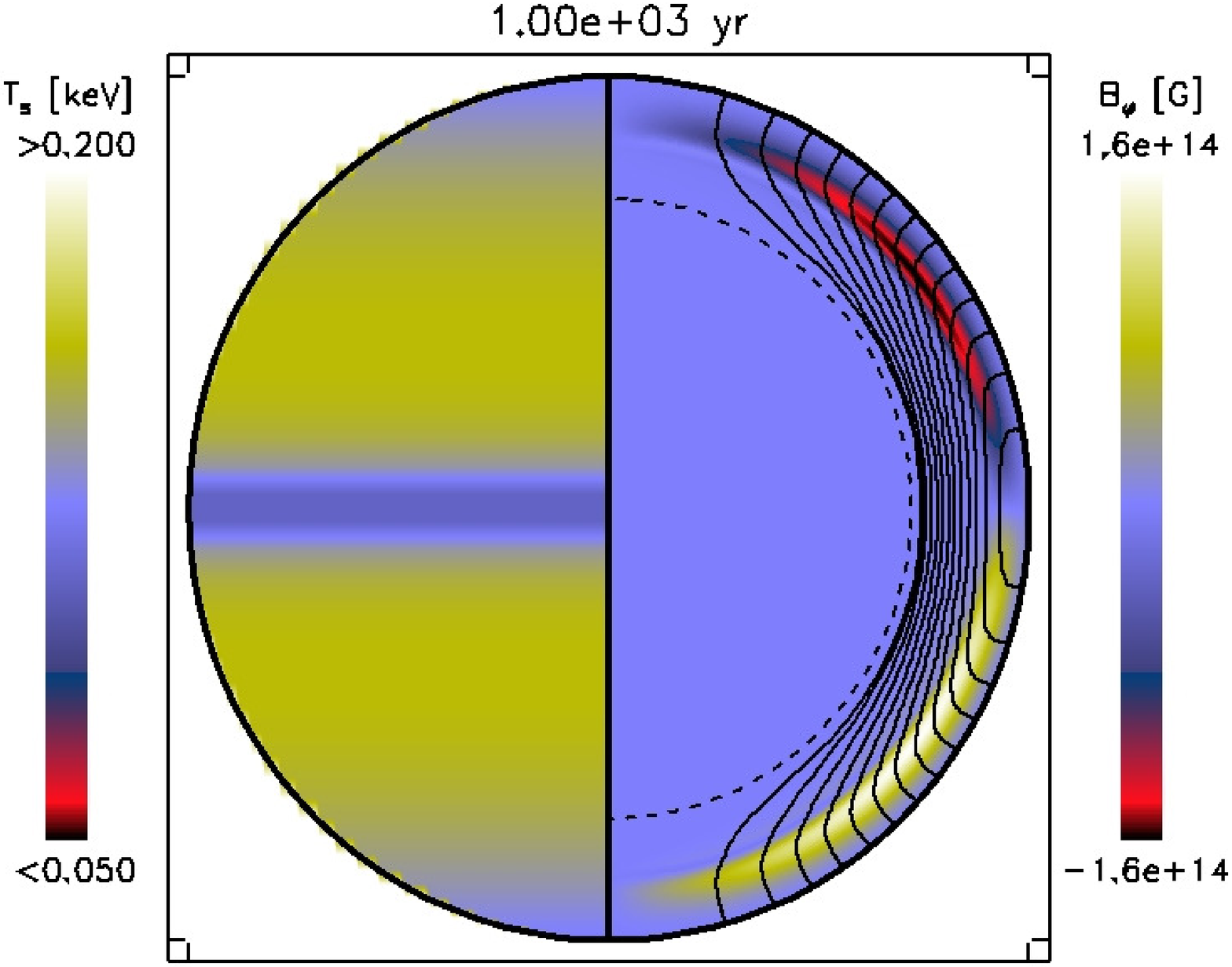}
\includegraphics[width=.3\textwidth]{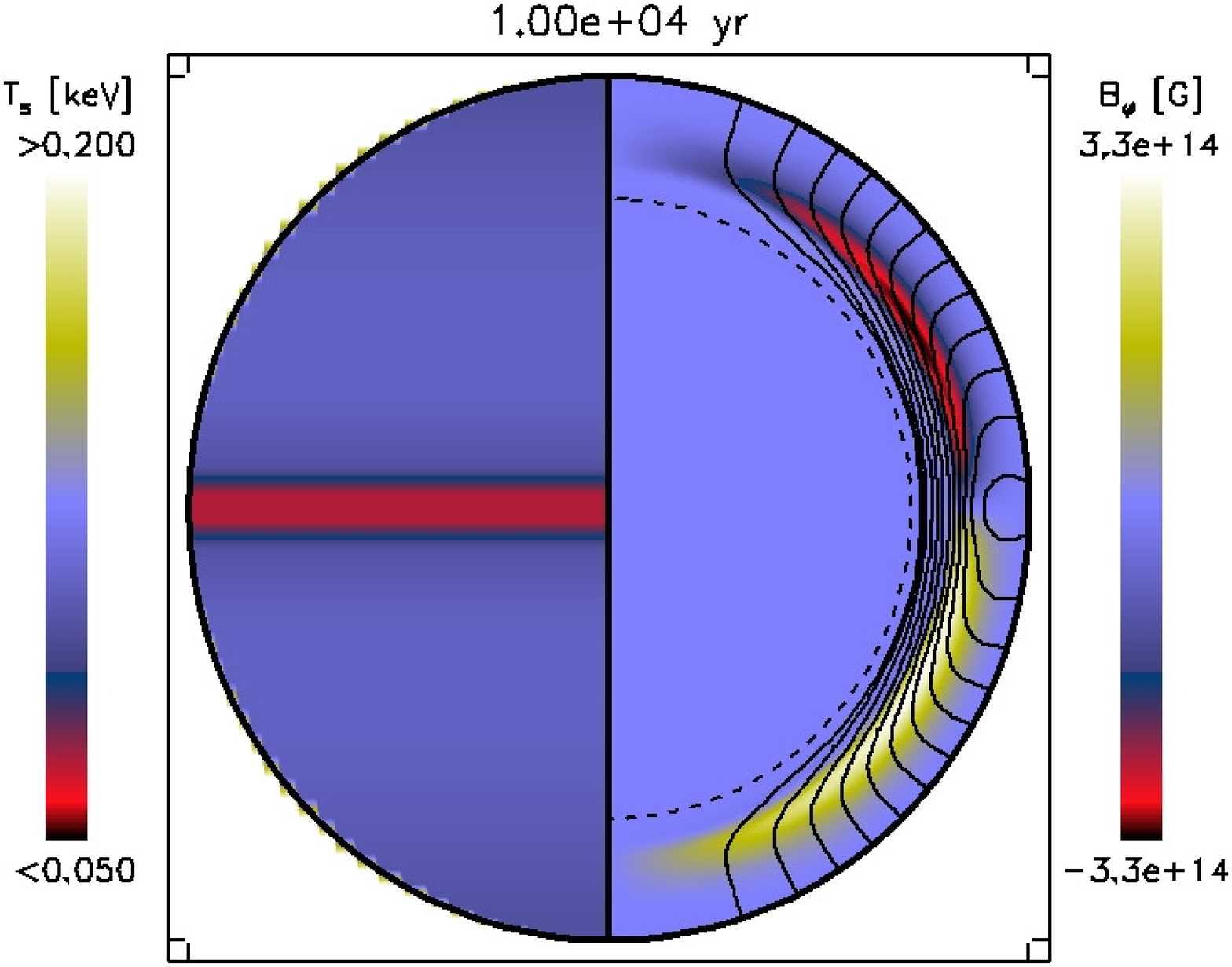}
\includegraphics[width=.3\textwidth]{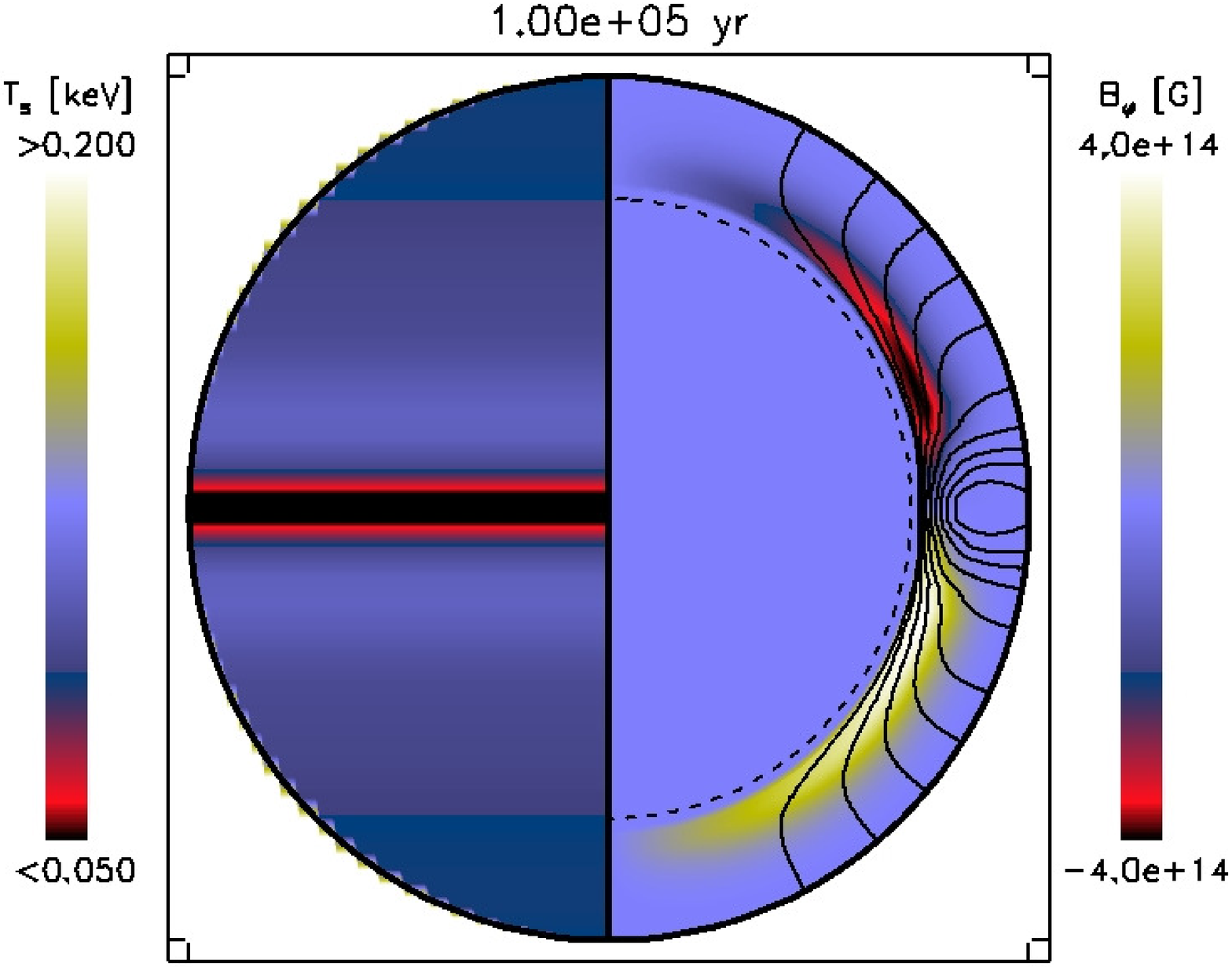}\\
\includegraphics[height=.25\textwidth,width=.27\textwidth]{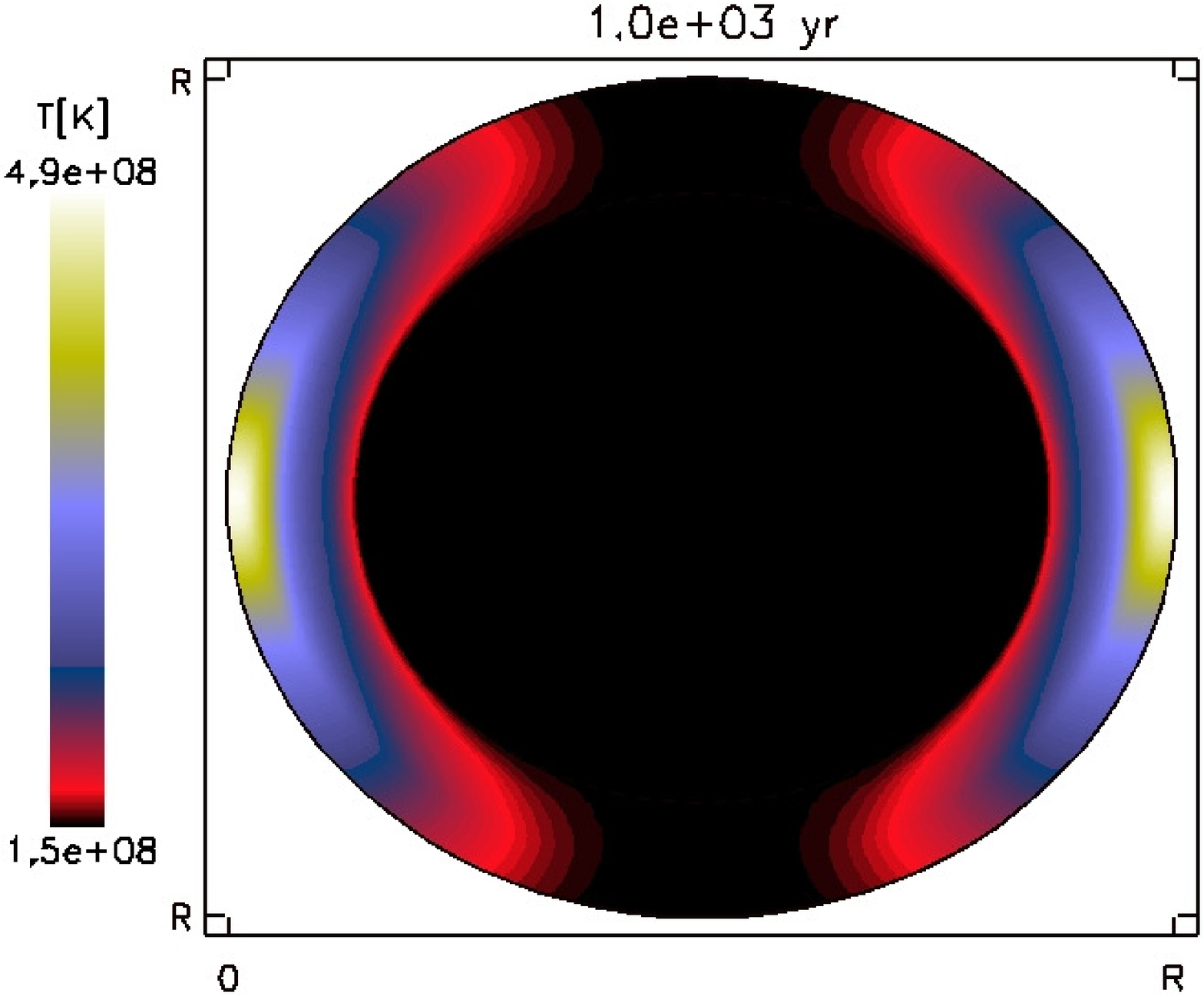}
\hspace{0.43cm}
\includegraphics[height=.25\textwidth,width=.27\textwidth]{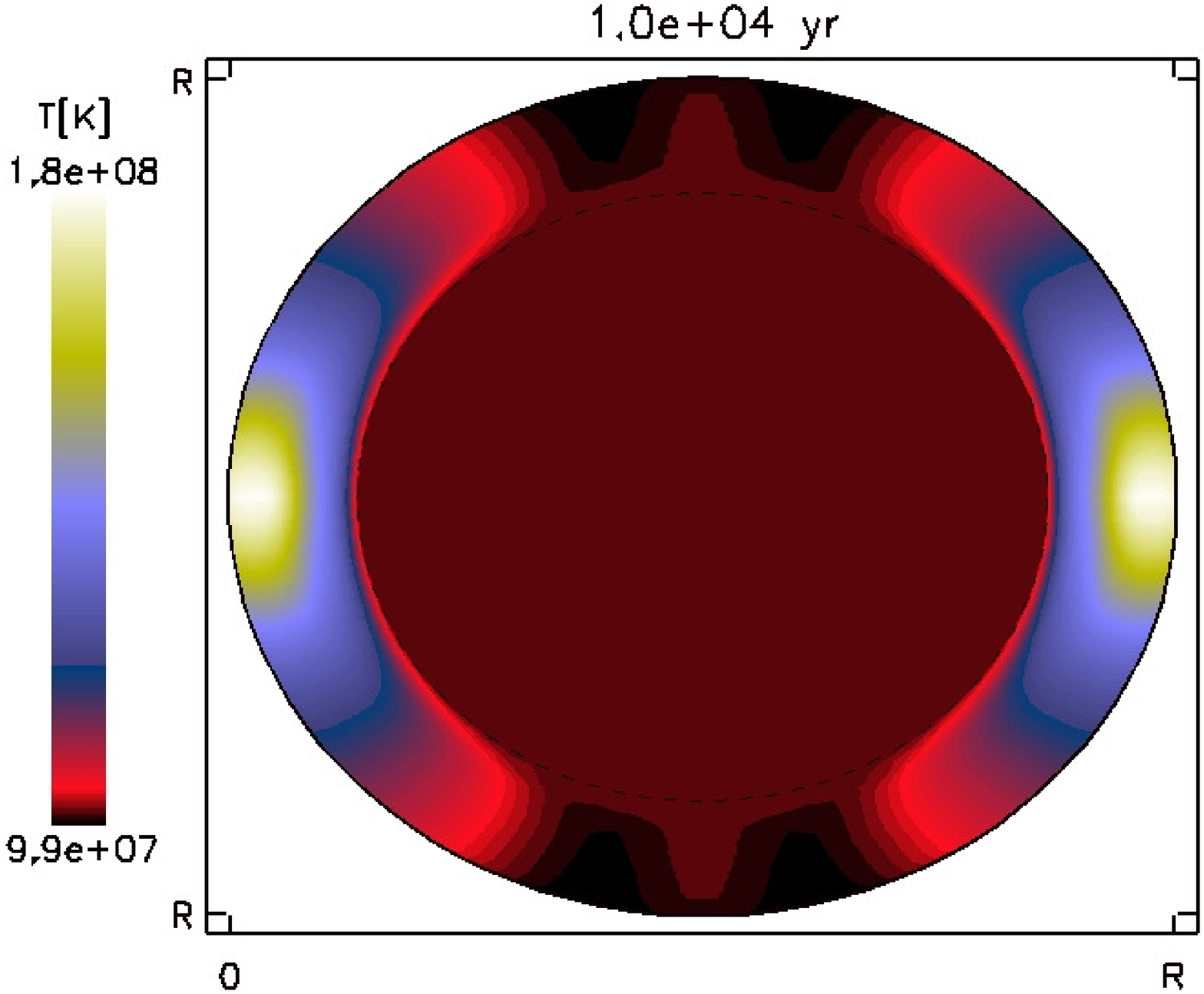}
\hspace{0.43cm}
\includegraphics[height=.25\textwidth,width=.27\textwidth]{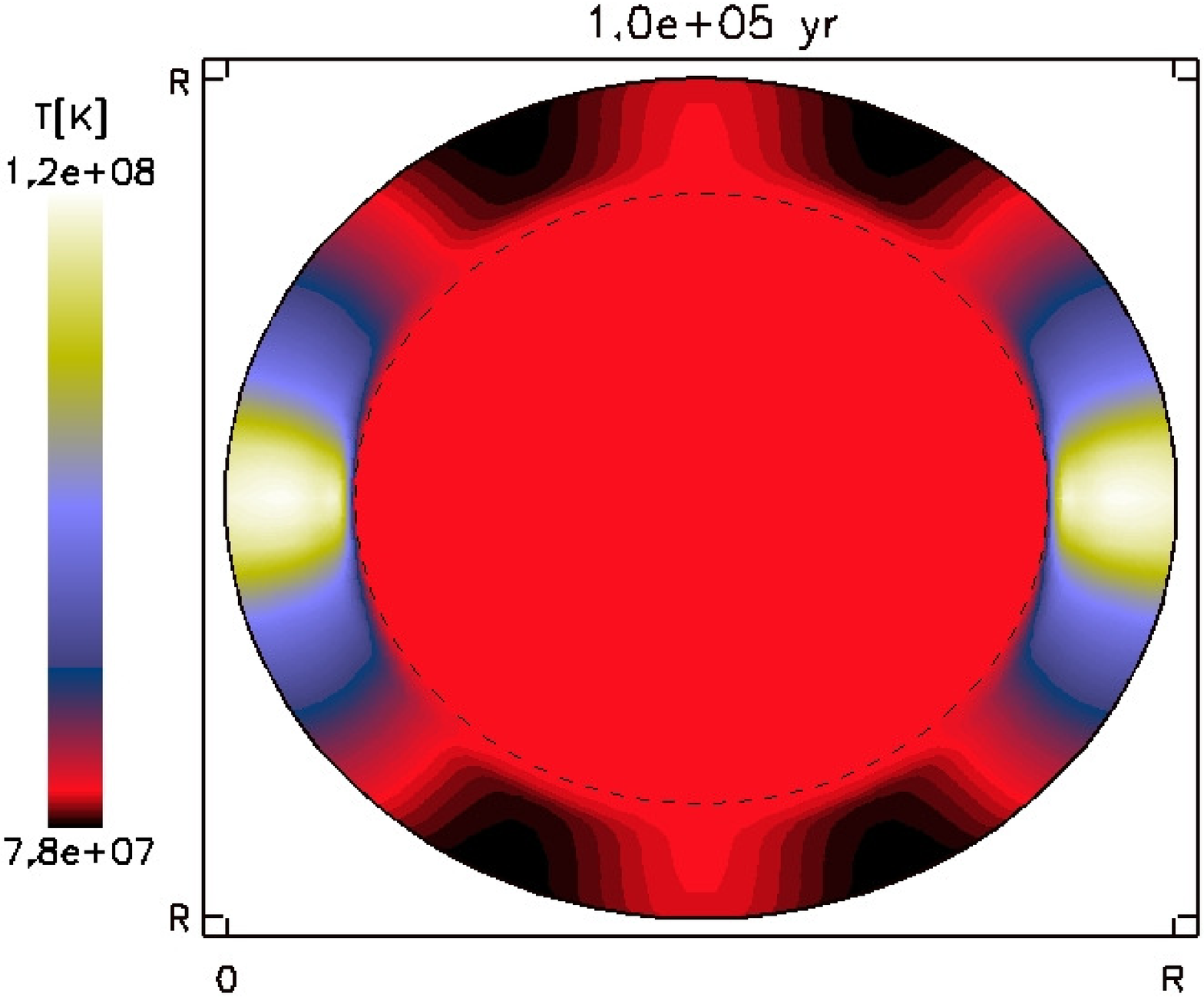}
\hspace{0.33cm}
\caption{Snapshots of the evolution of model A14 at $10^3, 10^4, 10^5$ yr, from left to right. {\it Top panels:} the left hemisphere shows in color scale the surface temperature, while the right hemisphere displays the magnetic configuration in the crust, whose thickness has been enlarged by a factor of 4 for visualization purposes. Black lines are the poloidal magnetic field lines, while color scale indicates the toroidal magnetic field intensity (yellow: positive, red: negative). {\it Bottom panels:} temperature map inside the star.}
 \label{fig:b14_evo}
\end{figure*}
%%%%%%%%%%%%%%%%%%%%%%%

%%%%%%%%%%
\section{Results}\label{sec:results}

%%%%%%%%%%
\subsection{Cooling of weakly magnetized neutron stars.}\label{sec:results_weak}

Before we discuss the effects of strong magnetic fields on the cooling history
of NSs, we begin by briefly revisiting the cooling of low magnetic
field sources, with our microphysical inputs. Comprehensive reviews
can be found in \cite{page04} and \cite{yakovlev04}.

In Fig.~\ref{fig:b0} we present the set of cooling curves for non-magnetised NSs 
with masses ranging between 1.10 and 1.76 $M_\odot$ (lines from top
to bottom), together with the available observational data presented
in \S\ref{sec:observations}.  We show the bolometric luminosity as a
function of time. The left/right panels correspond to models with
heavy element envelopes (iron) and light-element (hydrogen) envelopes,
respectively. Sources with estimates of the kinematic age are shown
with the associated error bar on the age. With arrows pointing towards
the left we indicate the sources with $\tau_c\gtrsim 50$ kyr and no
kinematic age. This implies a likely overestimate of the real age for
the middle-age and old objects; such an age overstimate could be even
orders of magnitude for the oldest objects (like SGR~0418+5729).

In general, after $\approx 100$ yr, low mass stars ($M \lesssim 1.4 M_\odot$) are brighter
than high mass stars.  For the high-mass family, $M\gtrsim 1.4
M_\odot$, the proton fraction in the center of the star (for our
choice of equation of state) is high enough to activate the direct URCA processes,
which results in fast cooling before one hundred years.  Within the
low-mass family, cooling curves are similar at early ages ($<100$
yr). The differences at $t\sim 10^2-10^3$ yr are due to the delayed
transition of neutrons in the core to a superfluid state, which
activates neutrino emission by means of Cooper pair formation and
breaking. This effect also depends on the equation of state employed:
a stiffer equation of state results in lower densities and the transition to
superfluidity is further delayed, assuming the same gap. After the
effect of the transition to a superfluid core is finished, at $\gtrsim 10^3$
yr, cooling curves for low mass NSs tend to converge again, following
the same curve independently of the mass. For models with heavy
element envelopes (left panel), the luminosity drops below $10^{33}$
erg/s at most a few thousand years after birth, while light-element envelope
models (right panel) predict luminosities around $10^{33}-10^{34}$
erg/s for quite a long period (up to several $10^4$ yr).

In young NSs, the evolution is dominated by neutrino cooling (the softest decay
of the curves), while at late times, the emission of thermal
photons from the star surface is the dominant factor (the steep
decline of luminosity in the log-log plot).  The turning point between
the neutrino-dominated cooling era and the photon-dominated cooling
era happens much earlier for light-element envelope models (a few times
$10^4$ yr) than for heavy envelope ones ($\approx 10^5$ yr).  During
the photon cooling, the evolution of the surface temperature of the
star can be roughly described by a power law $T_s \propto
t^{-1/8\alpha}$, with $\alpha\ll 1$, the exact value depending on the
envelope model, which relates $T_s\propto T_b^{0.5 + \alpha}$
\citep{page04}. For all weakly magnetised models, after $\sim 1$ Myr,
the luminosity has dropped below $10^{31}$ erg/s, the surface
temperature goes below $20-30$ eV and the star becomes invisible to
X-ray observations.  Note also that in the photon cooling era NSs with
light-element envelopes are much cooler than those with iron
envelopes.  \cite{page11} suggested that a star could in principle
have a light-element envelope at the beginning of its life, residual
of accretion of fall-back material after the supernova explosion, that
is progressively converted to heavier elements by nuclear reactions.

%%%%%%%%%%%%%%%TABLE 4 %%%%%%%
\begin{table}
\begin{center}
\begin{tabular}{l c c c}
\hline
\hline
Model & current  & $B_p^0$ & $B_t^0$  \\
 & location  & [G] &  [G]  \\
\hline
A14	& crust	& $10^{14}$ & 0 	\\
A15	& crust	& $10^{15}$ & 0 	\\
A14T	& crust	& $10^{14}$ & $5\times10^{15}$ \\
B14	& core	& $10^{14}$ & 0 	\\
C14	& crust+core & $10^{14}$ & 0 \\
\hline
\hline
\end{tabular}
\end{center}
\label{tab:models}
\caption{Summary of the initial magnetic field configurations.}
\end{table}
%%%%%%%%%%%%%%%%%%%%%%%

From our quick glance at the classical cooling theory of low magnetic
field NSs, the most interesting observation is that the
high-B objects are systematically hotter than what theoretical non-magnetised
cooling curves predict. This provides strong evidence in favor of the scenario in
which magnetic field decay powers their larger luminosity. Even
considering the (more than likely) overestimate of the NS
age by using its characteristic age, which can reconcile some of the objects 
with standard cooling curves, it is clear that most magnetars and some high-B pulsars
stars cannot be explained. Thus it is necessary to add the effects of
the magnetic field in the simulations, extending the four parameter
family of classical theoretical cooling models (compositional
differences in the core, superfluid properties, composition of the
envelope, and mass) with one more, the magnetic
field. This is the subject of the next sections.

%%%%%%%%%%
\subsection{Cooling of strongly magnetised neutron stars.}

%%%%%%%%%%
\subsubsection{Initial magnetic field.}

The initial configuration of the magnetic field in a newly born NS is
still unknown. MHD equilibria considering a hot and fluid, newly born
NS, point to a dipolar configuration with an equatorial torus where
the lines are twisted \citep[ for instance]{ciolfi09,lander13}. However, it is not clear whether the
complicated dynamics of the proto-NS actually leads to this
configuration, or if it is stable. We denote by ``initial'' the
configuration at the epoch of the freezing of the crust (from hours to
weeks after the SN explosion, depending on the density).

Lacking robust arguments in favor of specific initial
configurations, we consider three generic families with different
geometries, illustrated in Fig.~\ref{fig:initial_b} and mathematically
described in \S~2.1 of \cite{aguilera08b}. In the crustal confined model (type A geometry, left panel),
the magnetic field lines do not penetrate into the core ($B_r=0$ at the crust/core interface), and the
currents are entirely contained in the crust. The toroidal magnetic field is
extended through the crust, and its form is chosen to be dipolar:

\begin{equation}
  B_\phi=-k(r-r_c)^2(r-r_\star)^2\sin\theta/r~, 
\end{equation}
where $k$ is the normalization. In the type B configurations (central panel), the magnetic field
threads the core, where the bulk of the current circulates with very
slow dissipation.  Type C geometry (right panel) is a hybrid case
where a double system of currents supports a large-scale,
core-extended dipole and an additional, stronger crustal magnetic field.  Each
family is parametrised by the intensity of the dipolar component at
the pole, $B_p^0$, and the maximum intensity of the toroidal magnetic field at
any point, $B_t^0$.  In Table~\ref{tab:models} we list the models
studied in this work.

During the whole evolution we match the crustal magnetic field with a potential magnetic field at the surface, allowing for a general,
time-dependent combination of multipoles (see \citealt{vigano12a} for details about the outer boundary condition).
In particular, we follow the evolution of the external dipolar component $B_p$, 
which determines the timing properties (spin-down, braking index) of the star.

%%%%%%%%%%
\subsubsection{Evolution of a strongly magnetized NS.}

We begin the discussion of our results by describing the general
evolution of model A14.  We run simulations up to several Myr,
monitoring the temperature and magnetic field inside the star and at
its surface.  In the top panels of Fig.~\ref{fig:b14_evo} we show three snapshots at
$t=10^3, 10^4, 10^5$ yr. For each snapshot, we show in the left
hemisphere the surface temperature distribution, and in the right
hemisphere the evolution of the crustal magnetic field. We now
concentrate on the latter.

%%%%%%%%%%  Magnetic ENERGY
\begin{figure}
 \centering
\includegraphics[width=.45\textwidth]{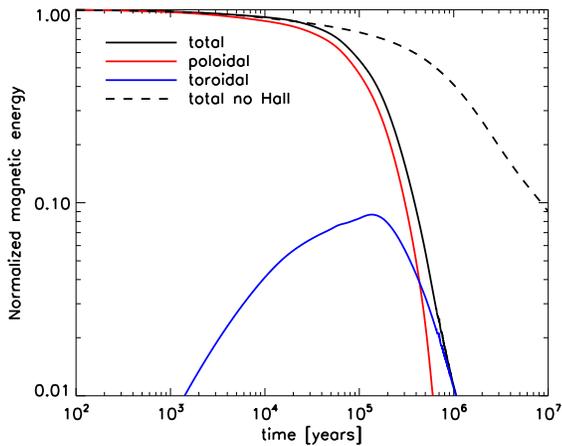}
\caption{Magnetic energy in the crust (normalized to the initial
  value) as a function of time, for model A14.  The solid lines
  correspond to: total magnetic energy (black), energy in the
  poloidal components (red), and energy in the toroidal magnetic field
  (blue). The dashed line shows the effect of switching off the Hall
  term (only Ohmic dissipation). }
 \label{fig:entransf}
\end{figure}
%%%%%%%%%%%%%%%%%%%%%%%%%%%%%%%%

The first effect of the Hall term in the induction equation is to
couple the poloidal and toroidal magnetic field, so that even if the latter is
zero at the beginning, it is soon created. After $\sim 10^3$ yr, the
poloidal dipolar magnetic field has generated a mainly quadrupolar toroidal magnetic
field, with a maximum strength of the same order of the poloidal
component, with being $B_\phi$ negative in the northern hemisphere and positive
in the southern hemisphere. An important effect of the Hall term is
that currents are gradually drained towards the inner crust, where
they can last more. After $\sim 10^5$ yr, the toroidal magnetic field and the
distorted poloidal lines are concentrated in the inner crust. In the
outer crust, the magnetic field approaches a potential configuration that
matches the external magnetic field.

In the bottom panels of Fig.~\ref{fig:b14_evo} we show three snapshots of the crustal temperature evolution. At $t=10^3$ yr, the equator is hotter than the poles by a factor of three. As the evolution proceeds and currents are dissipated,  
the temperature reflects the change of geometry of the poloidal lines
(see Fig.~\ref{fig:b14_evo}) and the anisotropy becomes weaker.

The presence of strong tangential components ($B_\theta$ and $B_\phi$) insulates the surface from the interior. For a dipolar geometry, the magnetic field is nearly radial at the poles, and these are thermally connected with the interior, while the equatorial region is insulated by tangential magnetic field lines. 
This has a twofold effect: if the core is hotter than the crust, the polar regions will be hotter than the equatorial regions; however, if Ohmic dissipation heats up the equatorial regions, they remain hotter than the poles. 
In addition, the insulating effect of the envelope  must be taken into account. Depending on the local conditions 
(temperature, magnetic field strength, angle of the magnetic field with the normal to the surface) it may invert the anisotropy: in the surface temperature (see Fig. \ref{fig:b14_evo}), the equator is cooler than the poles. 

In Fig.~\ref{fig:entransf} we show the evolution of the total magnetic energy, compared with the fraction stored in the different components.  We also compare our results to the purely resistive case, in which the Hall term has been switched off (and therefore there is no creation of toroidal magnetic field). When the Hall term is included, $\sim$99\% of the initial magnetic energy is dissipated in the first $\sim 10^6$ yr, compared to only the $\sim 60\%$ of the purely resistive case. At the same time, a $\sim 10\%$ of the initial energy is transferred to the toroidal component in $10^5$ yr, before it begins to decrease.
Note that the poloidal magnetic field, after $10^5$ yr, is dissipated faster than the toroidal component. The poloidal magnetic field is supported by  toroidal currents concentrated in the equator, which is hotter (see right bottom panel of Fig. \ref{fig:b14_evo}), while the toroidal field is supported by poloidal currents that circulate in a higher latitude region, where the temperature is lower. Since the resistivity in this regime is strongly dependent on temperature, this causes the more rapid dissipation of the poloidal magnetic field. As a result, at late times most of the magnetic energy is stored in the toroidal component.

%%%%%%%%%
\subsubsection{Dependence on mass and relevant microphysical parameters.}

%%%%%%%%%%%%%%%%%%%%%%%%%%%%%%%%%%%%%%%%%%%%%%%
\begin{figure}
 \centering
\includegraphics[width=.45\textwidth]{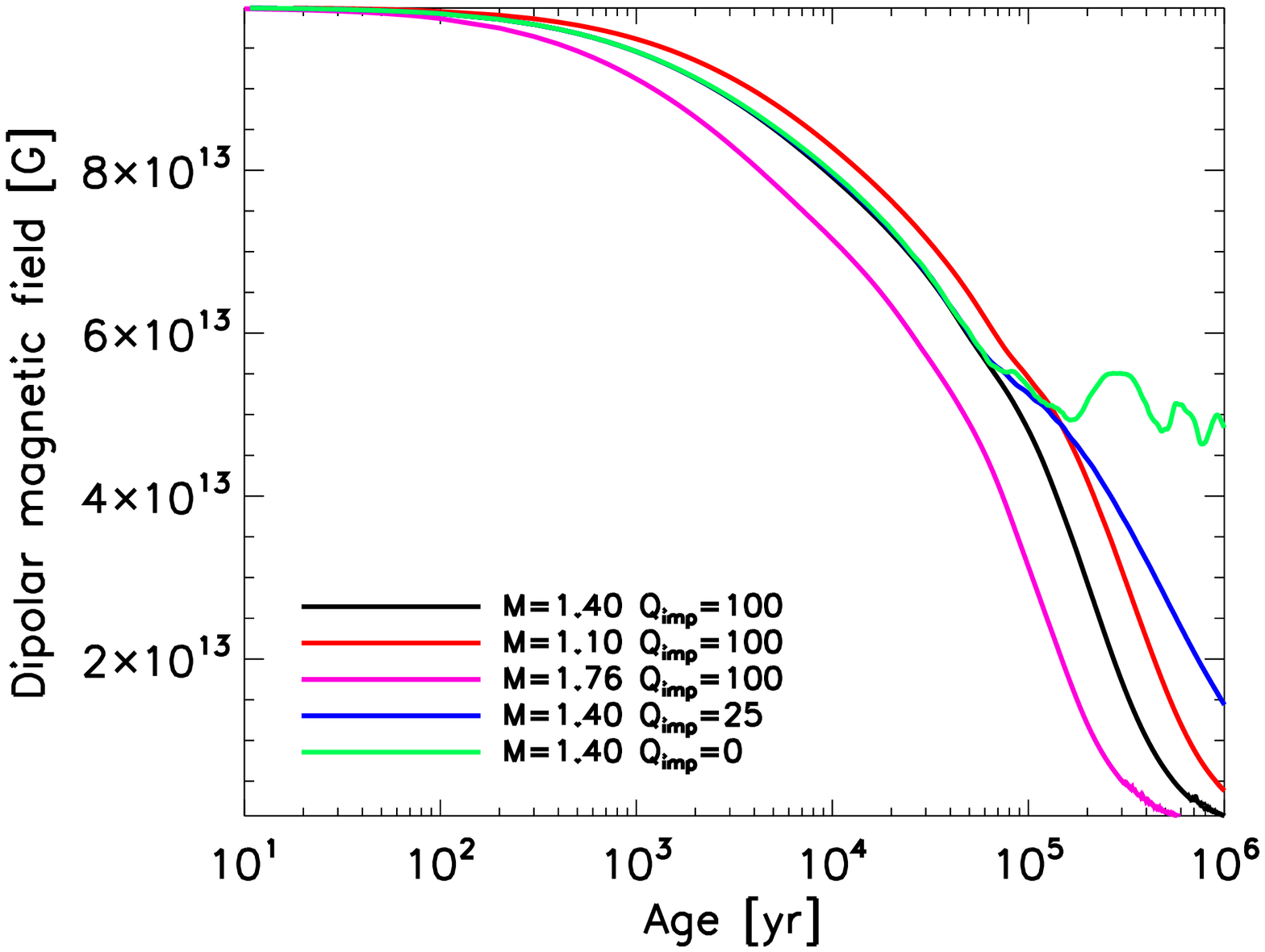}
\includegraphics[width=.45\textwidth]{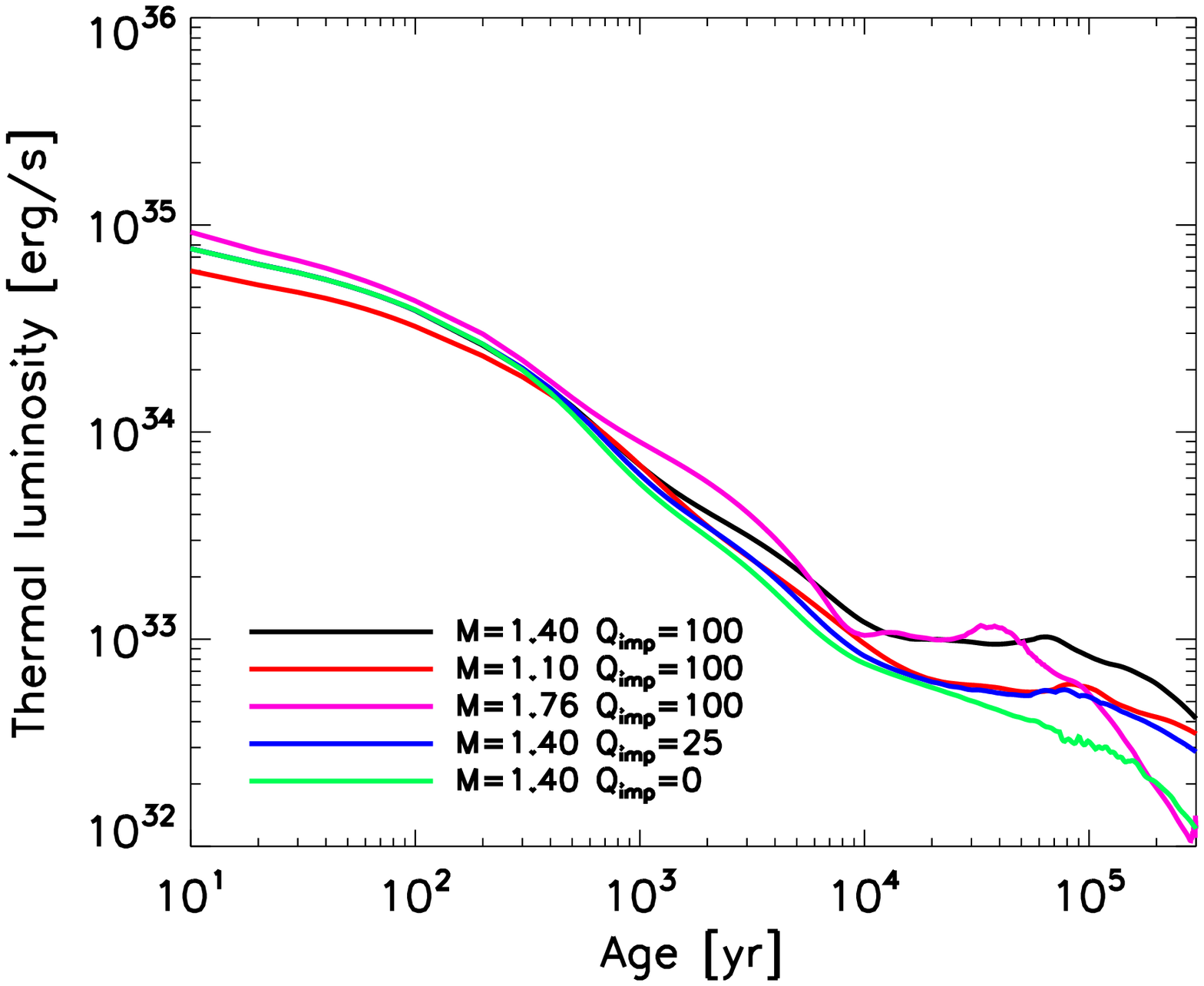}
\caption{Evolution of $B_p$ (top) and the luminosity (bottom) for model A14, but varying the NS mass and
the impurity parameter in the pasta region ($\rho>6 \times 10^{13}$ g/cm$^3$).}
 \label{fig:micro}
\end{figure}
%%%%%%%%%%%%%%%%%%%%%%%%%%%%%%%%%%%%%%%%%%%%%%%

%%%%%%%%%%%%%%%%%%%%%%%%%%%%%%%%%%%%%%%%%%%%%%%
\begin{figure*}
 \centering
\includegraphics[width=.3\textwidth]{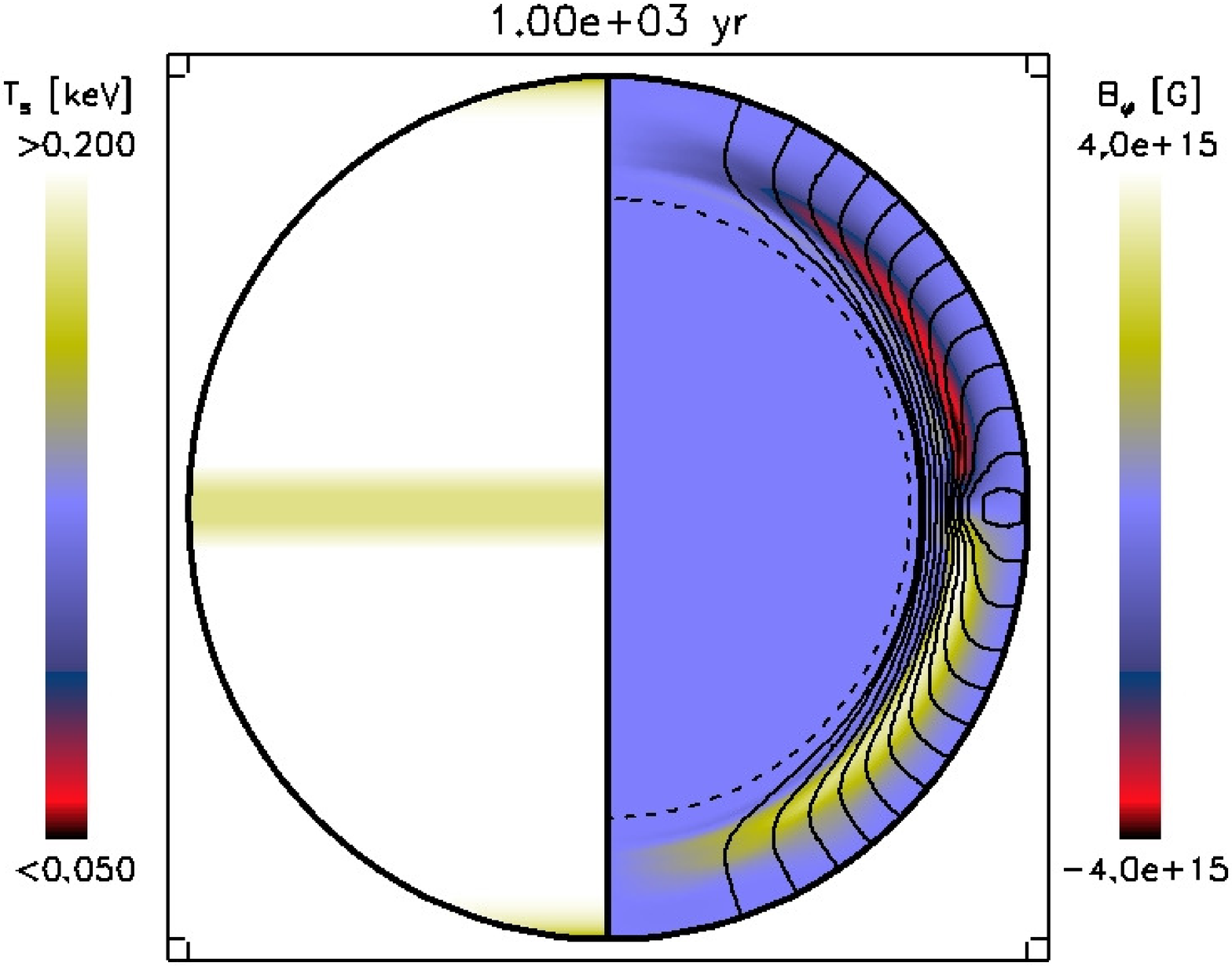}
\includegraphics[width=.3\textwidth]{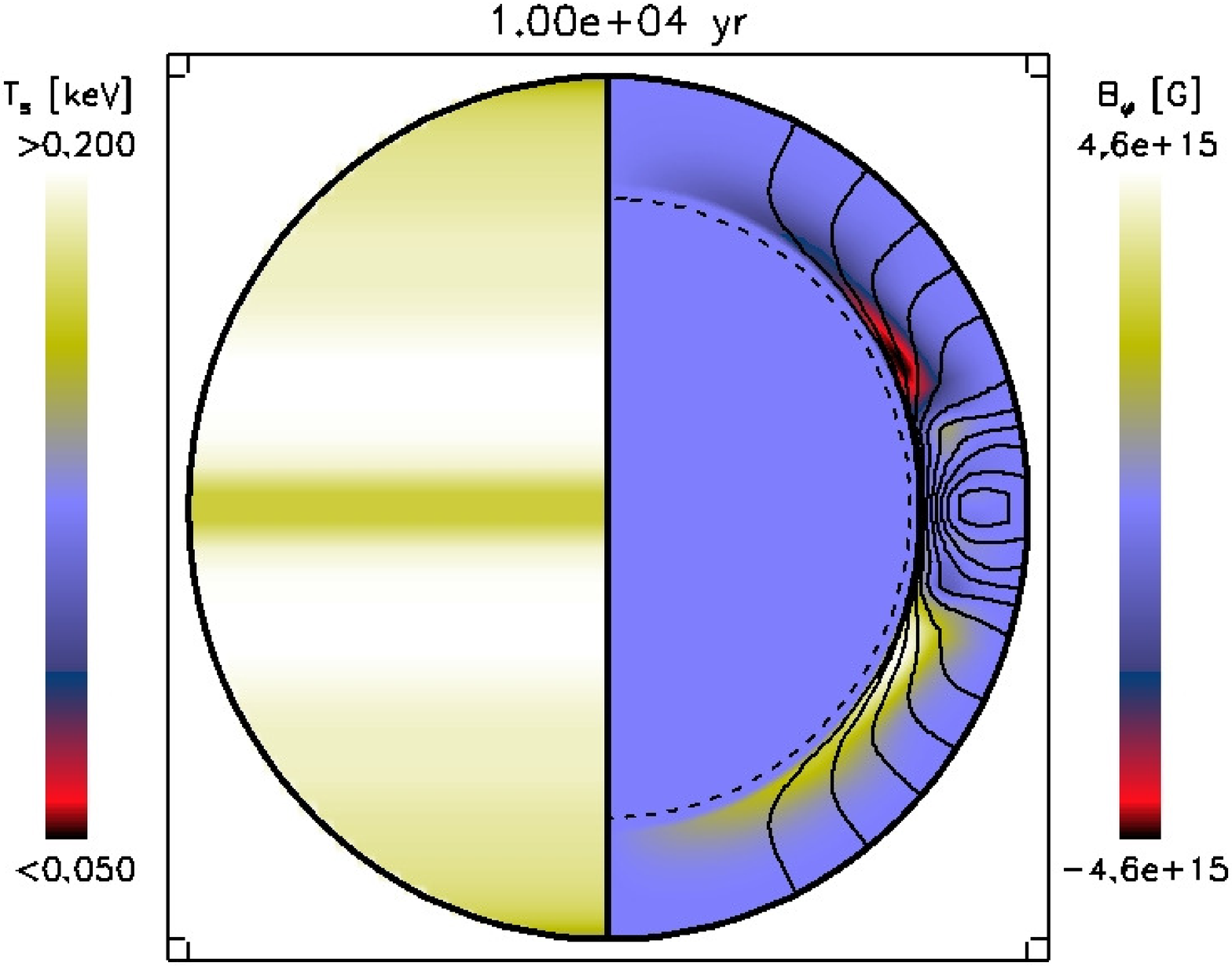}
\includegraphics[width=.3\textwidth]{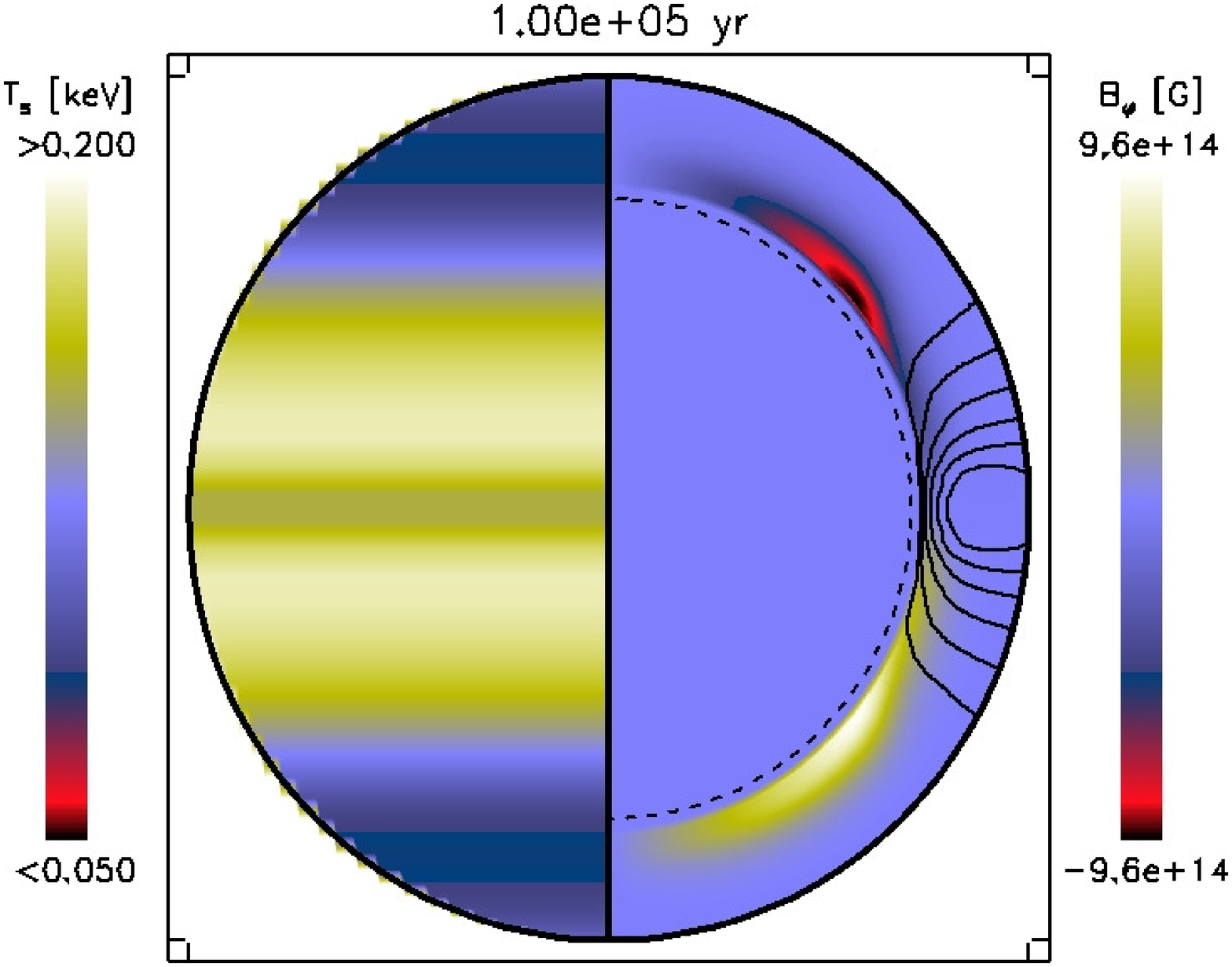}\\

\includegraphics[width=.3\textwidth]{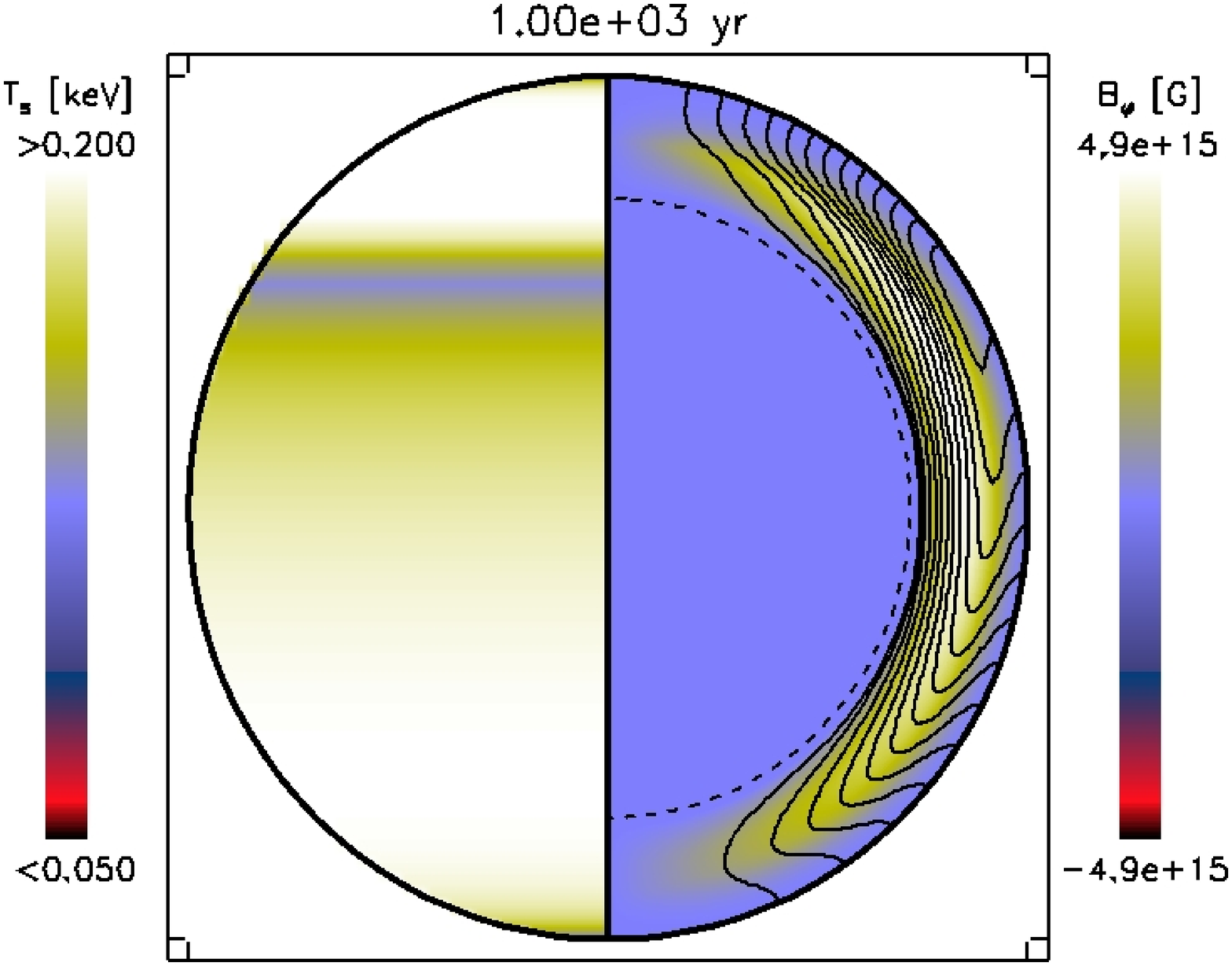}
\includegraphics[width=.3\textwidth]{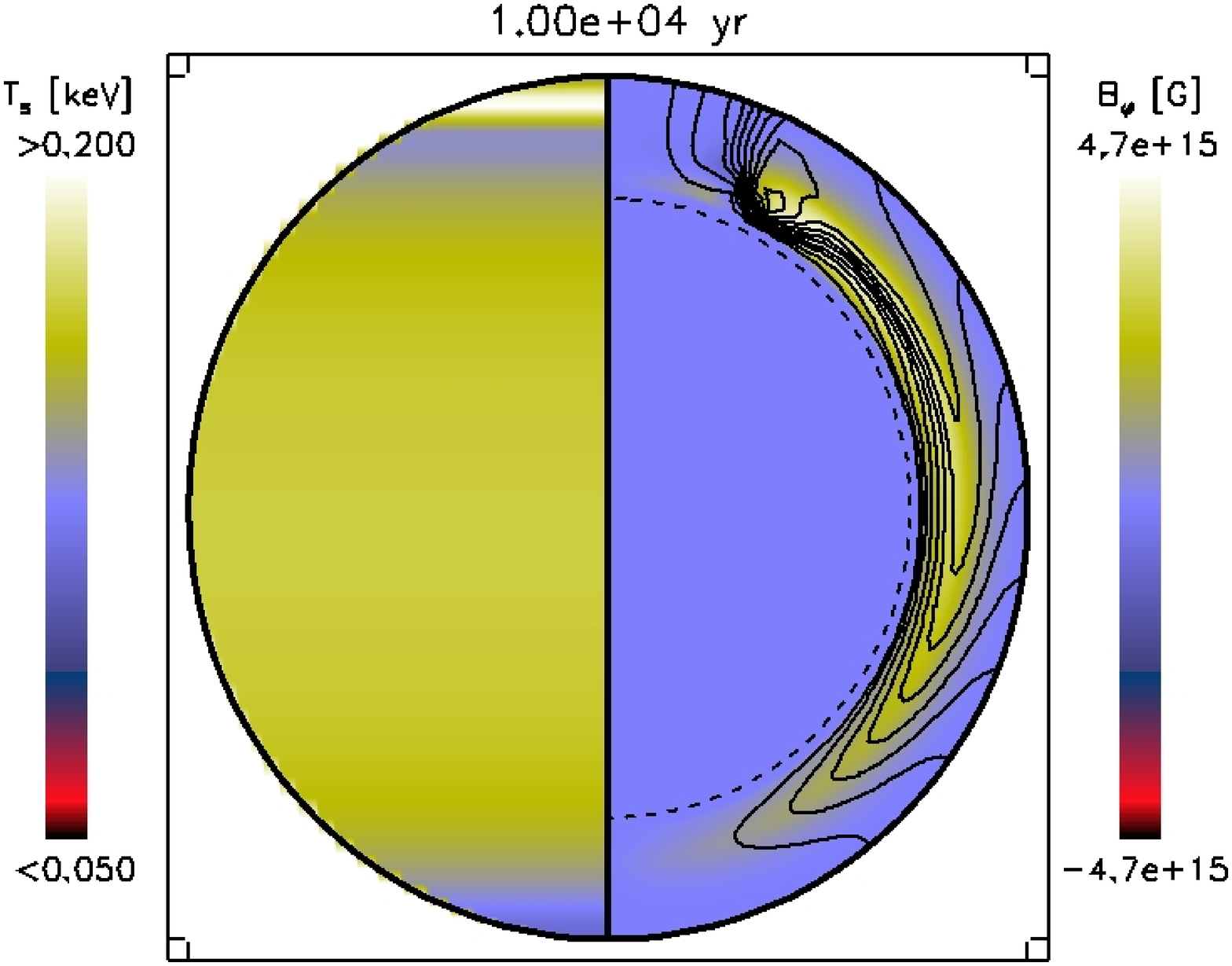}
\includegraphics[width=.3\textwidth]{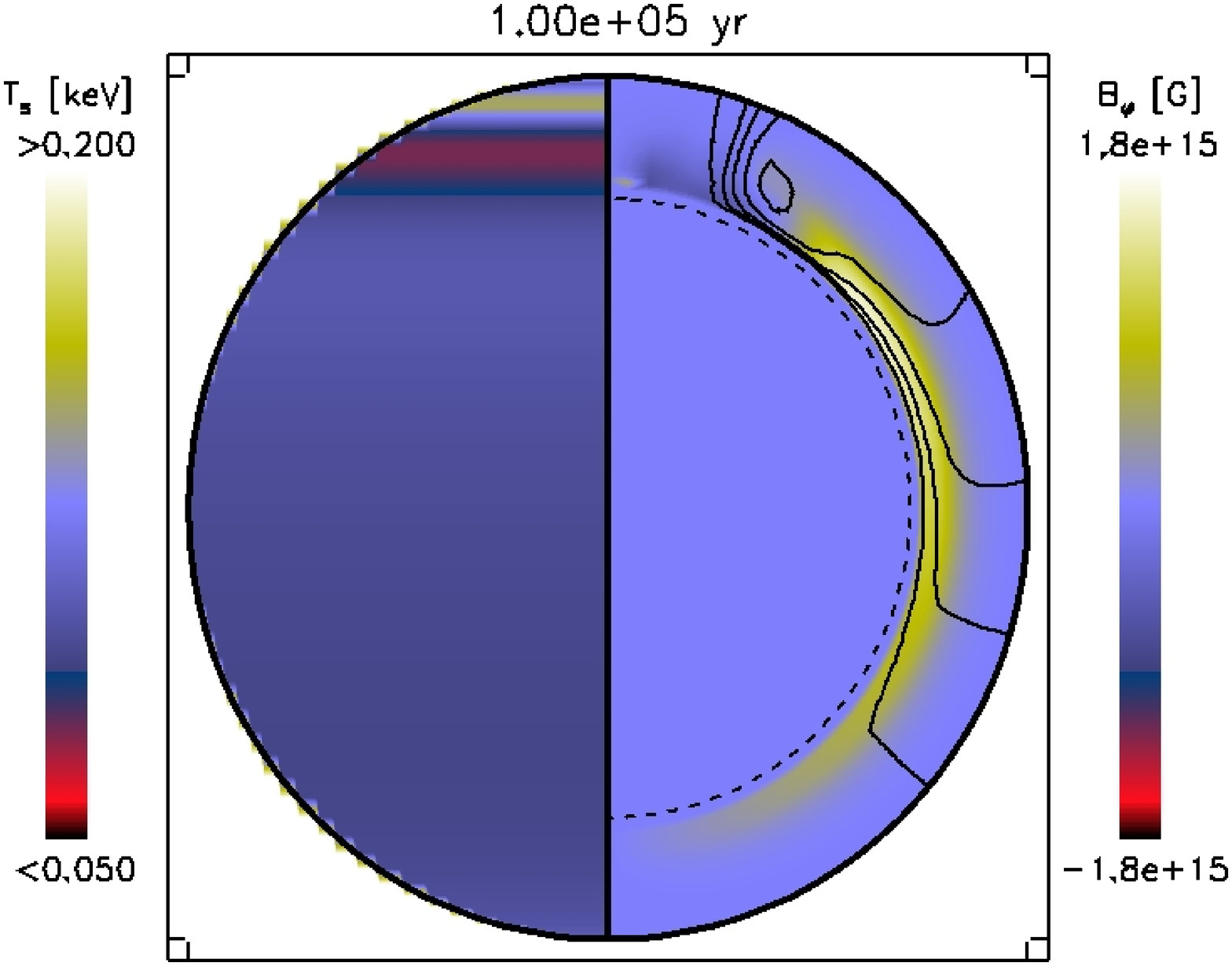}\\

\includegraphics[width=.3\textwidth]{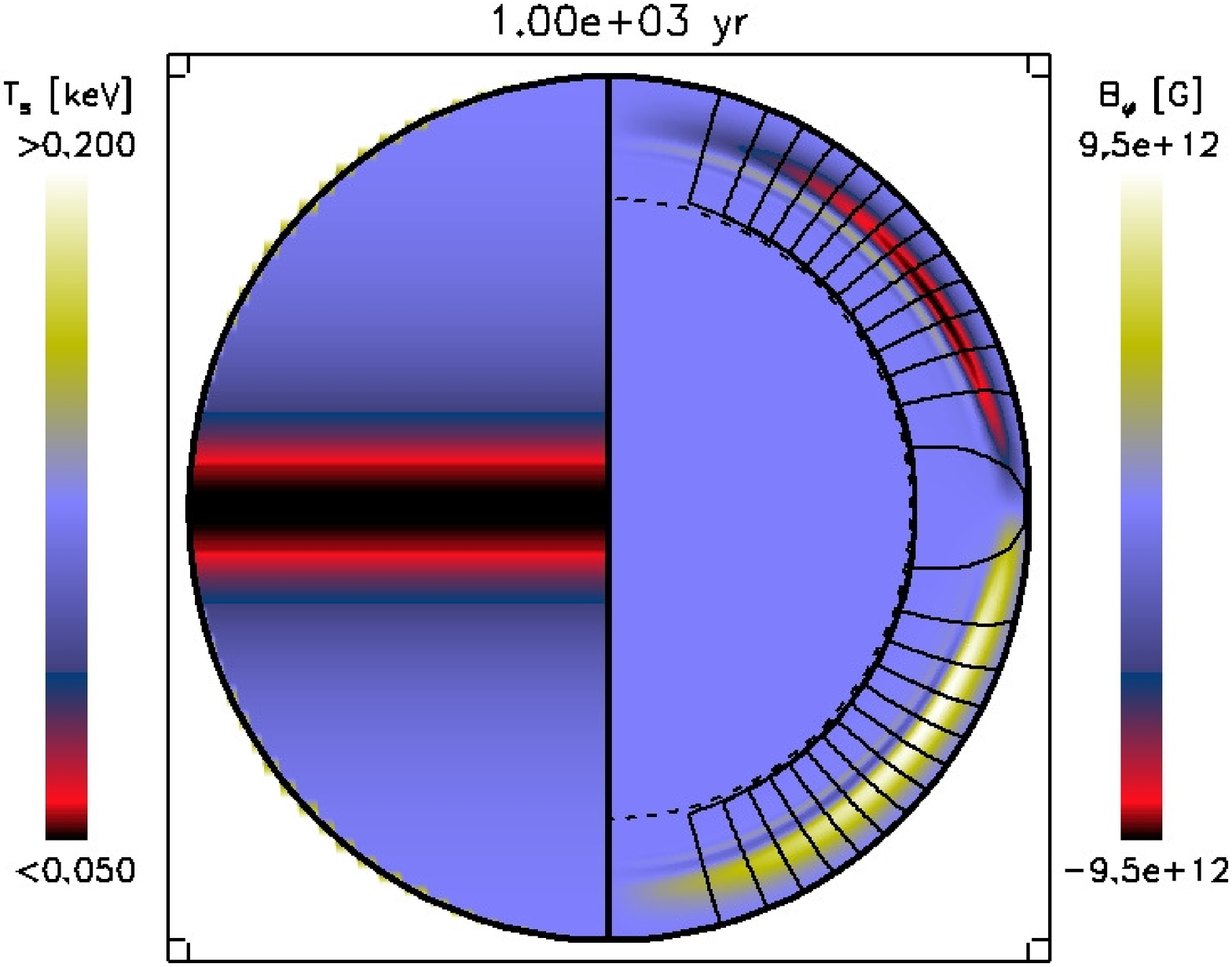}
\includegraphics[width=.3\textwidth]{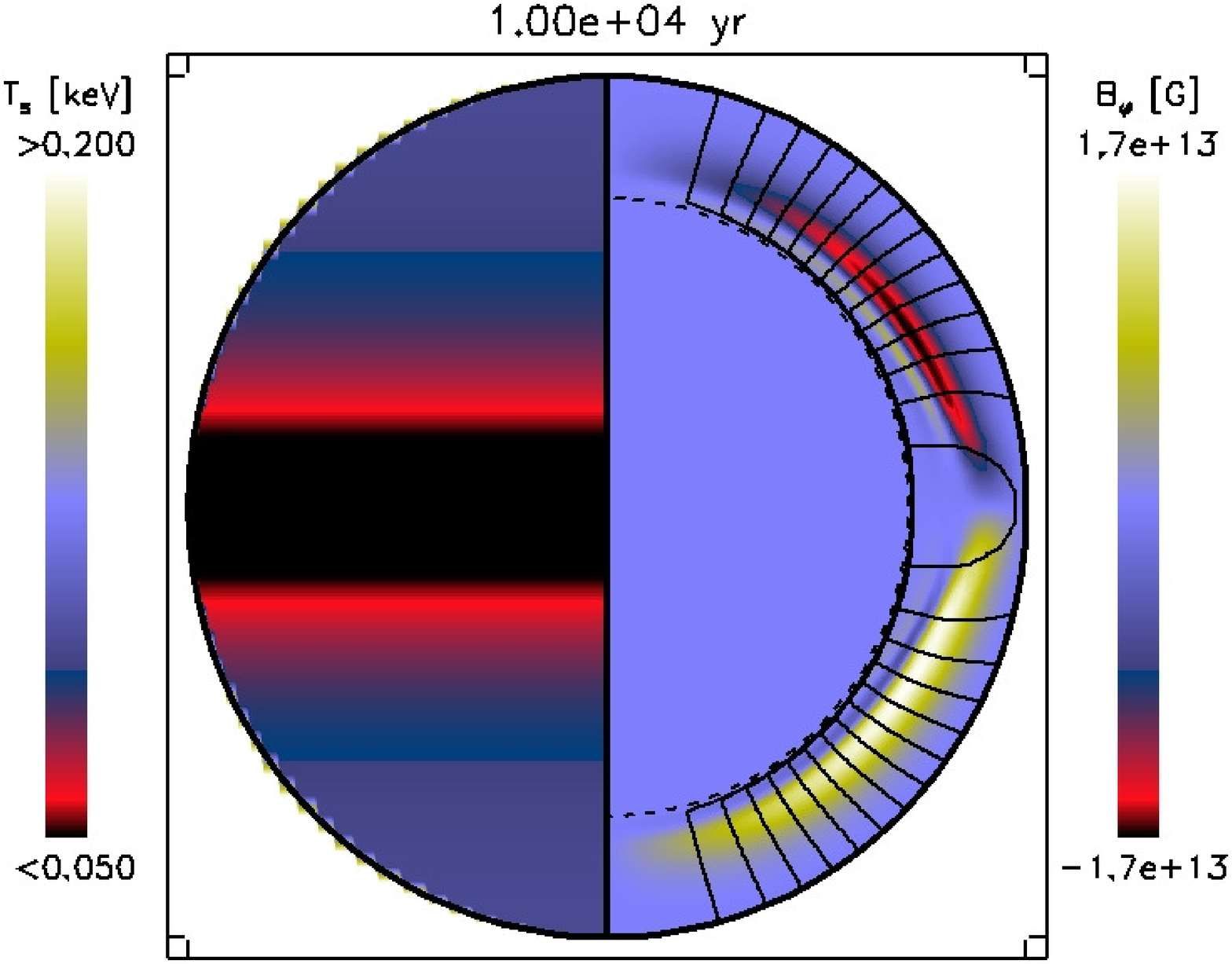}
\includegraphics[width=.3\textwidth]{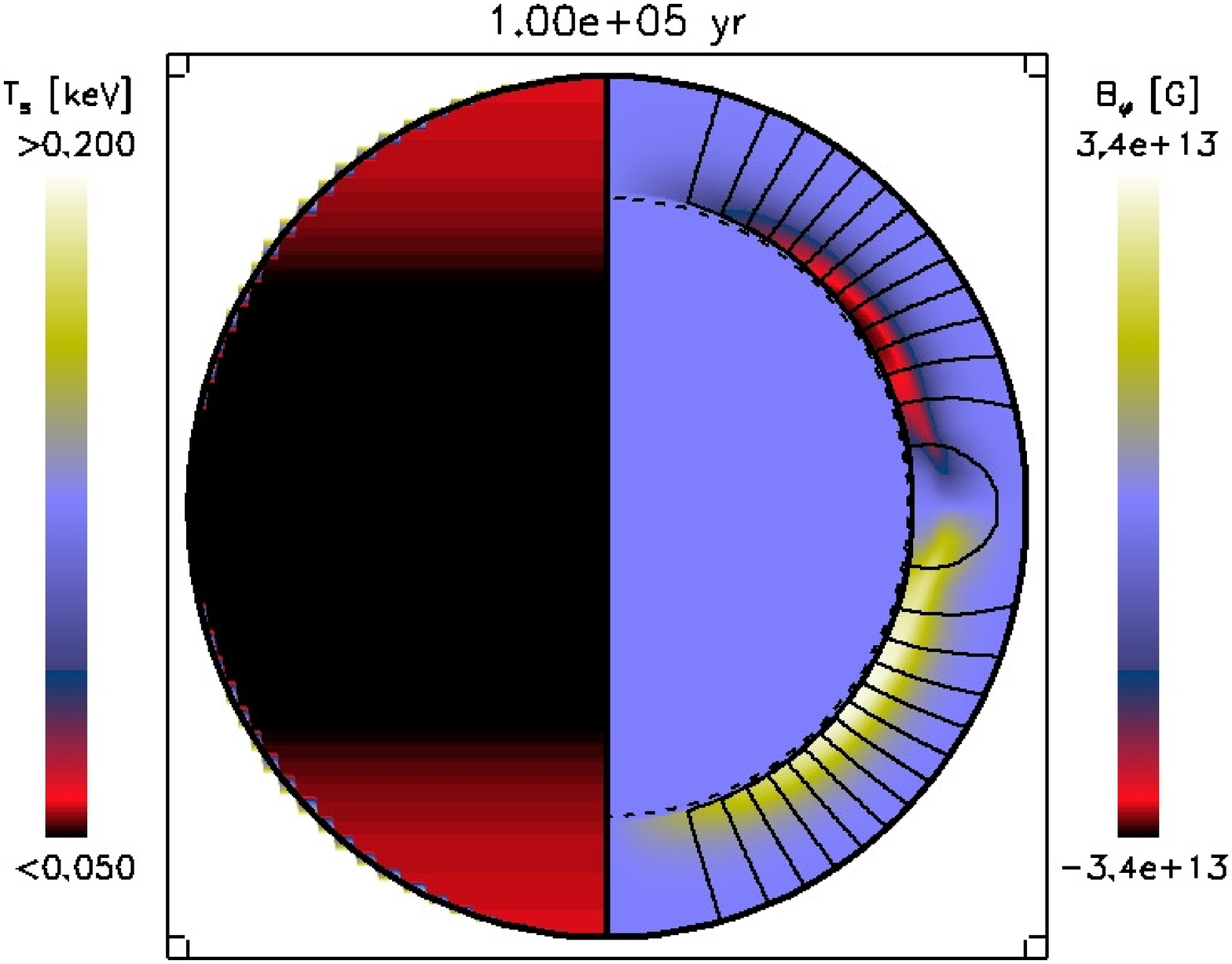}\\

\includegraphics[width=.3\textwidth]{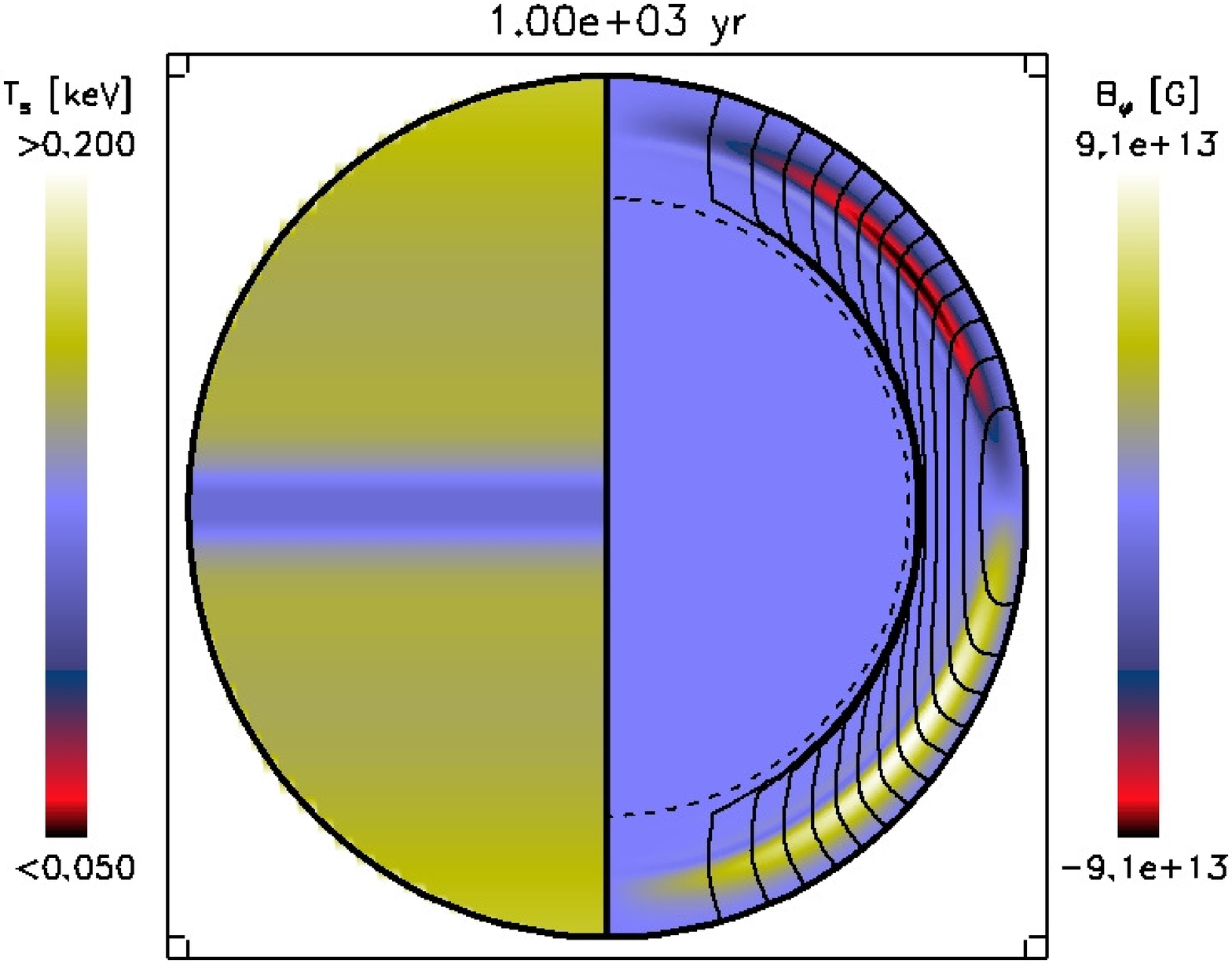}
\includegraphics[width=.3\textwidth]{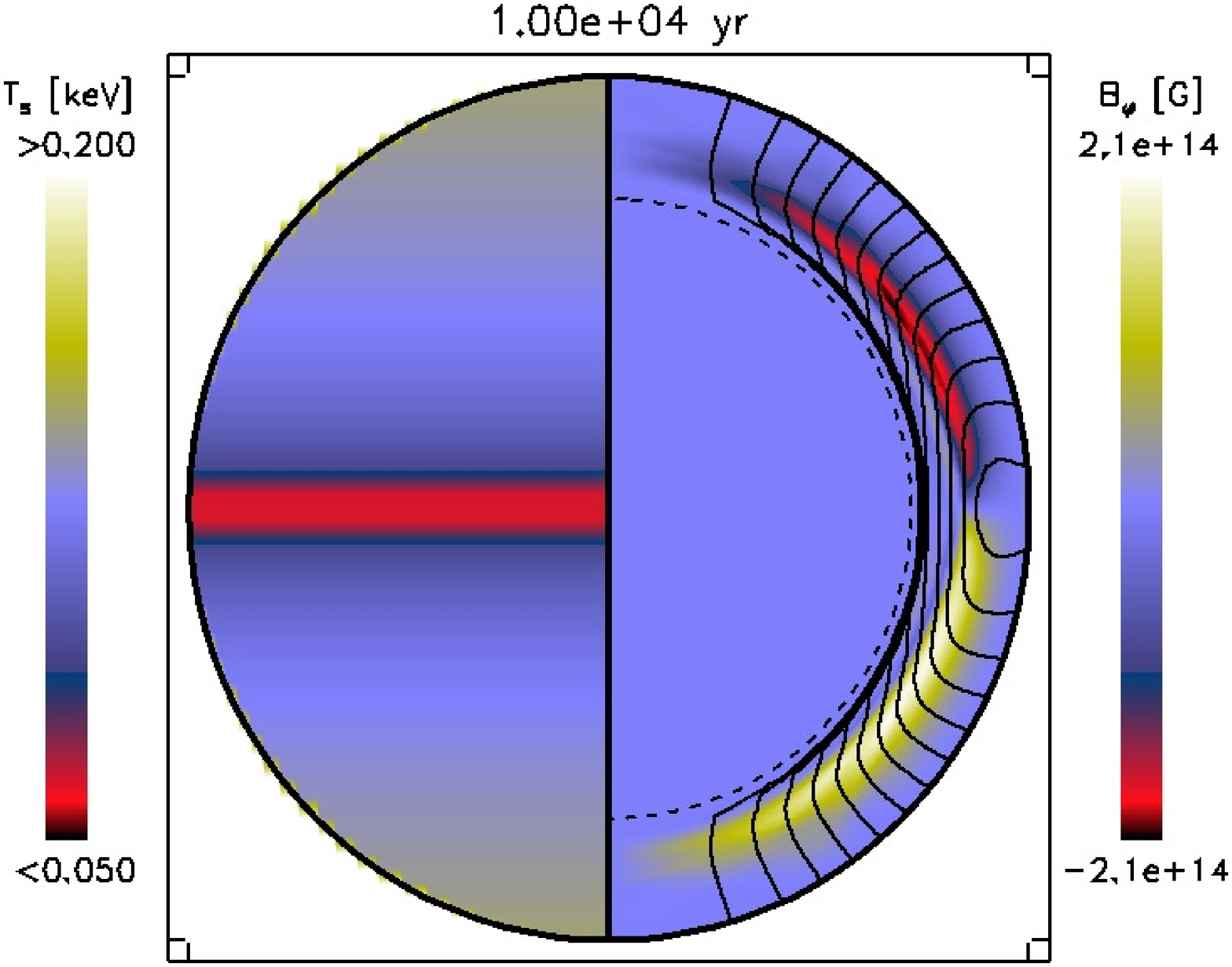}
\includegraphics[width=.3\textwidth]{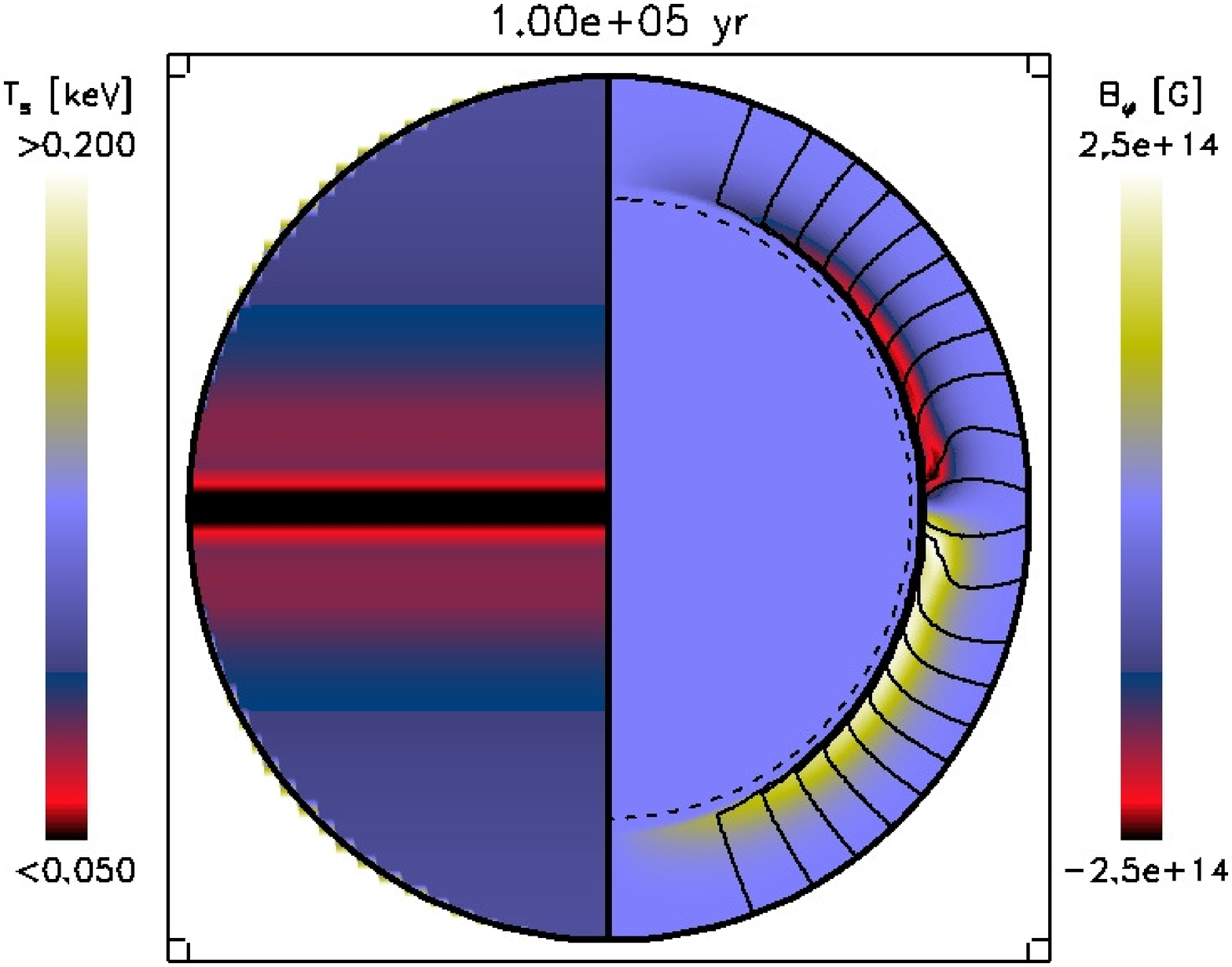}\\
% \caption{Same as Fig.~\ref{fig:b14_evo}, but for model A15.}
%  \label{fig:b15_evo}
% \caption{Same as Fig.~\ref{fig:b14_evo}, but for model A14T.}
%  \label{fig:b14t_evo}
% \caption{Same as Fig.~\ref{fig:b14_evo}, but for model B14. Magnetic field lines penetrate in the core, but they are not shown here.}
%  \label{fig:b14core_evo}
% \caption{Same as Fig.~\ref{fig:b14_evo}, but for model C14. Magnetic field lines penetrate in the core, but they are not shown here.}
%  \label{fig:b14mix_evo}
\caption{Same as the upper panels of Fig.~\ref{fig:b14_evo}, but for models: A15 (first row), A14T (second), B14 (third) and C14 (fourth). In models B14 and C14, the magnetic field lines penetrate in the core, but they are not shown here.}
 \label{fig:other_evo}
\end{figure*}
%%%%%%%%%%%%%%%%%%%%%%%%%%%%%%%%%%%%%%%%%%%%%%%%%%

It is worth to briefly explore the dependence on the NS mass and some of the main microphysical parameters, 
such as the impurity content of the innermost part of the crust (pasta region). 
The top panel of Fig.~\ref{fig:micro} illustrates the sensitivity of the time evolution of $B_p$ to microphysical parameters: 
we use the same initial magnetic field (model A14), but with different NS masses, and different value of $Q_{imp}$ (impurity parameter in the regions of the crust above $\rho>6 \times 10^{13}$ g~cm$^{-3}$). 
Comparing the evolution of the dipolar component
of the magnetic field (top panel) for different NS masses and the same $Q_{imp}$ in the pasta region, we do not see significant differences.
The small quantitative differences are caused by the different thickness of the crust, which varies with 
the NS mass (between $\sim 0.5$ km for M=1.76 $M_\odot$ and $\sim 1$ km for M=1.10 $M_\odot$). 
In the inner crust, the resistivity is likely dominated by electron-impurity scattering at low temperatures. As a consequence, the parameter $Q_{imp}$ controls the evolution at late time ($t\gtrsim 10^5$ yr), when the star is cold enough \citep{pons13}. 
If instead of assuming an amorphous pasta phase, a low value of $Q_{imp}$ in the pasta region is set, the dissipation at late times is much 
slower (compare models with $Q_{imp}=0, 25, 100$ in the pasta region for the same 1.4 $M_\odot$ NS).
In this low temperature, low resistivity regime, the magnetic field evolution becomes dominated by the Hall effect. 
The most extreme case ($Q_{imp}=0$ in the pasta region, cyan line in Fig.~\ref{fig:micro}) shows a sort of stationary oscillatory modes (see also \citealt{pons12}), with small twisted structures propagating meridionally. Another consequence of a low impurity parameter is that, since the magnetic field dissipation timescale is much longer than the star age, a similar asymptotic value of the magnetic field is reached for all NSs born as magnetars \citep{pons09}. On the contrary, in models with a high $Q_{imp}$ in the pasta region, the magnetic field dissipation is maintained for millions of years.

Another important effect is that, because of the extra heat provided
by the dissipation of magnetic energy, the dependence of the
luminosity on other parameters is strongly reduced, compared with non-magnetised NS cooling models.  In the bottom panel of
Fig. ~\ref{fig:micro}, we show that the cooling curves of models with
the same initial $B_p^0=10^{14}$ G, but different masses or impurity
content, are almost indistinguishable for at least $10^4$ yr. Compared
with non-magnetised models (Fig. \ref{fig:b0}), differences are
strongly reduced under the presence of Joule heating. We also note
that varying the superfluid gaps has a visible effect only when the
effect of magnetic field decay is negligible, i.e., for low values of $Q_{imp}$ in the pasta region
or weak magnetic fields.

Hereafter we will set the impurity paramter $Q_{imp}=100$ in the pasta region (in the regions of the crust where $\rho > 6\times10^{13}$ g\,cm$^{-3}$), leaving $Q_{imp}=0.1$ elsewhere in the crust, and the NS mass to 1.4
$M_\odot$ for the rest of the discussion.

%%%%%%%%%%%%%
\subsubsection{Dependence on initial magnetic field geometry and strength.}

We continue the discussion by considering how the initial magnetic field
configuration affects the evolution. In the top row of
Fig.~\ref{fig:other_evo}, we present the results for the evolution of
model A15.  The qualitative behaviour is the same as in model A14, but
since the Hall timescale goes as $\propto 1/B$, the dynamics is
accelerated by a factor of ten.  The more relevant role of the Hall term
for stronger magnetic field leads to the formation of a discontinuity in
$B_\phi$.  In this particular case the current sheet is located at the
equator (see the compressed lines and the ``colliding'' toroidal magnetic field
of opposite sign in the second panel), where the dissipation is
strongly enhanced \citep{vigano12a}. Note that the average surface
temperature is higher than for model A14 (Fig.~\ref{fig:b14_evo}).

The second row of Fig.~\ref{fig:other_evo} shows the evolution of model
A14T. The magnetic energy initially stored in the toroidal component
is $98\%$ of the total energy. Compared with model A14
(Fig.~\ref{fig:b14_evo}), there are significative differences. In particular, the symmetry with respect to the equator
is broken because the vertical drift of the toroidal component acts
towards the north pole. The initial poloidal magnetic field is distorted and
bent at intermediate latitudes, with the formation of a small scale bundle of magnetic field lines, which has been proposed to be necessary for the radio pulsar activity \citep{geppert13}. This again locally increases the
dissipation rate. At about $10^3$~yr, the northern hemisphere is, on
average, warmer than the southern hemisphere, and a characteristic hot
polar cap is observed. Later, the geometry of the magnetic field becomes more complicated, with the formation of localized bundles of nearly radial magnetic field lines. This results in a temperature pattern with hot and cool rings.

In the third row of Fig.~\ref{fig:other_evo}, the evolution of the core-extended
configuration (model B14) is shown. The magnetic field lines are
penetrating inside the core although, for clarity, the figure shows
only the configuration in the crust (enlarged for visualization, see
Fig. \ref{fig:initial_b} for the real scale). The magnetic field in the core is
almost frozen due to the high electrical conductivity of the interior (as discussed above).  Some
weak Hall activity is developed at the bottom of the crust, but the
maximum value of the toroidal field generated is about one order of
magnitude weaker than in model A14, and the poloidal lines do not
suffer any significant bending. In addition, the reduced heat
deposition in the crust results in a much cooler surface compared to
all the other models, and similar to the low field cooling models.

Between the two extremes (mostly crustal currents vs. mostly core
currents), we also consider the intermediate case of model C14, shown
in the bottom row of Fig.~\ref{fig:other_evo}. The presence of crustal currents
activates the Hall dynamics and leads to a similar evolution as for
model A14. The only relevant difference is that at very late times
there will be a long-lasting magnetic field in the core.

%%%%%%%%%%%%%%%%%%%%%%%%%%%%%%%%%%%%%%%%
\begin{figure}
 \centering
\includegraphics[width=.45\textwidth]{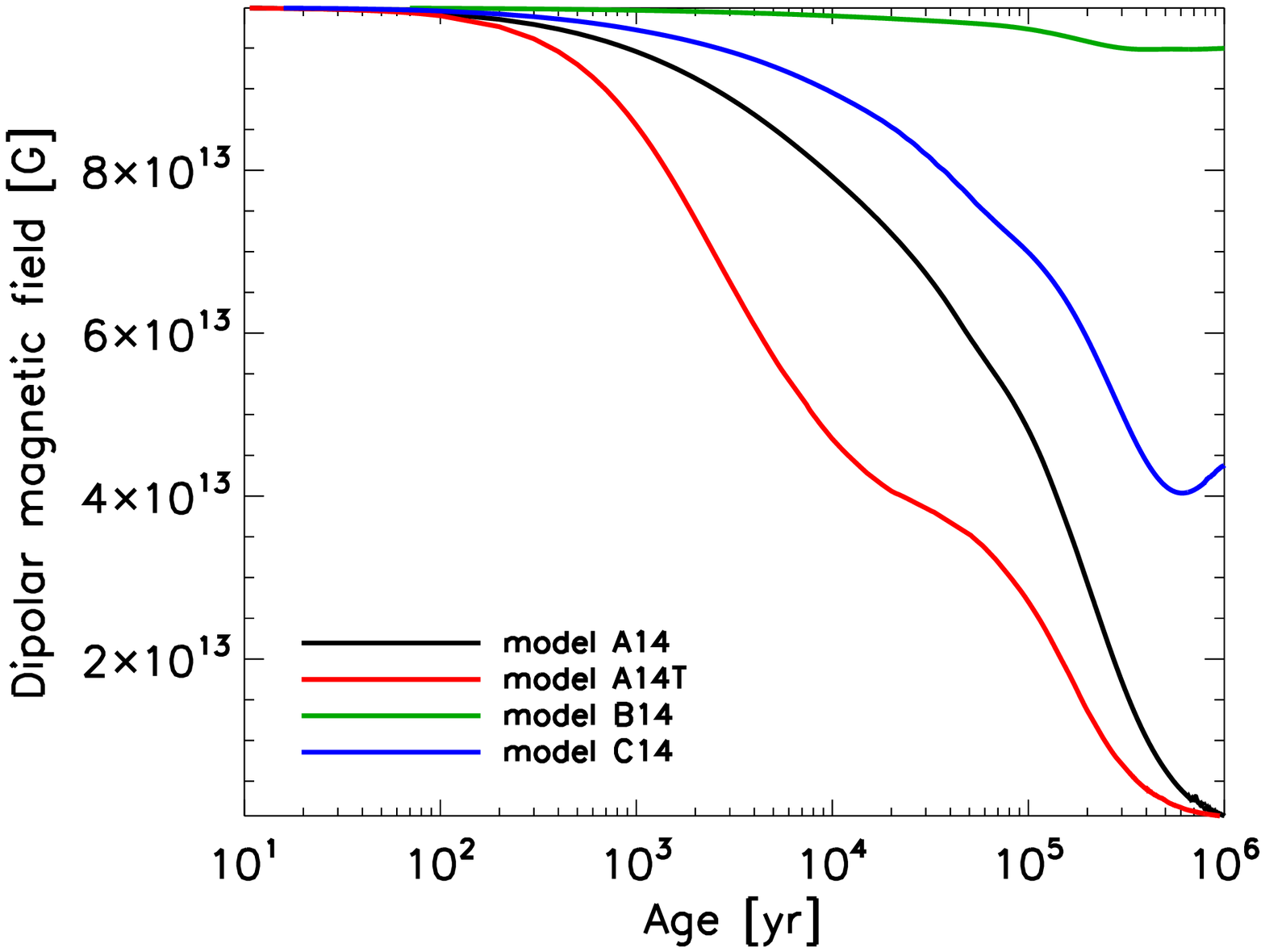}
\includegraphics[width=.45\textwidth]{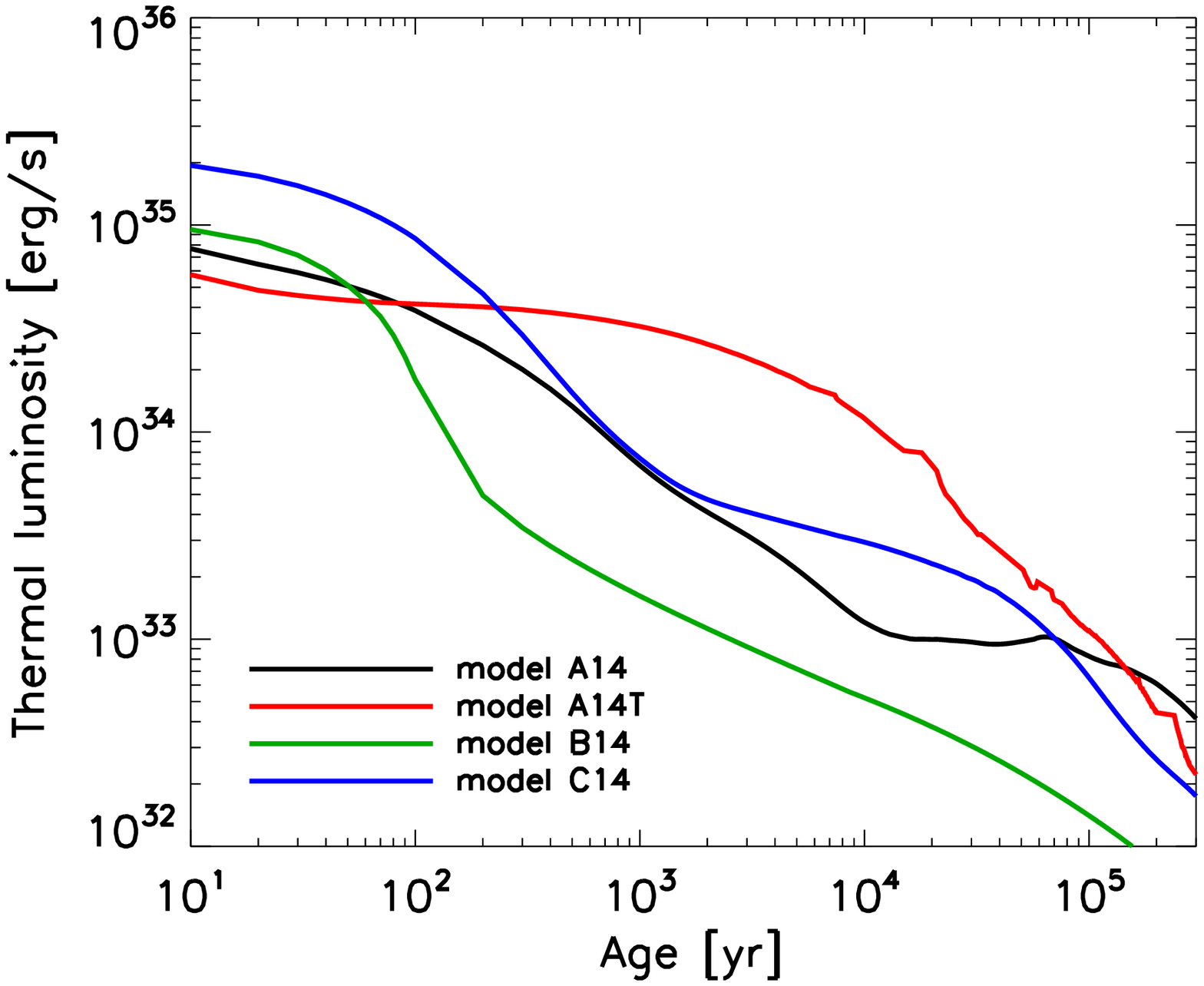}
\caption{Evolution of $B_p$ (top) and luminosity (bottom) for different magnetic configurations: 
model A14 (black); model A14T (red); model B14 (green); model C14 (blue).} 
 \label{fig:geometry}
\end{figure}
%%%%%%%%%%%%%%%%%%%%%%%%%%%%%%%%%%%%%%%%%%%

In the top panel of Fig.~\ref{fig:geometry} we compare the evolution of $B_p$ for four models, all with $B_p^0=10^{14}$ G.
The difference between crustal-confined (models A14 and A14T) and the core-extended configuration (model B14) is evident. In the latter, the magnetic field is almost constant during 1 Myr. The configuration with an extremely strong initial toroidal component shows a larger luminosity
(up to one order of magnitude at some age), due to the larger internal energy reservoir, but the higher temperatures also lead to a much faster decay of $B_p$ than for model A14. Moderate values of toroidal magnetic fields (e.g., with half of the magnetic energy stored in) lead to an intermediate decay, between models A14 and A14T. In the hybrid case, model C14, $B_p$ dissipates the crustal currents before oscillating around the value given by the currents circulating in the core.
In the bottom panel of Fig.~\ref{fig:geometry}, we compare the luminosity for the same four models. Model B14 is similar to
the non-magnetised case (Fig. \ref{fig:b0}), because most of the current is in the core and does not decay. On the other hand, when currents are totally or partially placed in the crust, extra heat is deposited in the outer layers, resulting in 
 higher luminosities.
One important result, not discussed in previous works, is that the Hall term causes the self-regulation of the
ratio of the toroidal and poloidal components. The differences due to particular initial ratios are much less important than for
the Ohmic case \citep{pons09}, except when the initial toroidal component is significantly larger than the poloidal magnetic field. 
If the initial energy stored in poloidal and toroidal components are similar, the difference between models with and without initial toroidal fields is 
much smaller. 

%%%%%%%%%%%%%%%%%%%%%%%%%%%%%%%%%%%%%%%%
\begin{figure}
 \centering
\includegraphics[width=.45\textwidth]{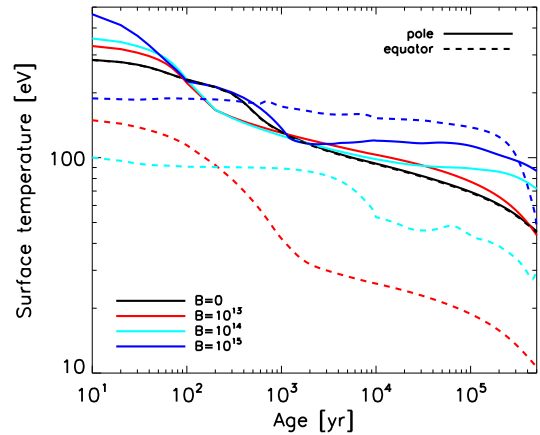}
\caption{Surface temperature at the pole (solid) and equator (dashed) for $B_p^0=0,10^{13},10^{14},10^{15}$ G (black, red, blue, green lines), in all cases for type A configurations.} 
 \label{fig:tevo_m140}
\end{figure}
%%%%%%%%%%%%%%%%%%%%%%%%%%%%%%%%%%%%%%%%

To conclude this section, we discuss the dependence of the surface
temperature on magnetic field strength.  In
Fig.~\ref{fig:tevo_m140}, we compare the surface temperatures at the
pole and equator for $B_p^0=10^{13},10^{14},10^{15}$ G. At early
times, little difference is seen in the evolution of the temperature
at the pole, regardless of the magnetic field strength. For the stronger
magnetic fields, the polar temperatures can be kept high for longer times. The
largest differences, however, are seen in the equatorial
temperatures. For weakly magnetised models, the anisotropic blanketing effect of the magnetised
envelope leads to much cooler equatorial regions. Conversely, for the
strongly magnetised models, the high energy deposition rate by magnetic field decay
is able to compensate this effect and the equator is actually warmer
than the pole after a few hundred years. A dedicated work will explore the temperature anisotropies and their observational consequences \citep{perna13}.

%%%%%%%%%%%%%%
\subsection{Expected outburst rates.}\label{sec:bursts}

The evolution of the magnetic field in the NS causes continuous
stresses on the crust. Generally, magnetic stresses are balanced by
elastic stresses. However, during the evolution, local magnetic
stresses can occasionally become too strong to be balanced by the
elastic restoring forces of the crust, which hence breaks, and the extra
stored magnetic/elastic energy becomes available for powering the
observed bursts and flares \citep{thompson95,thompson96}. 

The maximum stress that a NS crust can sustain has been estimated
with analytical arguments and through molecular dynamics simulations. In particular, 
\cite{chugunov10} obtained the fit 
\begin{equation}
\sigma_b^{\rm max} = \left(0.0195-\frac{1.27}{\Gamma-71}\right)\; n_i\,
\frac{Z^2 e^2}{a}\;,
\label{eq:sigma}
\end{equation}
where $\Gamma=Z^2e^2/aT$ is the Coulomb coupling parameter,
$a=[3/(4\pi n_i)]^{1/3}$ is the ion sphere radius, $n_i$ the ion
number density, $Z$ the charge number, $e$ the electron charge, and
$T$ the temperature.

\cite{perna11} combined results from the magnetothermal evolution of NSs \citep{pons09},
with a computation of the magnetic stresses exerted on the NS crust.
This allowed them to estimate the frequency and the location (depth and latitude)
of {\em starquakes}, i.e. NS crust failures. The released energies were
also estimated from the stored elastic energy.  
Their work showed that the classification of objects as AXPs, or SGRs, 
or high-$B$ NSs, or 'normal' radio pulsars, etc. does not correspond
to an underlying intrinsic physical difference: outbursts can occur also in objects outside of the 
traditional magnetar range, albeit with a lower probability. 
The follow up study by \cite{pons11} further highlighted the 
importance that the toroidal magnetic field has on the NS observed phenomenology:
two objects with similar inferred dipolar $B$-field (as measured by $P$ and
$\dot{P}$) can display a very different behaviour depending on the strength
of the (unmeasured) internal toroidal component. The stronger the latter, the higher the
luminosity of the object, and the more likely starquakes are to occur.
In order to follow numerically the long term
evolution of NSs, these works used an approximation to treat the
Hall term, which was only included when
studying the short term evolution of the magnetic field.
In the current work, the above approximation has been released, and the
effect of the Hall term is fully implemented. Hence, in the following we update their results 
by using a similar formalism to compute the outburst statistics from the results of our last
simulations.

%%%%%%%%%%%%%%%%%%%%%%%%%%%%%%%%%%%%
\begin{figure}
 \centering
\includegraphics[width=.45\textwidth]{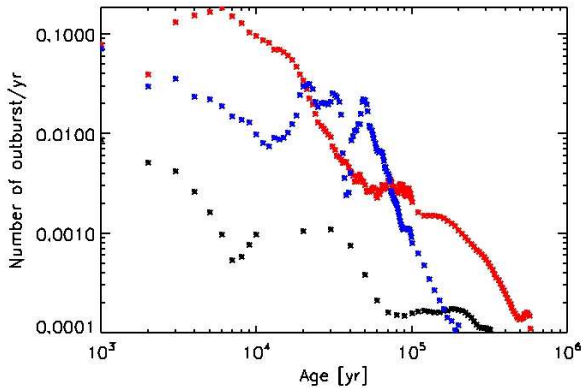}
\caption{Expected outburst rate for model A with $B_p^0=3\times 10^{14}$ G (black), $B_p^0=10^{15}$ G (red) and $B_p^0=10^{14}$ G with initial strong toroidal field ($B_t^0=5\times 10^{15}$ G).} 
 \label{fig:outburst}
\end{figure}
%%%%%%%%%%%%%%%%%%%%%%%%%%%%%%%%%%%%

In Fig.~\ref{fig:outburst} we show the evolution of the predicted outburst rate for models A15 (red), a model of type A with $B_p^0=3\times 10^{14}$ (black), and the model A14T (blue). While young NSs may undergo energetic outbursts every few years (for model A15) or tens of years (for the model with a weaker magnetic field), middle age sources are expected to show transient activity only every few thousands of years, and with less energy involved. Other models with initial lower magnetic fields, like model A14, do not give an appreciable event rate (less than $10^{-3}$ events per year even for young NSs). However, in presence of strong toroidal magnetic field, model A14T, the outburst rate is strongly enhanced.

Assuming a NS birth rate of $10^{-2}$ per year, there must be $\approx 10^4$ NSs in the Galaxy with ages of $\lesssim 1$ Myr. If 10\% of them are born with $B\gtrsim 10^{14}$ G, a naive extrapolation of the estimated event rate at that age ($\lesssim 10^{-4}$ per year) leads to an outburst rate by old magnetars of one every few years. Therefore, we expect that more and more objects of this class will be discovered in the upcoming years. It is interesting to note that such old magnetars are expected to be detected only via their outburst activation, being too faint in the quiescent state even for very deep X-ray surveys. Since the launch of the {\em Swift} satellites in 2004, our capability of detecting and monitoring galactic transients has largely increased, and 5 new magnetars have been discovered in the past years, two of those being relatively old and displaying low inferred magnetic field: SGR~0418 \citep{rea10,rea13} and Swift~J1822 \citep{rea12,scholz12}.

On the other hand, the observational detection of an outburst event depends on the quiescent flux. The brightest sources, with $L\gtrsim 10^{35}$ erg/s, will experience a barely detectable flux enhancement, while for dimmer magnetars ($L\sim 10^{33}-10^{34}$ erg/s) the flux increase by 2 or 3 orders of magnitude during the outburst and are much easier to detect \citep{pons12b}.

%%%%%%%%%%%%%%
\section{The unification of the NS zoo.}
\label{sec:unification}

%%%%%%%%%%%%%%%%%%%%%%%%%%%%%%%%%%%%%%%%%%%
\begin{figure*}
% \centering
\includegraphics[width=.7\textwidth]{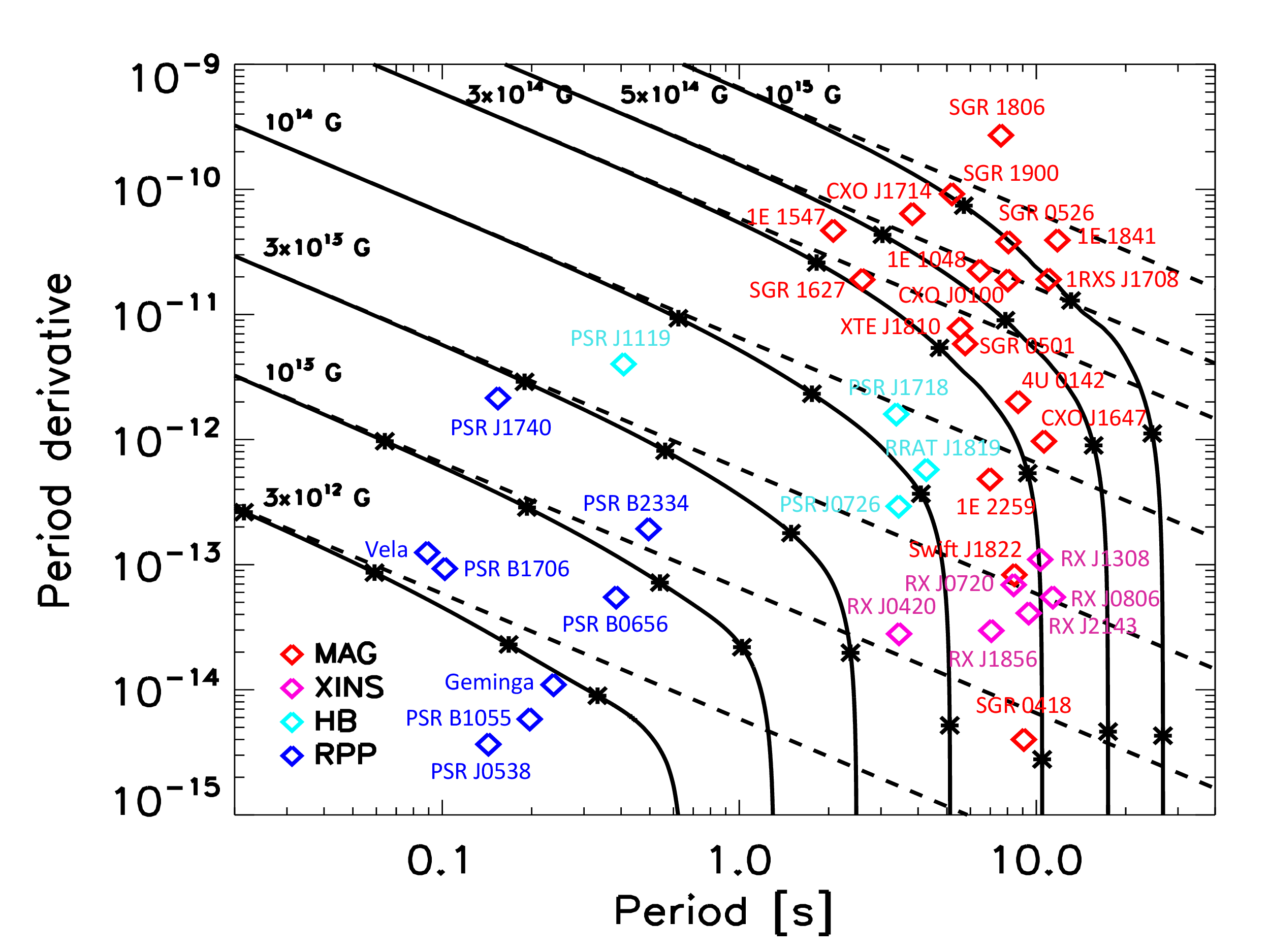}
\caption{Evolutionary tracks in the $P-\dot{P}$ diagram with mass and radius of our model, with $B_p^0= 3\times10^{12}, 10^{13}, 3\times10^{13}, 10^{14}, 3\times10^{14}, 10^{15}$ G. Asterisks mark the real ages $t=10^3,10^4,10^5,5\times10^5$ yr, while dashed lines show the tracks followed in absence of magnetic field decay.} 
 \label{fig:ppdot_m140}
\end{figure*}
%%%%%%%%%%%%%%%%%%%%%%%%%%%%%%%%%%%%%%%%%%%

After having thoroughly reviewed the available observations in
section \S 2, and having presented in \S 4 
magneto-thermal evolution models that include the effects of magnetic field in the
cooling history of NSs, we now proceed to compare 
the data with theoretical models.

The spin--down behavior of a rotating NS is governed by the
energy balance equation relating the loss of rotational energy to the
energy loss due to emission of electromagnetic radiation, 
gravitational radiation, or winds.  In the following we will assume
that the dominant contribution to the torque is electromagnetic radiation. 
The spin-down formula for force-free magnetospheres, and valid for oblique rotators \citep{spitkovsky06}, gives 
the following relation
\begin{equation}
\label{eq:spindown}
I\Omega \dot{\Omega} \approx  - \frac{B_p^2R^6\Omega^4}{4 c^3} (1+\sin^2{\chi})
%  + \dot{E}_{other}
\end{equation}
where $R$ is the NS radius, $\chi$ is the angle between the rotational and the magnetic axis, $c$ is the speed of light, 
$\Omega=2 \pi/P$ is the angular velocity, and $I$ is the moment of
inertia of the star. 
In our model with $M=1.4 M_\odot$ and $R=11.6$ km, the moment of inertia is $I=1.33\times 10^{45}$ g
cm$^2$. If a star with a different mass is considered, the value of
$I$ can change by $\sim 30\%$ at most. Note that, compared with the
standard formula for vacuum, $I\Omega \dot{\Omega} =
-\frac{B_p^2R^6\Omega^4}{6 c^3}\sin^2{\chi}$, the torque in
Eq.~(\ref{eq:spindown}) is larger by a factor of $1.5-3$ compared with the
orthogonal vacuum rotator and it is non-zero even for an aligned rotator
($\chi=0$).

We have numerically integrated Eq.~(\ref{eq:spindown}) with the time
dependent $B_p(t)$ obtained from our simulations. We assumed an aligned rotator (constant $\chi=0$), and an initial period 
$P_0=0.01$~s, low enough for the results not to be sensitive
to the particular value of the initial period. The integrated period at a given age can change by a factor $\sim 2$ if the uncertainties in the inclination angle are considered in the spin-down formula. Here we do not consider two possible effects that can contribute to the evolution of timing properties: the variation with time of the angle, $\chi=\chi(t)$, and the growth of the superfluid region in the core 
during the early stages, resulting in a change of the effective moment of inertia, $I=I(t)$.

In Fig.~\ref{fig:ppdot_m140} we show the evolutionary tracks in the
$P-\dot{P}$ diagram for a $1.4 M_\odot$ NS with type A geometry and
different initial values of the magnetic field strength, compared
to the measured timing properties of X-ray pulsars. Asterisks
mark different ages ($t=10^3,10^4,10^5,5\times10^5$ yr), while dashed lines
show evolutionary tracks that the star would follow if there were no
magnetic field decay. $B_p$ is almost constant during an initial epoch,
$t\lesssim 10^3-10^5$ yr, which depends on the initial $B_p^0$: stronger
initial magnetic fields decay before  weaker ones. After that initial phase, tracks
bend downwards due to the dissipation of the magnetic field under the
combined action of the Hall effect with the large resistivity in the innermost part of the crust. 
Therefore, an asymptotic value of $P$ is reached.  The age at which this vertical
drop in the track begins depends on the initial magnetic field, being roughly
$\sim 10^5$ yr for strong magnetic fields ($B_p^0\gtrsim 10^{14}$ G), and up to
several Myr for weaker magnetic fields. The specific magnitude of the asymptotic value
of $P$ depends on the initial magnetic field, the resistivity in the innermost part of the crust, 
the mass of the star and the angle $\chi$ (see Eq.~\ref{eq:spindown}).

An inspection of the data reveals that there may be common evolutionary links
among groups of isolated NSs.  One of such links roughly includes nine
magnetars (1E~1547, SGR~1627, SGR~0501, XTE~J1810, CXO~J1647, 1E~2259, 4U~0142 and the low B magnetars Swift~1822, SGR~0418) and five of the XINS (excluding only the faint source RX~J0420 and RX~J1605, that has no measure of $P$ and $\dot{P}$). For our particular model, this
track corresponds to a NS born with $B_p^0=3 \times 10^{14}$ G.  In the upper right corner,
we identify a second group of eight {\it extreme magnetars},
characterized by a larger $\dot{P}$ (CXOU~J1714, SGR~1900, 1E~1048, SGR~0526, CXOU~J0100, 1RXS~J1708, 1E~1841, SGR~1806). These would be consistent with 
NSs born with a higher initial magnetic field $B_p^0 \sim 10^{15}$ G.

In light of these results, we can revisit the classical comparison
between cooling curves and data reviewed in \S 4.1 with the
predictions of the full magneto-thermal evolution, for a range of
magnetic field strengths.
In Fig.~\ref{fig:cooling_data_all} we show the luminosity as a function of time for
different values of the initial magnetic field up to $B_p^0=3\times
10^{15}$ G, and we compare these cooling curves with the observational
data. Compared with the non-magnetised cooling curves
(Fig.~\ref{fig:b0}), the most relevant difference is that the inclusion
of the magnetic field allows to generally explain objects with high luminosities.
Several authors already pointed out the observed correlation between inferred magnetic field and luminosity or temperature in magnetars (e.g., \citealt{pons07a,an12}). We are now able to confirm and quantify the trend. 
Magnetic fields above $B\gtrsim 10^{14}$ G are strong enough to noticeably 
heat up the crust and power the observed X-ray radiation. The
cooling timescale for strongly magnetised objects is about one order
of magnitude larger than for the weakly magnetised NSs.

For clarity of discussion, we now comment separately on the sources classified in three groups:
\begin{itemize}
\item {\it NSs with initial magnetic field $B_p^0\lesssim 10^{14}$ G.} This group includes all RPPs, two high-$B$ PSRs (PSR~J0726 and PSR~J1119), and two of the XINS (RX~J0420 and RX~J1605). For these relatively low magnetic fields, the luminosity
  is not expected to be significantly affected by the presence of the
  magnetic field.
\item {\it NSs with initial magnetic field $B_p^0\sim 1-5\times 10^{14}$ G.} This group includes two high B pulsars
  (PSR~J1819 and PSR~J1718), 11 magnetars and 5 XINS.
\item {\it NSs with initial magnetic field $B_p^0\gtrsim 5\times 10^{14}$ G.} The 8 magnetars in the upper-right corner of Fig. \ref{fig:ppdot_m140}. 
\end{itemize}                                                                                                                                                                                         
We emphasize that this classification does not reflect intrinsic differences between groups; it is simply an arbitrary grouping that helps to highlight the evolutionary paths as a function of the initial magnetic field strength.

\subsection{NSs with initial magnetic field $\lesssim 10^{14}$ G.}
\begin{figure*}
 \centering
\includegraphics[width=.7\textwidth]{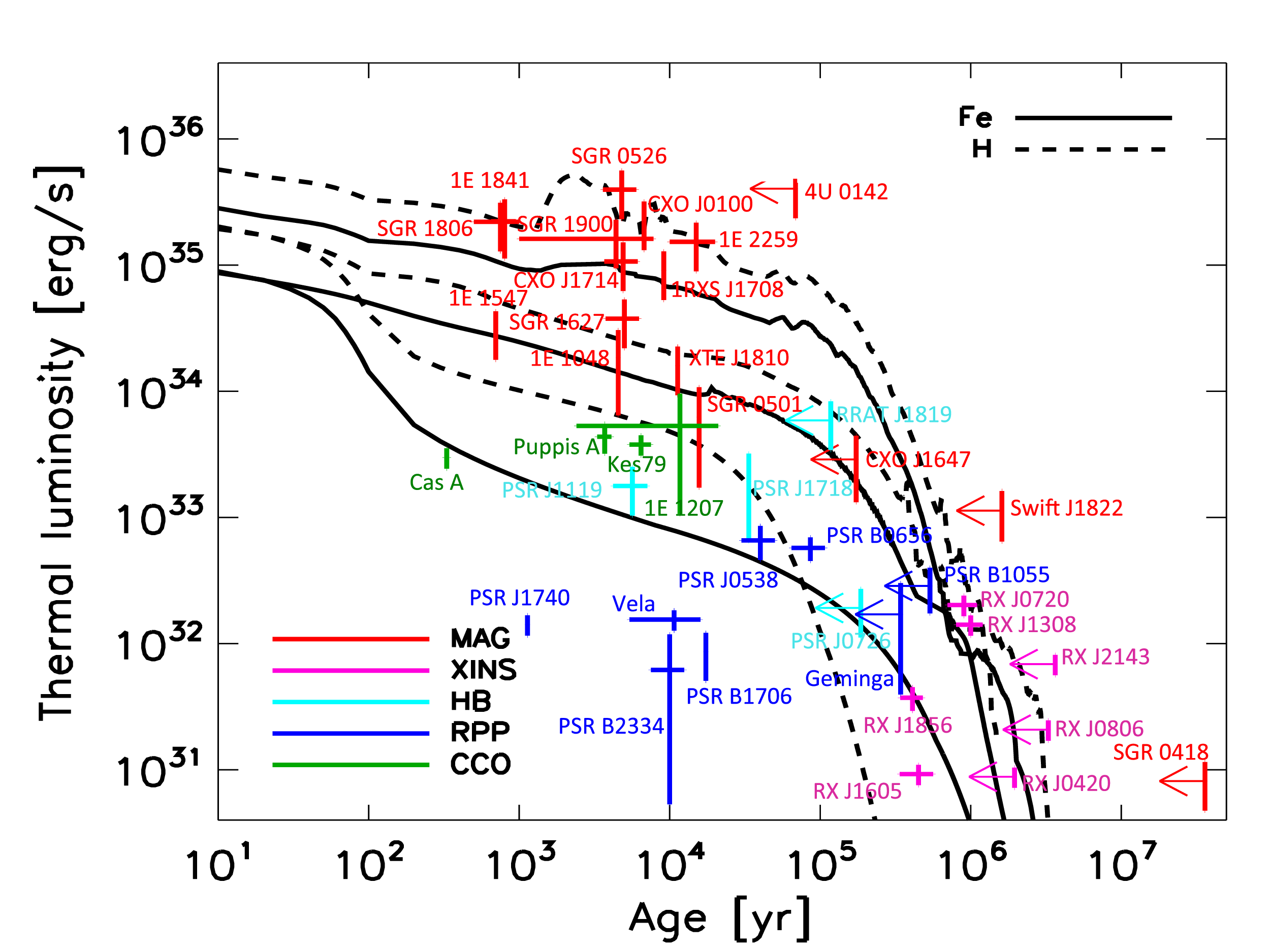} \\
\caption{Comparison between observational data and theoretical cooling curves. 
Models A with $B=0, 3\times 10^{14}, 3\times 10^{15}$ G are shown for Fe envelopes (solid) and light-element envelopes (dashed).}
 \label{fig:cooling_data_all}
\end{figure*}

As already discussed in Sec.~\ref{sec:theory}, the standard cooling curves can account for all
the observed sources by simply varying the star mass. We note that
there is actually more freedom in the theoretical models than in the
particular NS model shown in Fig.~\ref{fig:cooling_data_all}, in which the microphysics input (e.g. gaps) and the NS mass has been fixed. 
Considering the uncertainties in the inferred ages and luminosities, the cooling curves for weakly magnetised NSs are consistent with all the observational data, with a few particular cases which are worth discussing. 

First, CCOs have very low $\dot{P}$ (below the shown range), implying a weak external magnetic field. Due to the small
spin-down rate, their periods have not changed appreciably since birth, and no information about their real ages can be inferred from timing properties. The weak inferred magnetic field apparently contrasts with the observed surface anisotropies and high luminosities (significantly higher than for standard RPPs) of the sources in the SNRs Kes 79 and Puppis A. Light-element envelope models (see Fig. \ref{fig:b0}) can reconcile their age and large luminosity, but not the large anisotropy. The latter can be explained by is the hidden magnetic field scenario \citep{young95,muslimov95,geppert99,ho11,shabaltas12,vigano12b}, in which the object has actually a strong subsurface magnetic field, which has been screened by the initial fallback of the supernova debris onto the NS. The slow reemergence of the surface magnetic field happens on timescales of $10^3-10^5$ yr. This scenario physically justifies the models with light-element envelope.

On the other hand, a few sources (Vela pulsar, PSR~B2334 and PSR~J1740) show evidence for enhanced cooling
by direct URCA processes (or alternative exotic fast cooling neutrino processes), as already discussed by many authors.  
With our equation of state and microphysics setup, this can happen only for high masses, $M\gtrsim 1.5 M_\odot$.

%%%%%%%%%%%%%%%%%%%%%%%%%%%%%%%%%%%%%%%%%%%

\subsection{NSs with initial magnetic field $\sim 1-5\times10^{14}$ G.}

Comparing the cooling curves to the group of NSs with initial magnetic field $B_p^0\sim 1-5\times10^{14}$ G, we see that,
although a few objects could still be compatible with non-magnetised curves, many others are clearly too luminous and need an
additional heating source (see Fig.~\ref{fig:cooling_data_all}). Our results show that most of them can be
reconciled with the theoretical models for a narrow range of initial
magnetic fields, between $B_p^0\sim3-5 \times 10^{14}$ G.  Note also that,
for the few cases where we have both kinematic and characteristic ages (e.g., some XINSs), the latter overestimates the real age. 
For this reason,  it is likely that the two XINSs and SGR 0418, placed on the right hand side of the cooling curves, 
have in fact real ages $\approx$0.5-1 Myr.

For the most luminous object that would belong to this group, 4U~0142, we do not have any alternative estimate of the age. However note that it is quite similar in both timing properties and luminosity to 1E~2259, whose kinematic age inferred for the SNR~CTB109 associated with 1E~2259 is about $10^4$ yr \citep{castro12}, more than one order of magnitude smaller than the characteristic age. We note that for these two objects it is difficult to reconcile the observed timing properties and luminosity, even with more extreme models (very strong toroidal magnetic field). While their luminosity is compatible with very high magnetic fields,  the timing properties are more consistent with initial $B_p^0 \sim 3\times 10^{14}$ G.
  
Interestingly, these are the two cases in which some weak evidence for the presence of
fallback disks has been reported (see \citealt{wang06} for 4U~0142, and
\citealt{kaplan09} for 1E~2259). In both cases, the IR
measurements are consistent with passive disks, i.e. disks in which
the viscosity has been substantially reduced when they become neutral,
as expected after $\sim 10^3-10^4$~yr \citep{menou01}. These disks are
no longer interacting with the pulsar magnetosphere, and hence the
measured characteristic age would again overestimate the actual NS age,
since in the past there was a higher torque than there is today.

\subsection{NSs with initial magnetic field $\gtrsim 5\times10^{14}$ G.}

For these sources, the cooling curves with iron envelopes for
$B_p^0=10^{15}$~G are barely reaching their large luminosities, which may be an indication that these, still young, sources possess
light-element envelopes,  or perhaps these objects are simply born with even higher magnetic
fields of a few times $10^{15}$~G.

We need to make some more considerations about these extreme magnetars. First, we note from the
evolutionary tracks in the $P-\dot{P}$ diagram that this group does not appear to have older
descendants, in contrast to the first group of magnetars, which
evolves towards the XINSs and SGR~0418. The expected descendants
of the extreme magnetars should have periods of few tens of seconds and
would be bright enough to be seen. No selection effect in X-ray
observations would prevent to observe such sources.

Second, four of them
(SGR~1900, SGR~1806, 1E~1841 and 1RXS~J1708) show
a strong, non-thermal component in the hard X-ray band ($20-100$ keV),
whose contribution to the luminosity is up to $10^{36}$ erg/s \citep{kuiper04,kuiper06,gotz06}. Together with 4U~0142 and
1E~2259, they are the only magnetars showing this hard X-ray emission persistently (i.e., not
connected to outburst activity as in SGR~0501 and 1E~1547; \citealt{rea09,enoto10}). We stress again that the tails seen in soft and hard X-ray are ultimately due to the magnetospheric plasma, which can provide a significant amount of energy to the seed thermal spectrum emerging from the surface.

Third, in most of the objects for which the kinematic age is available, a reverse, unusual mismatch with characteristic age, $\tau_k>\tau_c$, is observed (see Table~\ref{tab:timing}). This implies that the current value of $\dot{P}$ is larger than in the past. Note also that SGR~1900 and SGR~1806 are the magnetars with the most variable timing properties (see Table~2 of McGill catalog).

This facts, i.e. the possible overestimate of both magnetic torque and thermal luminosity, and the absence of
descendants, could point to some additional torque, apart from the dipole braking, acting temporarily during the early stages.  
One possibility discussed in the literature is wind braking \citep{tong13}, which can be effective during some epochs of the NS
life, explaining the anomalous high values of $\dot{P}$ of some magnetars. Another compelling possibility, supported by the timing noise, is the contribution of magnetosphere. In such scenarios, this group of most extreme magnetars would be born with $B_p^0\lesssim 10^{15}$ G, but they would experience extra torque and luminosity due to temporary effects.

%%%%%%%%%%%%
\section{Summary}\label{sec:conclusion}

We have presented a comprehensive study of the magneto-thermal
evolution of isolated NSs, exploring the influence of their 
initial magnetic field strength and geometry, their mass, envelope
composition, and relevant microphysical parameters such as the
impurity content of the innermost part of the crust (the pasta
region). Our state-of-the-art, 2D magneto-thermal code, is the first
able to work with large magnetic field strengths, while
consistently including the Hall term throughout the full evolution. 
The Hall term plays a very important role in the overall magnetic field evolution, strongly enhancing the dissipation of energy over the first $\sim 10^6$~yr
of NS life, with respect to the purely resistive case. This is due to two main effects: the generation of smaller structures and currents sheets, and the gradual compression of currents and toroidal magnetic field towards the crust/core interface. Hence, the rate of magnetic field dissipation strongly depends on the resistivity given by the
amount of impurities in the innermost region of the crust. A highly impure or amorphous inner crust produces a significant increase in the magnetic field
dissipation on timescales $\gtrsim 10^5$~yr.

We also found that, while for weakly magnetized objects low mass stars ($M \lesssim 1.4\,M_\odot$) are
systematically brighter than high mass stars, this separation is smeared out
for highly magnetized stars.
We confirm that light-element envelopes are able to maintain a higher
luminosity (up to an order of magnitude) than iron envelopes for a long
period of time, $\gtrsim 10^4$~yr, regardless of the magnetic field strength. 

The initial magnetic field configuration plays an important role
in the observational properties of the NS. If the currents sustaining the magnetic field flow in the core, their dissipation is negliglible, comparable with models in which (most of) currents flow in the crust. In particular, the presence
of an initial strong dipolar, toroidal component in the crust breaks the symmetry with respect to the equator,
resulting in a warmer hemisphere, and a
characteristic hot ring replaces the traditional hot polar cap. If the magnetic
field is allowed to penetrate into the core (rather than remaining
confined within the crust), then the reduced heat deposition in the
crust results in a much cooler surface compared to the case in which
the magnetic field lives in the crust only. 

The estimated outburst rate, resulting from breaking of the crust by the
strong magnetic stresses, is found to be an increasing function of the
initial magnetic field strength and a decreasing function of age (in
agreement with the results of \citealt{perna11,pons11}).
While this qualitative trend is consistent with the data (younger and
more strongly magnetized objects tend to have more frequent outbursts),
a more quantitative comparison between the simulations and the observations
is not possible at this stage, due to the lack of sufficient statistics
in the data. 

For a given choice of the model parameters summarized above, our
simulations allow us to predict, for any age of the NS, the
distribution of surface temperature, luminosity, or timing
parameters. In order to compare theoretical models with observations,
we have studied all currently known isolated NSs with a detected
surface thermal emission. These objects are representative of
different observational classes, including the classical magnetars
(AXPs and SGRs), high-$B$ pulsars, standard radio pulsars, XINSs, and CCOs. The sample includes only the objects with clearly seen
thermal emission.  We have performed for all of them an homogeneous,
systematic analysis of the X-ray spectra, inferring the thermal luminosity, and trying to reduce at the minimum the systematic
effects due to different data analysis and modelling.  We published
on--line the results of our analysis, including detailed references
for every source$^1$.

A comparison between a range of theoretical models and the
observations (both timing and luminosities), has shown that, for the
weakly magnetised NSs ($B_p\lesssim 10^{14}$~G in the
$P-\dot{P}$ diagram), the magnetic field has little effect on the
luminosity. These objects, of which the RPPs are the most notable
representatives, have luminosities which are compatible with the ones
predicted by standard cooling models, with the dimmest ones in the
sample (Vela pulsar, PSR~B2334 and PSR~J1740) requiring an iron envelope and an NS
mass $\gtrsim 1.4 M_\odot$. 

The bulk of the magnetars, with inferred magnetic fields identified in the range
$B_p\sim$ a few $\times 10^{14}$~G in the $P-\dot{P}$~diagram,
displays luminosities generally too high to be compatible with
standard cooling alone. The magneto-thermal evolutionary models
with $B_p^0\sim 3-5 \times 10^{14}$~G can account for their
range of luminosities at the corresponding inferred ages. As these
objects evolve and their magnetic fields dissipate, their observational
properties (both timing and luminosities) appear compatible with those
of the XINSs.

Finally, the most extreme magnetars, endowed with magnetic fields of strength $\gtrsim
10^{15}$~G as inferred from their timing parameters, are characterized by
the highest luminosities, barely compatible with the $B_p^0=10^{15}$~G cooling
curve with an iron envelope. However, for the same initial magnetic
field, a light-element envelope is able to account for the luminosity of even
the brightest objects. Note that most of these extreme objects show hard X-ray, non-thermal emission and unreliable estimate of the dipolar magnetic fields from timing properties, due to the large timing noise. We suggest that, at least for some of them, the magnetosphere plays an important role in influencing timing properties and luminosity of these objects, whose magnetic field may be not so larger than the bulk of magnetars.

In summary, our magneto-thermal simulations have begun to paint a
unified picture of the variety of the observational properties of
isolated NSs, and their evolutionary paths. Our results can account
for the overall phenomenology, whereas in-depth testing of each
specific magnetic field configuration will require detailed spectral
and timing modeling of the thermal component of each source. This is
reserved to future work.

Further, deeper observations and discoveries of magnetars will allow
in the next years to refine the comparison with data and constrain the current uncertainties of the
theoretical models, such as the properties of the inner crust and the initial magnetic field geometry.

\section*{Acknowledgements}
This research was supported by the grants AYA 2010-21097-C03-02 (DV, JAP, JAM); AYA2009-07391, AYA2012-39303, SGR2009-811, and iLINK 2011-0303 (NR); CONICET and PIP-2011-00170 (DNA); NSF grant No. AST 1009396 and NASA grants AR1-12003X, DD1-12053X, GO2-13068X, GO2-13076X (RP). DV is supported by a fellowship from the \textit{Prometeo} program for research groups of excellence of the Generalitat Valenciana (Prometeo/2009/103) and NR is supported by a Ramon y Cajal Fellowship. DV thanks JILA (Boulder, CO, USA) and the MAP group (IEEC-CSIC, Barcelona, Spain) for their kind hospitality during the time that some of this work was carried out. 

\bibliography{magnetothermal}

\end{document}